\documentclass[a4paper,12pt]{article}
\usepackage{dcolumn}
\usepackage{bm}
\usepackage{amsmath,amssymb}
\usepackage{booktabs,subfigure}
\usepackage{graphicx,psfrag}
\usepackage{epsfig}
\usepackage{color} 
\usepackage{rotating}
\usepackage{axodraw}
\usepackage{slashed}
\usepackage{geometry}
\usepackage{rotating}
\usepackage{cite}
\allowdisplaybreaks[3]
\newcommand{\ie}{\ensuremath{\mathrm{i}}}

\newcommand{\cw}[1][{}]{\ensuremath{\cos^{#1} \theta_{w}}}
\newcommand{\sw}[1][{}]{\ensuremath{\sin^{#1} \theta_{w}}}
\newcommand{\tw}[1][{}]{\ensuremath{\tan^{#1} \theta_{w}}}

\newcommand{\vv}{\ensuremath{\bar{v}}}
\newcommand{\uu}{\ensuremath{\bar{u}}}
\newcommand{\real}{\ensuremath{\mathfrak{Re}}}
\newcommand{\imag}{\ensuremath{\mathfrak{Im}}}
\newcommand{\mink}[2]{\ensuremath{\,#1\! \cdot \! #2 \,}}

\newcommand{\dif}{\ensuremath{\mathrm{d}}}

\newcommand{\M}{\ensuremath{\mathcal{M}}}

\newcommand{\ssl}[1]{\ensuremath{\slashed{#1}}}
\newcommand{\eps}{\ensuremath{\epsilon_{\kappa\lambda\mu\nu}k_1^\kappa p_1^\lambda p_2^\mu q^\nu}}
\newcommand{\half}{\ensuremath{\frac{1}{2}}}

\newcommand{\signal}[1][{}]{\ensuremath{e^+e^- \rightarrow \tilde\chi^0_1 \tilde\chi^0_1 \gamma}#1}
\newcommand{\GeV}{\ensuremath{\enspace\mathrm{GeV}}}
\newcommand{\fb}{\ensuremath{\enspace\mathrm{fb}}}
\newcommand{\pb}{\ensuremath{\enspace\mathrm{pb}}}

 \numberwithin{equation}{section} 
\def\lsim{\raise0.3ex\hbox{$\;<$\kern-0.75em\raise-1.1ex\hbox{$\sim\;$}}}
\def\gsim{\raise0.3ex\hbox{$\;>$\kern-0.75em\raise-1.1ex\hbox{$\sim\;$}}}
\begin{document}

\title{\hfill \phantom{Hallo}\\[-30mm]
  \hfill\mbox{\small BONN-TH-2006-05}\\[-3mm]
  \hfill\mbox{\small hep-ph/0610020}\\[5mm]
  Discovery Potential of Radiative Neutralino Production at the ILC}
\author{{ Herbi K.~Dreiner, Olaf Kittel, Ulrich Langenfeld\vspace*{5mm}}\\
       {\it \small Physikalisches Institut der Universit\"at Bonn, 
        Nu{\ss}allee 12, 53115 Bonn, Germany}}
\date{}
\thispagestyle{myheadings}

\maketitle
 
\begin{abstract}
\noindent
We study radiative neutralino production $e^+e^- \to \tilde\chi^0_1
\tilde\chi^0_1\gamma$ at the linear collider with longitudinally
polarised beams.  We consider the Standard Model background from
radiative neutrino production $e^+e^- \to \nu \bar\nu \gamma$, and the
supersymmetric radiative production of sneutrinos $e^+e^- \to
\tilde\nu \tilde\nu^\ast \gamma$, which can be a background for
invisible sneutrino decays.  We give the complete tree-level formulas
for the amplitudes and matrix elements squared. In the Minimal
Supersymmetric Standard Model, we study the dependence of the cross
sections on the beam polarisations, on the parameters of the
neutralino sector, and on the selectron masses.  We show that
for bino-like neutralinos
longitudinal polarised beams enhance the signal and simultaneously
reduce the background, such that statistics is significantly enhanced.
We point out that there are parameter regions where
radiative neutralino production is 
the only channel to study SUSY particles, since heavier neutralinos, charginos 
and sleptons are too heavy to be pair-produced in the first stage of the 
linear collider with $\sqrt{s} =500\GeV$.

\end{abstract}

\section{Introduction}
\label{sec:intro}
Supersymmetry (SUSY) is an attractive concept for theories beyond the
Standard Model (SM) of particle physics. SUSY models like the Minimal
Supersymmetric Standard Model 
(MSSM)~\cite{Haber:1984rc,Gunion:1984yn,Nilles:1983ge}
predict SUSY partners of the SM particles with masses of the order of
a few hundred GeV.  Their discovery is one of the main goals of
present and future colliders in the TeV range.  In particular, the
international $e^+e^-$ linear collider (ILC) will be an {excellent}
tool to determine the parameters of the SUSY model with high
precision~\cite{Aguilar-Saavedra:2001rg,Abe:2001nn,Abe:2001gc,
Weiglein:2004hn,Aguilar-Saavedra:2005pw}.  Such a machine provides
high luminosity ${\mathcal L}=500\fb^{-1}$, a center-of-mass 
energy of $\sqrt s = 500\GeV$ in the first stage, and a polarised
electron beam with the option of a polarised positron
beam~\cite{Moortgat-Pick:2005cw}.

The neutralinos are the fermionic SUSY partners of the neutral gauge
and CP-even Higgs bosons.  Since they are among the lightest particles
in many SUSY models, they are expected to be also the first states to
be observed.  At the ILC, they can be directly produced in pairs
\begin{equation}
e^+ + e^-\to\tilde\chi_i^0 + \tilde\chi_j^0\,, 
\label{neut-pairs}
\end{equation}
which proceeds via $Z$ boson and selectron exchange~\cite{Bartl:1986hp,
  neutralino-pair}. At tree level, the neutralino sector depends only
on the four parameters $M_1$, $M_2$, $\mu$, and $\tan\beta$, which are the
$U(1)_Y$ and $SU(2)_L$ gaugino masses, the higgsino mass parameter,
and the ratio of the vacuum expectation values of the two Higgs
fields, respectively. These parameters can be determined by measuring
the neutralino production cross sections and decay
distributions~\cite{Weiglein:2004hn,Choi:2001ww,Choi:2005gt,
  Barger:1999tn,Kneur:1999nx}.  In the MSSM with R-parity (or proton
hexality, $P_6$, \cite{Dreiner:2005rd}) conservation, the lightest
neutralino $\tilde\chi_1^0$ is typically the lightest SUSY particle
(LSP) and as such is stable and a good dark matter 
candidate~\cite{Goldberg:1983nd, Ellis:1983ew}.
In collider experiments the LSP
escapes detection such that the direct production of the lightest
neutralino pair $e^+e^-\to\tilde\chi_1^0\tilde\chi_1^0$ is invisible.
Their pair production can only be observed indirectly via radiative
production $e^+e^-\to\tilde\chi_1^0\tilde\chi_1^0\gamma$, where the
photon is radiated off the incoming beams or off the exchanged
selectrons.  Although this higher order process is suppressed by the
square of the additional photon-electron coupling, it might be the
lightest state of SUSY particles to be observed at colliders.  The
signal is a single high energetic photon and missing energy, carried
by the neutralinos.

As a unique process to search for the first SUSY signatures at
$e^+e^-$ colliders, the radiative production of neutralinos has been
intensively studied in the literature~\cite{Fayet:1982ky,Ellis:1982zz,
  Grassie:1983kq,Kobayashi:1984wu,Ware:1984kq,Bento:1985in,
  Chen:1987ux,Kon:1987gi,Bayer,Choi:1999bs,Baer:2001ia,Weidner,
  Fraas:1991ky,Datta:1994ac,Datta:1996ur,Datta:2002jh,
  Ambrosanio:1995it}.\footnote{In addition we found two 
  references~\cite{Ahmadov:2006xr, Ahmadov:2005ci}, which are however 
  almost identical in wording and layout to Ref.~\cite{Fraas:1991ky}.}
Early investigations focus on LEP energies and discuss special
neutralino mixing scenarios only, in particular the pure photino
case~\cite{Fayet:1982ky,Ellis:1982zz,Grassie:1983kq,Kobayashi:1984wu,
  Ware:1984kq,Bento:1985in,Chen:1987ux,Kon:1987gi}.  More recent
studies assume general neutralino mixing~\cite{Bayer,Choi:1999bs,
  Baer:2001ia,Weidner,Fraas:1991ky,Datta:1994ac,Datta:1996ur,
  Datta:2002jh,Ambrosanio:1995it} and some of them underline the
importance of longitudinal~\cite{Bayer,Choi:1999bs,Baer:2001ia,
  Weidner} and even transverse beam polarisations~\cite{Bayer,
  Weidner}.  The transition amplitudes are given in a generic
factorised form~\cite{Choi:1999bs}, which allows the inclusion of
anomalous $WW\gamma$ couplings.  Cross sections are calculated with
the program {\tt CompHEP}~\cite{Baer:2001ia}, or in the helicity
formalism~\cite{Weidner}.
Some of the studies~\cite{Fraas:1991ky,Datta:1994ac,Datta:1996ur,
  Datta:2002jh,Ambrosanio:1995it} however do not include longitudinal
beam polarisations, which might be {essential} for measuring radiative
neutralino production at the ILC.  Special scenarios are considered,
where besides the sneutrinos also the heavier neutralinos~\cite{
  Datta:1994ac,Datta:1996ur,Datta:2002jh}, and even charginos~\cite{
  Chen:1995yu,Kane:1997yr,Datta:1998yw} decay invisibly or almost
invisibly.  However, a part of such unconventional signatures are by
now ruled out by LEP2 data~\cite{Datta:1994ac,Kane:1997yr,
  Abbiendi:2002vz}.  For the ILC, such ``effective'' LSP scenarios
have been analysed~\cite{Datta:1996ur}, and strategies for detecting
invisible decays of neutralinos and charginos have been
proposed~\cite{Chen:1995yu,Datta:1998yw}.
Moreover, the radiative production of neutralinos can serve as a direct 
test, whether neutralinos are dark matter candidates. See for example
Ref.~\cite{Birkedal:2004xn}, which presents a model independent calculation for 
the cross section of radiatively produced dark matter candidates at 
high-energy colliders, including polarised beams for the ILC.

The signature ``photon plus missing energy'' has been studied
intensively by the LEP collaborations ALEPH~\cite{Heister:2002ut},
DELPHI~\cite{Abdallah:2003np}, L3~\cite{Achard:2003tx}, and
OPAL~\cite{Abbiendi:2002vz,Abbiendi:2000hh}.  In the SM, $e^+e^- \to
\nu \bar\nu \gamma$ is the leading process with this signature.  Since
the cross section depends on the number $N_\nu$ of light neutrino
generations~\cite{Gaemers:1978fe}, it has been used to measure $N_\nu
$ consistent with three.  In addition, the LEP collaborations have
tested physics beyond the SM, like non-standard neutrino interactions,
extra dimensions, and SUSY particle productions.  However, no
deviations from SM predictions have been found, and only bounds on
SUSY particle masses have been set, e.g.  on the gravitino 
mass~\cite{Heister:2002ut,Abdallah:2003np,Achard:2003tx,Abbiendi:2000hh}. 
This process is also important in determining collider
bounds on a very light neutralino~\cite{lightneutralino}.  For a
combined short review, see for example Ref.~\cite{Gataullin:2003sy}.

Although there are so many theoretical studies on radiative neutralino
production in the literature, a thorough analysis of this process is
still missing in the light of the ILC with a high center-of-mass
energy, high luminosity, and longitudinally polarised beams. As noted
above, most of the existing analyses discuss SUSY scenarios with
parameters which are ruled out by LEP2 already, or without taking beam
polarisations into account. In particular, the question of the role of
the positron beam polarisation has to be addressed. If both beams are
polarised, the discovery potential of the ILC might be significantly
extended, especially if other SUSY states like heavier neutralino, chargino or
even slepton pairs are too heavy to be produced at the first stage of
the ILC at $\sqrt s = 500$~GeV.  Moreover the SM background photons
from radiative neutrino production $e^+e^- \to\nu\bar\nu\gamma $ have
to be included in an analysis with beam polarisations.  Proper beam
polarisations could enhance the signal photons and reduce those from
the SM background at the same time, which enhances the statistics
considerably. In this respect also the MSSM background photons from
radiative sneutrino production $e^+e^-
\to\tilde\nu\tilde\nu^\ast\gamma $ have to be discussed, if sneutrino
production is kinematically accessible and if the sneutrino decay is
invisible.

Finally the studies which analyse beam polarisations do not give
explicit formulas for the squared matrix elements, but only for the
transition amplitudes~\cite{Bayer,Choi:1999bs,Weidner}.  Other authors
admit sign errors~\cite{Datta:2002jh} in some interfering amplitudes
in precedent works~\cite{Datta:1996ur}, however do not provide the
corrected formulas for radiative neutrino and sneutrino production.
Additionally, we found inconsistencies and sign errors in the $Z$
exchange terms in some works~\cite{Bayer,Weidner}, which yield wrong
results for scenarios with dominating $Z$ exchange.  Thus we will give
the complete tree-level amplitudes and the squared matrix elements
including longitudinal beam polarisations, such that the formulas can
be used for further studies on radiative production of neutralinos,
neutrinos and sneutrinos.

In Sec.~\ref{sec:xsection}, we discuss our signal process, radiative
neutralino pair production, as well as the major SM and MSSM
background processes.  In Sec.~\ref{sec:results}, we define cuts on the
photon angle and energy, and define a statistical significance for
measuring an excess of photons from radiative neutralino production
over the backgrounds.  We analyse numerically the dependence of cross
sections and significances on the electron and positron beam
polarisations, on the parameters of the neutralino sector, and on the
selectron masses.  We summarise and conclude in
Sec.~\ref{sec:conclusion}.  In the Appendix we define neutralino
mixing and couplings, and give the tree-level amplitudes as well as
the squared matrix elements with longitudinal beam polarisations for
radiative production of neutralinos, neutrinos and sneutrinos.  In
addition we give details on the parametrisation of the phase space.

\section{Radiative Neutralino Production and Backgrounds}
  \label{sec:xsection}	

\subsection{Signal Process}
\begin{figure}[t!]
{%
\unitlength=1.0 pt
\SetScale{1.0}
\SetWidth{0.7}      
\scriptsize    
\allowbreak
\begin{picture}(95,79)(0,0)
\Text(15.0,70.0)[r]{$e^-$}
\ArrowLine(16.0,70.0)(58.0,70.0) 
\Text(80.0,70.0)[l]{$\gamma$}
\Photon(58.0,70.0)(79.0,70.0){1.0}{5} 
\Text(54.0,60.0)[r]{$e^-$}
\ArrowLine(58.0,70.0)(58.0,50.0) 
\Text(80.0,50.0)[l]{$\widetilde\chi^0_1$}
\Line(58.0,50.0)(79.0,50.0) 
\Text(54.0,40.0)[r]{$\widetilde{e}_R$}
\DashArrowLine(58.0,50.0)(58.0,30.0){1.0} 
\Text(15.0,30.0)[r]{$e^+$}
\ArrowLine(58.0,30.0)(16.0,30.0) 
\Text(80.0,30.0)[l]{$\widetilde\chi^0_1$}
\Line(58.0,30.0)(79.0,30.0) 
\Text(47,0)[b] {diagr. 1/4}
\end{picture} \ 
{} \qquad\allowbreak
\begin{picture}(95,79)(0,0)
\Text(15.0,70.0)[r]{$e^-$}
\ArrowLine(16.0,70.0)(58.0,70.0) 
\Text(80.0,70.0)[l]{$\widetilde\chi^0_1$}
\Line(58.0,70.0)(79.0,70.0) 
\Text(54.0,60.0)[r]{$\widetilde{e}_R$}
\DashArrowLine(58.0,70.0)(58.0,50.0){1.0} 
\Text(80.0,50.0)[l]{$\widetilde\chi^0_1$}
\Line(58.0,50.0)(79.0,50.0) 
\Text(54.0,40.0)[r]{$e^-$}
\ArrowLine(58.0,50.0)(58.0,30.0) 
\Text(15.0,30.0)[r]{$e^+$}
\ArrowLine(58.0,30.0)(16.0,30.0) 
\Text(80.0,30.0)[l]{$\gamma$}
\Photon(58.0,30.0)(79.0,30.0){1.0}{5} 
\Text(47,0)[b] {diagr. 2/5}
\end{picture} \ 
{} \qquad\allowbreak
\begin{picture}(95,79)(0,0)
\Text(15.0,70.0)[r]{$e^-$}
\ArrowLine(16.0,70.0)(58.0,70.0) 
\Text(80.0,70.0)[l]{$\widetilde\chi^0_1$}
\Line(58.0,70.0)(79.0,70.0) 
\Text(54.0,60.0)[r]{$\widetilde{e}_R$}
\DashArrowLine(58.0,70.0)(58.0,50.0){1.0} 
\Text(80.0,50.0)[l]{$\gamma$}
\Photon(58.0,50.0)(79.0,50.0){1.0}{5} 
\Text(54.0,40.0)[r]{$\widetilde{e}_R$}
\DashArrowLine(58.0,50.0)(58.0,30.0){1.0} 
\Text(15.0,30.0)[r]{$e^+$}
\ArrowLine(58.0,30.0)(16.0,30.0) 
\Text(80.0,30.0)[l]{$\widetilde\chi^0_1$}
\Line(58.0,30.0)(79.0,30.0) 
\Text(47,0)[b] {diagr. 3/6}
\end{picture} \ 
{} \qquad\allowbreak
\begin{picture}(95,79)(0,0)
\Text(15.0,60.0)[r]{$e^-$}
\ArrowLine(16.0,60.0)(37.0,60.0) 
\Photon(37.0,60.0)(58.0,60.0){1.0}{5} 
\Text(80.0,70.0)[l]{$\gamma$}
\Photon(58.0,60.0)(79.0,70.0){1.0}{5} 
\Text(33.0,50.0)[r]{$e^-$}
\ArrowLine(37.0,60.0)(37.0,40.0) 
\Text(15.0,40.0)[r]{$e^+$}
\ArrowLine(37.0,40.0)(16.0,40.0) 
\Text(47.0,41.0)[b]{$Z$}
\DashLine(37.0,40.0)(58.0,40.0){3.0} 
\Text(80.0,50.0)[l]{$\widetilde\chi^0_1$}
\Line(58.0,40.0)(79.0,50.0) 
\Text(80.0,30.0)[l]{$\widetilde\chi^0_1$}
\Line(58.0,40.0)(79.0,30.0) 
\Text(47,0)[b] {diagr. 7}
\end{picture} \ 
{} \qquad\allowbreak
\begin{picture}(95,79)(0,0)
\Text(15.0,60.0)[r]{$e^-$}
\ArrowLine(16.0,60.0)(37.0,60.0) 
\Text(47.0,61.0)[b]{$Z$}
\DashLine(37.0,60.0)(58.0,60.0){3.0} 
\Text(80.0,70.0)[l]{$\widetilde\chi^0_1$}
\Line(58.0,60.0)(79.0,70.0) 
\Text(80.0,50.0)[l]{$\widetilde\chi^0_1$}
\Line(58.0,60.0)(79.0,50.0) 
\Text(33.0,50.0)[r]{$e^-$}
\ArrowLine(37.0,60.0)(37.0,40.0) 
\Text(15.0,40.0)[r]{$e^+$}
\ArrowLine(37.0,40.0)(16.0,40.0) 
\Photon(37.0,40.0)(58.0,40.0){1.0}{5} 
\Text(80.0,30.0)[l]{$\gamma$}
\Photon(58.0,40.0)(79.0,30.0){1.0}{5} 
\Text(47,0)[b] {diagr. 8}
\end{picture} \ 
{} \qquad\allowbreak
\begin{picture}(95,79)(0,0)
\Text(15.0,70.0)[r]{$e^-$}
\ArrowLine(16.0,70.0)(58.0,70.0) 
\Text(80.0,70.0)[l]{$\gamma$}
\Photon(58.0,70.0)(79.0,70.0){1.0}{5} 
\Text(54.0,60.0)[r]{$e^-$}
\ArrowLine(58.0,70.0)(58.0,50.0) 
\Text(80.0,50.0)[l]{$\widetilde\chi^0_1$}
\Line(58.0,50.0)(79.0,50.0) 
\Text(54.0,40.0)[r]{$\widetilde{e}_L$}
\DashArrowLine(58.0,50.0)(58.0,30.0){1.0} 
\Text(15.0,30.0)[r]{$e^+$}
\ArrowLine(58.0,30.0)(16.0,30.0) 
\Text(80.0,30.0)[l]{$\widetilde\chi^0_1$}
\Line(58.0,30.0)(79.0,30.0) 
\Text(47,0)[b] {diagr. 9/12}
\end{picture} \ 
{} \qquad\allowbreak
\begin{picture}(95,79)(0,0)
\Text(15.0,70.0)[r]{$e^-$}
\ArrowLine(16.0,70.0)(58.0,70.0) 
\Text(80.0,70.0)[l]{$\widetilde\chi^0_1$}
\Line(58.0,70.0)(79.0,70.0) 
\Text(54.0,60.0)[r]{$\widetilde{e}_L$}
\DashArrowLine(58.0,70.0)(58.0,50.0){1.0} 
\Text(80.0,50.0)[l]{$\widetilde\chi^0_1$}
\Line(58.0,50.0)(79.0,50.0) 
\Text(54.0,40.0)[r]{$e^-$}
\ArrowLine(58.0,50.0)(58.0,30.0) 
\Text(15.0,30.0)[r]{$e^+$}
\ArrowLine(58.0,30.0)(16.0,30.0) 
\Text(80.0,30.0)[l]{$\gamma$}
\Photon(58.0,30.0)(79.0,30.0){1.0}{5} 
\Text(47,0)[b] {diagr. 10/13}
\end{picture} \ 
{} \qquad\allowbreak
\begin{picture}(95,79)(0,0)
\Text(15.0,70.0)[r]{$e^-$}
\ArrowLine(16.0,70.0)(58.0,70.0) 
\Text(80.0,70.0)[l]{$\widetilde\chi^0_1$}
\Line(58.0,70.0)(79.0,70.0) 
\Text(54.0,60.0)[r]{$\widetilde{e}_L$}
\DashArrowLine(58.0,70.0)(58.0,50.0){1.0} 
\Text(80.0,50.0)[l]{$\gamma$}
\Photon(58.0,50.0)(79.0,50.0){1.0}{5} 
\Text(54.0,40.0)[r]{$\widetilde{e}_L$}
\DashArrowLine(58.0,50.0)(58.0,30.0){1.0} 
\Text(15.0,30.0)[r]{$e^+$}
\ArrowLine(58.0,30.0)(16.0,30.0) 
\Text(80.0,30.0)[l]{$\widetilde\chi^0_1$}
\Line(58.0,30.0)(79.0,30.0) 
\Text(47,0)[b] {diagr. 11/14}
\end{picture} \ 
}
\caption{Diagrams for radiative neutralino production $e^+e^- \to
  \tilde\chi_1^0\tilde\chi_1^0\gamma$~\cite{Boos:2004kh}. For the
  calculation in Appendix~\ref{sec:app:chifore}, the first number of the
  diagrams labels $t$-channel, the second one $u$-channel exchange of
  selectrons, where the neutralinos are crossed.}
\label{fig:diagrams}
\end{figure}
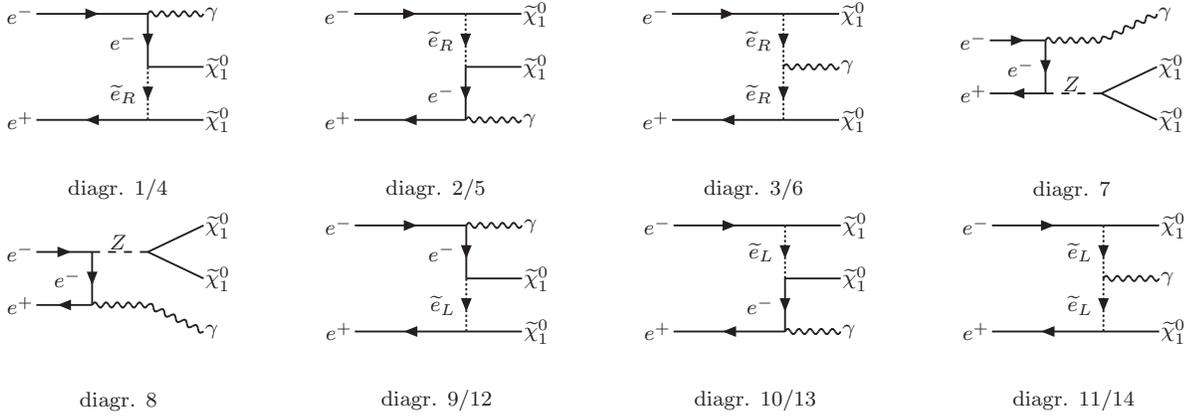

Within the MSSM, radiative neutralino production~\cite{Fayet:1982ky,
Ellis:1982zz,Grassie:1983kq,Kobayashi:1984wu,Ware:1984kq,Bento:1985in,
Chen:1987ux,Kon:1987gi,Bayer,Choi:1999bs,Baer:2001ia,Weidner,
Fraas:1991ky,Datta:1994ac,Datta:1996ur,Datta:2002jh,Ambrosanio:1995it}
\begin{equation}
e^++e^- \to \tilde\chi_1^0+\tilde\chi_1^0+\gamma
\label{productionChi}
\end{equation}
proceeds at tree-level via $t$- and $u$-channel exchange of right and
left selectrons $\tilde e_{R,L}$, as well as $Z$ boson exchange in the
$s$-channel. The photon is radiated off the incoming beams or the
exchanged selectrons; see the contributing diagrams in
Fig.~\ref{fig:diagrams}. We give the relevant Feynman rules for
general neutralino mixing, the tree-level amplitudes, and the complete
analytical formulas for the amplitude squared, including longitudinal
electron and positron beam polarisations, in
Appendix~\ref{sec:app:chifore}. We also summarise the details of the
neutralino mixing matrix there. For the calculation of cross sections
and distributions we use cuts, as defined in Eq.~(\ref{cuts}). An
example of the photon energy distribution and the $\sqrt s$ dependence
of the cross section is shown in Fig.~\ref{plotEdist}.

\subsection{Neutrino Background}
\noindent
Radiative neutrino production~\cite{Datta:1996ur,Gaemers:1978fe,Berends:1987zz,
Boudjema:1996qg,Montagna:1998ce}
\begin{equation}
e^+ +e^- \to \nu_\ell+\bar\nu_\ell+\gamma\,,\;\;\qquad \ell=e,\mu,\tau
\label{productionNu}
\end{equation}
is a major SM background. Electron neutrinos $\nu_e$ are produced via
$t$-channel $W$ boson exchange, and $\nu_{e,\mu,\tau}$ via $s$-channel
$Z$ boson exchange.  We show the corresponding diagrams in
Appendix~\ref{sec:app:nuback}, where we also give the tree-level
amplitudes and matrix elements squared including longitudinal beam
polarisations.

\subsection{MSSM Backgrounds}
Next we consider radiative sneutrino
production~\cite{Datta:1996ur,Franke:thesis, Franke:1994ph}
\begin{equation}
e^+ +e^- \to \tilde\nu_\ell+\tilde\nu^\ast_\ell+\gamma\,, 
\;\qquad \ell=e,\mu,\tau\,.
\label{productionSneut}
\end{equation}
We present the tree-level Feynman graphs as well as the amplitudes and
amplitudes squared, including beam polarisations, in Appendix~\ref
{sec:app:snuback}.  The process has $t$-channel contributions via
virtual charginos for $\tilde\nu_e\tilde\nu_e^\ast $-production, as
well as $s$-channel contributions from $Z$ boson exchange for
$\tilde\nu_{e, \mu,\tau}\tilde\nu_{e, \mu,\tau}^\ast $-production, see
Fig.~\ref{fig:sneutrino}.  Radiative sneutrino production,
Eq.~(\ref{productionSneut}), can be a major MSSM background to
neutralino production, Eq.~(\ref{productionChi}), if the sneutrinos
decay mainly invisibly, e.g., via $\tilde\nu\to\tilde \chi^0_1\nu$.
This leads to so called ``virtual LSP'' scenarios~\cite{Datta:1994ac,
  Datta:1996ur, Datta:2002jh}.  However, if kinematically allowed,
other visible decay channels like $\tilde\nu\to\tilde\chi^\pm_1\ell^
\mp$ reduce the background rate from radiative sneutrino production.
For example in the SPS~1a scenario~\cite{Ghodbane:2002kg,
  Allanach:2002nj} we have ${\rm  BR}(\tilde\nu_e\to\tilde
\chi_1^0\nu_e)=85\%$, see Table~\ref{scenarioSPS1}.

In principle, also neutralino production $e^+e^- \to \tilde\chi_1^0
\tilde\chi^0_2$ followed by the subsequent radiative neutralino
decay~\cite{Haber:1988px} $\tilde\chi^0_2 \to \tilde\chi^0_1 \gamma$
is a potential background.  However, significant branching ratios
${\rm BR}(\tilde\chi^0_2 \to \tilde\chi^0_1 \gamma)>10\%$ are only
obtained for small values of $\tan\beta<5$ and/or $M_1\sim
M_2$~\cite{Ambrosanio:1995it,Ambrosanio:1995az,Ambrosanio:1996gz}.
Thus we neglect this background in the following.  For details see
Refs.~\cite{Ambrosanio:1995az,Ambrosanio:1996gz,Baer:2002kv}.

\section{Numerical Results}
\label{sec:results}

We present numerical results for the tree-level cross section for
radiative neutralino production, Eq.~(\ref{productionChi}), and the
background from radiative neutrino and sneutrino production,
Eqs.~(\ref{productionNu}) and (\ref{productionSneut}), respectively.
We define the cuts on the photon energy and angle, and define the statistical
significance. 
We study the dependence of the cross sections and the significance
on the beam polarisations $P_{e^-}$ and $P_{e^+}$, the supersymmetric
parameters $\mu$ and $M_2$, and on the selectron masses. In order to
reduce the number of parameters, we assume the SUSY GUT relation
\begin{eqnarray}
\label{eq:gutrelation}
M_1 = \frac{5}{3}\tw[2]M_2\,.
\end{eqnarray}
Therefore the mass of the lightest neutralino is $m_{\chi^0_1}\gsim50\,
\mathrm{GeV}$~\cite{aleph}.  We also use the approximate renormalisation
group equations (RGE) for the slepton
masses~\cite{Ibanez:1983di,Ibanez:1984vq,Hall:1985zn},
\begin{eqnarray}
\label{sleptonR}
m_{\tilde e_R}^2 &=& m_0^2 +0.23 M_2^2-m_Z^2\cos 2 \beta \sw[2],\\
\label{sleptonL}
m_{\tilde e_L  }^2 &=& m_0^2 +0.79 M_2^2+
              m_Z^2\cos 2 \beta\Big(-\frac{1}{2}+ \sw[2]\Big),\\
m_{\tilde\nu_e  }^2 &=& m_0^2 +0.79 M_2^2+ \frac{1}{2}m_Z^2\cos 2 \beta,
\label{sneutrino}
\end{eqnarray}
with $m_0$ the common scalar mass parameter. 
Since in our scenarios the dependence on $\tan\beta$ is rather mild,
we fix $\tan\beta = 10$.

\begin{table}
\renewcommand{\arraystretch}{1.2}
\caption{Parameters and masses for  SPS~1a 
scenario~\cite{Ghodbane:2002kg,Allanach:2002nj}.}
\begin{center}
        \begin{tabular}{|c|c|c|c|}
\hline
$\tan\beta=10$ & $\mu= 352$~GeV &$M_2= 193$~GeV &$m_0=100$~GeV \\
\hline\hline
$m_{\chi^0_{1}}=94$~GeV & 
$m_{\chi^\pm_{1}}=178$~GeV & 
$m_{\tilde e_{R}}=143$~GeV & 
$m_{\tilde\nu_e}=188$ GeV \\
\hline
$m_{\chi^0_{2}}=178$~GeV & 
$m_{\chi^\pm_{2}}=376$~GeV & 
$m_{\tilde e_{L}}=204$~GeV &
${\rm BR}(\tilde\nu_e\to\tilde\chi_1^0\nu_e)=85\%$ \\
\hline
\end{tabular}
\end{center}
\renewcommand{\arraystretch}{1.0}
\label{scenarioSPS1}
\end{table}

\subsection{Cuts on Photon Angle and Energy}
\label{subsec:cuts}
To regularise the infrared and collinear divergencies of the
tree-level cross sections, we apply cuts on the photon scattering
angle $\theta_\gamma$ and on the photon energy $E_\gamma$
\begin{equation}
 -0.99 \le \cos\theta_\gamma \le 0.99,\quad \quad
0.02 \le x\le \;1-\frac{m_{\chi_1^0}^2}{E_{\rm beam}^2}, \quad \quad
x = \frac{E_\gamma}{E_{\rm beam}}, 
\label{cuts}
\end{equation}
with the beam energy $E_{\rm beam}=\sqrt{s}/2$.  The cut on the
scattering angle corresponds to $\theta_\gamma \in [8^\circ,172^\circ]
$, and reduces much of the background from radiative Bhabha
scattering, $e^+e^-\to e^+e^-\gamma$, where both leptons escape close
to the beam pipe~\cite{Abdallah:2003np,Heister:2002ut}.  The lower cut
on the photon energy is $E_\gamma= 5 \GeV$ for $\sqrt{s}=500\GeV$. The
upper cut on the photon energy $x^{\mathrm{max}}=1-m_{\chi_1^0}^2/E_
{\rm beam}^2$ is the kinematical limit of radiative neutralino
production. At $\sqrt{s}=500\GeV$ and for $m_{\chi_1^0}\gsim70\GeV$,
this cut reduces much of the on-shell $Z$ boson contribution to
radiative neutrino production, see Refs.~\cite{Baer:2001ia,
  Datta:1994ac,Franke:1994ph,Vest:2000ad} and Fig.~\ref{plotEdist}(a).
We assume that the neutralino mass $m_{\chi_1^0}$ is known from LHC or
ILC measurements~\cite{Weiglein:2004hn}.  If $m_{\chi_1^0}$ is
unknown, a fixed cut, e.g., $E_\gamma^{\rm max}=175$~GeV at $\sqrt{s}=500
\GeV$, could be used instead~\cite{Vest:2000ad}.

\subsection{Theoretical Significance}
In order to quantify whether an excess of signal photons from 
neutralino production, $N_{\mathrm{S}}=\sigma {\mathcal L}$, 
for a given integrated luminosity $\mathcal{L}$, can be measured
over the SM background photons, $N_{\rm B}=\sigma_{\rm B}{\mathcal L}$,
from radiative neutrino production,
we define the theoretical significance 
\begin{equation}
S  =  \frac{N_{\rm S}}{\sqrt{N_{\rm S} + N_{\rm B}}}=
\frac{\sigma}{\sqrt{\sigma + \sigma_{\rm B}}} \sqrt{\mathcal L}.
\label{significance}
\end{equation}
A theoretical significance of, e.g., $S = 1$ implies that the signal
can be measured at the statistical 68\% confidence level.  Also the
the signal to background ratio $N_{\mathrm{S}}/N_{\mathrm{B}}$ should
be considered to judge the reliability of the analysis.  For example,
if the background cross section is known experimentally to 1\%
accuracy, we should have $N_{\mathrm{S}}/N_{\mathrm{B}}>1/100$.

We will not include additional cuts on the missing mass or on the
transverse momentum distributions of the photons~\cite{Baer:2001ia,
  Vest:2000ad}. Detailed Monte Carlo analyses, including detector
simulations and particle identification and reconstruction
efficiencies, would be required to predict the significance more
accurately, which is however beyond the scope of the present work.
Also the effect of beamstrahlung should be included in such an
experimental analysis~\cite{Vest:2000ad,Ohl:1996fi,Hinze:2005xt}.
Beamstrahlung distorts the peak of the beam energy spectrum to lower
values of $E_{\rm beam}=\sqrt{s}/2$, and is more significant at
colliders with high luminosity.  In the processes we consider, the
cross sections for $e^+e^- \to \tilde\chi^0_1 \tilde\chi^0_1\gamma$
and $e^+e^- \to \nu \bar\nu \gamma$ depend significantly on the beam
energy only near threshold. In most of the parameter space we
consider, for $\sqrt{s}= 500\GeV$ the cross sections are nearly
constant, see for example Fig.~\ref{plotEdist}(b), so we expect that
the effect of beamstrahlung will be rather small. However, for $M_2,\,
\mu\gsim 300\,\mathrm{GeV}$, $\,\signal$ is the only SUSY production,
which is kinematically accessible, see Fig.~\ref{CrossSectionMuM2}.
In order to exactly determine the kinematic reach of the ILC
beamstrahlung must be taken into account.

\subsection{Energy Distribution and $\sqrt{s}$ Dependence}

\begin{figure}[t!]
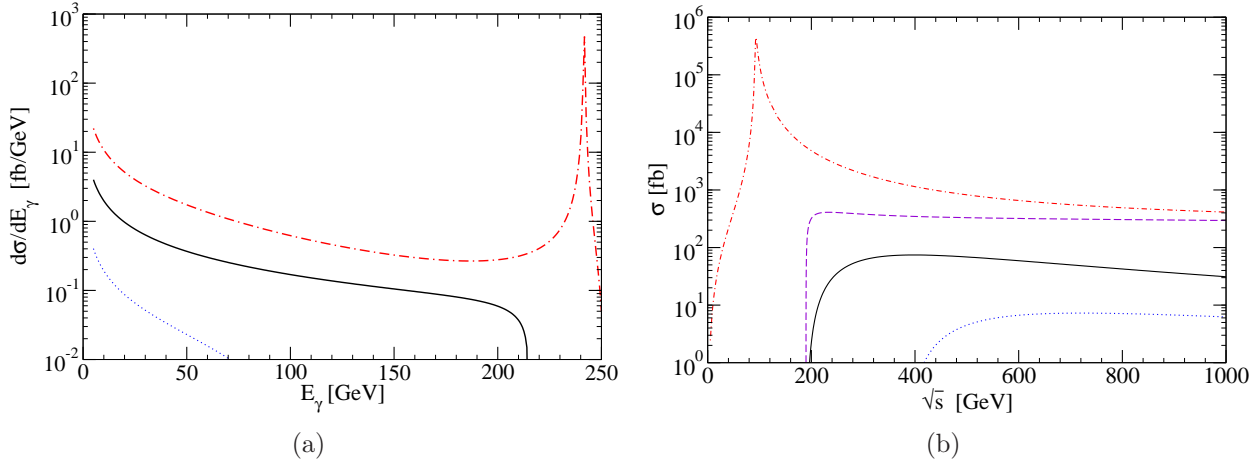

\setlength{\unitlength}{1cm}
\subfigure[
\label{fig:sps1adiff}]%
{\scalebox{0.3}{\includegraphics{sps1adiffnewcut.eps}}}
\hspace{1mm}
\subfigure[
\label{fig:sps1a}]{\scalebox{0.3}{\includegraphics{sps1anewcut.eps}}}
\caption{ (a) Photon energy distributions for $\sqrt s = 500\GeV$, 
        and (b) $\sqrt s$ dependence of the cross sections $\sigma$
        for radiative neutralino production $e^+e^- \to \tilde\chi^0_1
        \tilde\chi^0_1\gamma$ (black, solid), neutrino  production 
        $e^+e^- \to \nu\bar\nu\gamma$ (violet, dashed) and sneutrino  
        production $e^+ e^- \to \tilde\nu\tilde\nu^\ast\gamma$ (blue, dotted)
        for scenario  SPS~1a~\cite{Ghodbane:2002kg,Allanach:2002nj}, 
        see Table~\ref{scenarioSPS1}, with $(P_{e^-},P_{e^+})=(0.8,-0.6)$.
        The red dot-dashed line is in~(a) the photon energy distribution
        for radiative neutrino production $e^+e^- \to \nu\bar\nu\gamma$,
        and in~(b) the cross section without the upper cut
        on the photon energy $E_\gamma$, see Eq.~(\ref{cuts}).        
        \label{plotEdist}}
\end{figure}
In Fig.~\ref{plotEdist}(a) we show the energy distributions of the
photon from radiative neutralino production, neutrino production, and
sneutrino production for scenario SPS~1a~\cite{Ghodbane:2002kg,
  Allanach:2002nj}, see Table~\ref{scenarioSPS1}, with $\sqrt s =
500\GeV$, beam polarisations $(P_{e^-},P_{e^+})=(0.8,-0.6)$, and cuts
as defined in Eq.~(\ref{cuts}). The energy distribution of the photon
from neutrino production peaks at $E_\gamma= (s -m_Z^2)/(2\sqrt{s})
\approx242$~GeV due to radiative $Z$ return, which is possible for
$\sqrt s > m_Z$.  Much of this photon background from radiative
neutrino production can be reduced by the upper cut on the photon
energy $x^{\rm max}=E_\gamma^{\rm max}/E_{\rm beam}=1-m_{\chi_1^0}^2
/E_{\rm beam}^2$, see Eq.~(\ref{cuts}), which is the kinematical
endpoint $E_\gamma^\mathrm{max}\approx215$~GeV of the energy
distribution of the photon from radiative neutralino production, see
the solid line in Fig.~\ref{plotEdist}(a). Note that in principle the
neutralino mass could be determined by a measurement of this 
endpoint $E_\gamma^\mathrm{max} = E_\gamma^\mathrm{max}(m_{\chi^0_1})$
\begin{eqnarray}
 m_{\chi_1^0}^2 = \frac{1}{4}\left(s - 2\sqrt{s}E_\gamma^\mathrm{max}
\right).
\end{eqnarray}
For this one would need to be able to very well separate the signal
and background processes. This might be possible if the neutralino is
heavy enough, such that the endpoint is sufficiently removed from the
$Z^0$-peak of the background distribution.

In Fig.~\ref{plotEdist}(b) we show the $\sqrt s$ dependence of the
cross sections.  Without the upper cut on the photon energy $x^{\rm
  max}$, see Eq.~(\ref{cuts}), the background cross section from
radiative neutrino production $e^+e^- \to \nu\bar\nu\gamma$, see the
dot-dashed line in Fig.~\ref{plotEdist}(b), is much larger than the
corresponding cross section with the cut, see the dashed line.
However with the cut, the signal cross section from radiative
neutralino production, see the solid line, is then only about one
order of magnitude smaller than the background.

\subsection{Beam Polarisation Dependence
\label{beampoldep}}

\begin{figure}
\setlength{\unitlength}{1cm}
 \begin{picture}(20,20)(0,-2)
   \put(2.0,16.5){\fbox{$\sigma(e^+e^- \to \tilde\chi^0_1\tilde\chi^
    0_1\gamma)$ in fb}}
     \put(-4.5,-4.5){\includegraphics{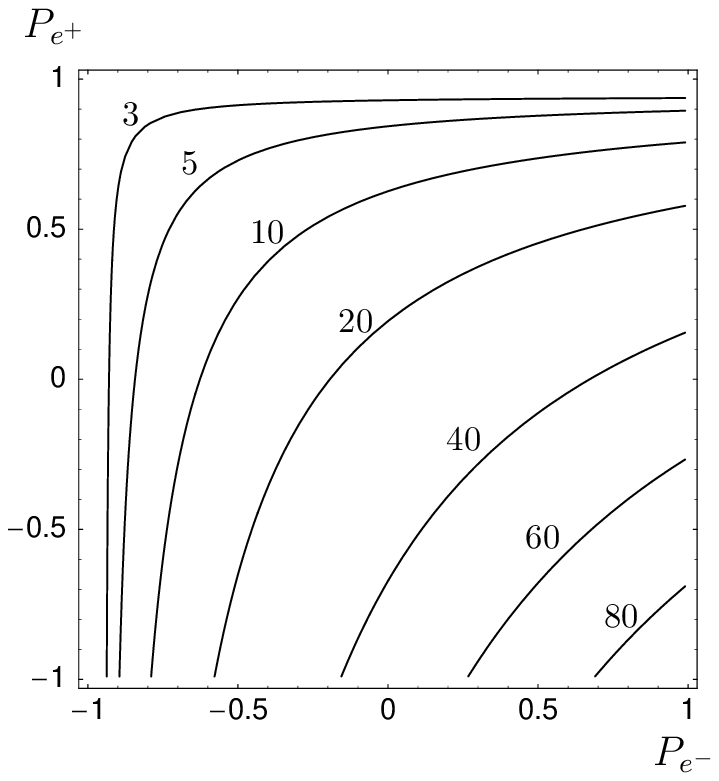}}
         \put(1.,8.8){(a)} 
   
        \put(3.5,-4.5){\includegraphics{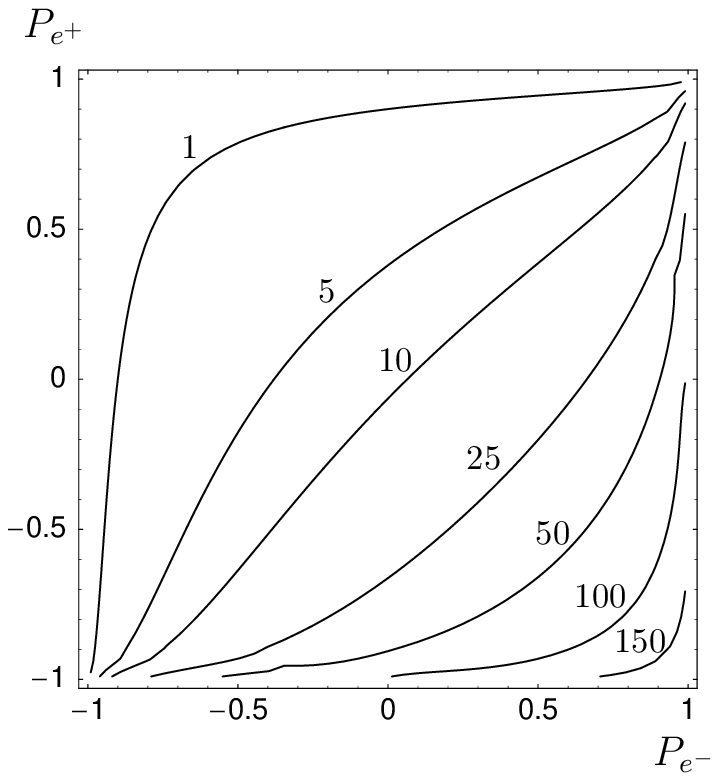}}
        \put(11.,16.5){\fbox{$S=\frac{\sigma}{\sqrt{\sigma+
       \sigma_{\rm B}}}\sqrt{\mathcal{L}} $} }
        \put(9.,8.8){(b)} 
        \put(-4.5,-13.5){\includegraphics{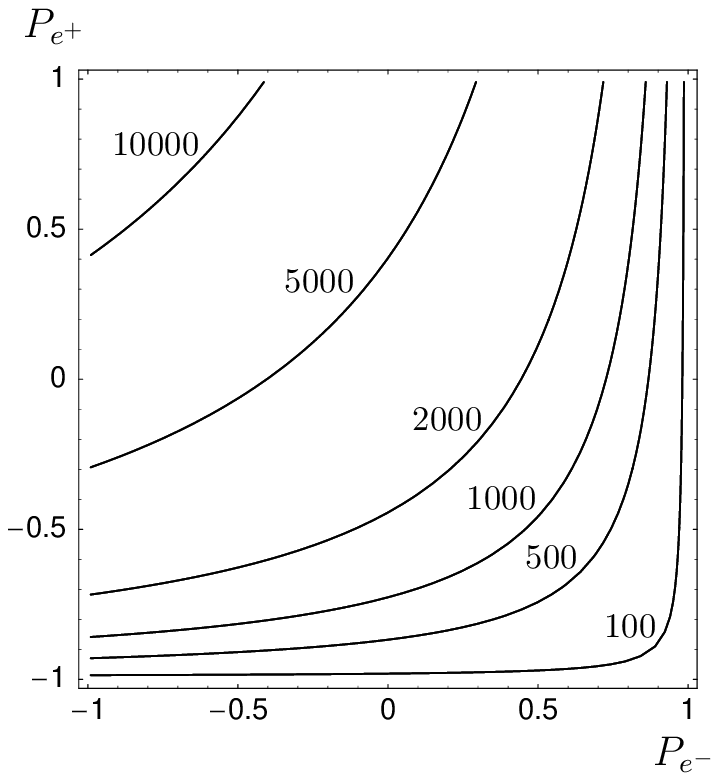}}
        \put(2.0,7.5){\fbox{$\sigma_{\rm B}(e^+e^-\to\nu\bar\nu
          \gamma)$ in fb }}
        \put(1.0,-0.3){(c)} 
        \put(3.5,-13.5){\includegraphics{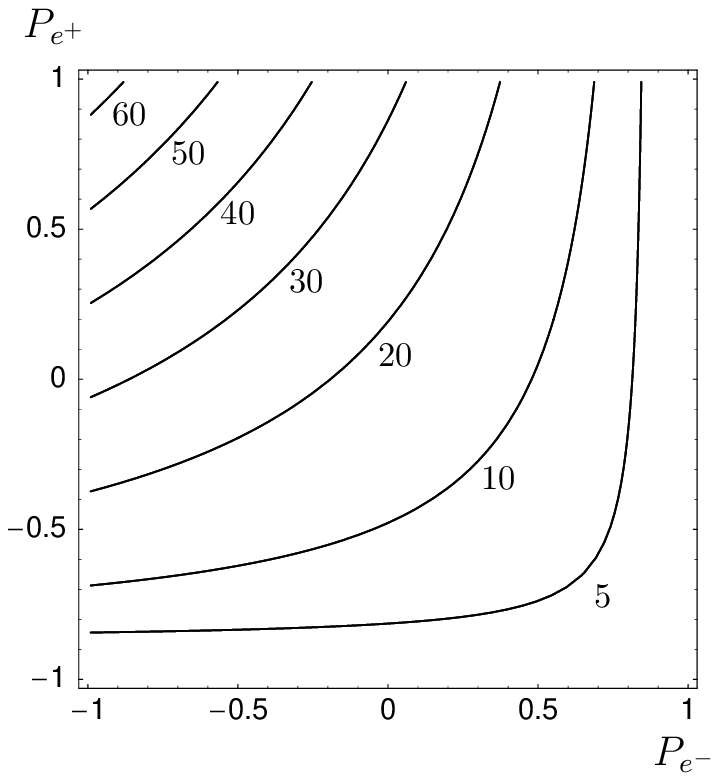}}
        \put(10.,7.5){\fbox{$\sigma(e^+e^-\to\tilde\nu\tilde
          \nu^\ast\gamma)$ in fb}}
          \put(9.0,-0.3){(d)} 
\end{picture}
\vspace*{-1.5cm}
\caption{%
        (a) Contour lines of the cross section and (b) the significance $S$
        for  $e^+e^- \to \tilde\chi^0_1\tilde\chi^0_1\gamma$ at $\sqrt{s}
        =500$~GeV and ${\mathcal{L}}=500~{\rm fb}^{-1}$ for scenario 
        SPS~1a~\cite{Ghodbane:2002kg,Allanach:2002nj}, see 
        Table~\ref{scenarioSPS1}. The beam polarisation dependence of 
        the cross section for radiative neutrino and sneutrino production are shown in~(c) 
        and~(d), respectively. 
        \label{varBeamPol}}
\end{figure}

%
In Fig.~\ref{varBeamPol}(a) we show the beam polarisation dependence
of the cross section $\sigma(e^+e^-\to\tilde\chi^0_1\tilde\chi^0_1
\gamma)$ for the SPS~1a scenario~\cite{Ghodbane:2002kg,
  Allanach:2002nj}, where radiative neutralino production proceeds
mainly via right selectron $\tilde e_R$ exchange.  Since the
neutralino is mostly bino, the coupling to the right selectron is more
than twice as large as to the left selectron.
Thus the contributions from right selectron exchange to the cross section are about
a factor 16 larger than the $\tilde{e}_L$ contributions.
In addition the $\tilde{e}_L$ contributions are suppressed compared 
to  the $\tilde{e}_R$ contributions by a factor of about 2 
since $m_{\tilde{e}_R }<m_{\tilde{e}_L} $, see 
Eqs.~(\ref{sleptonR})-(\ref{sleptonL}). The $Z$ boson exchange is
negligible. The background process, radiative neutrino production,
mainly proceeds via $W$ boson exchange, see the corresponding diagram
in Fig.~\ref{fig:neutrino}. Thus
positive electron beam polarisation $P_{e^-}$ and negative positron
beam polarisation $P_{e^+}$ enhance the signal cross section and
reduce the background at the same time, see Figs.~\ref{varBeamPol}(a)
and \ref{varBeamPol}(c), which was also observed in 
Refs.~\cite{Choi:1999bs,Vest:2000ad}. 
The positive electron beam polarisation
and negative positron beam polarisation 
enhance $\tilde e_R$ exchange and suppress $\tilde e_L$
exchange, such that it becomes negligible. Opposite
polarisations would lead to comparable contributions from both selectrons.
In going from unpolarised beams $(P_{e^-},P_{e^+})=(0,0)$ to polarised
beams, e.g., $(P_{e^-},P_{e^+})=(0.8,-0.6)$, the signal cross section
is enhanced by a factor $\approx 3$, and the background cross section
is reduced by a factor $\approx 10$.  The signal to background ratio
increases from $N_{\rm S}/N_{\rm B}\approx 0.007$ to $N_{\rm S}/N_{\rm
  B}\approx 0.2$, such that the statistical significance $S$, shown in
Fig.~\ref{varBeamPol}(b), is increased by a factor $\approx 8.5$ to
$S\approx 77$.  If only the electron beam is polarised,
$(P_{e^-},P_{e^+})=(0.8,0)$, we still have $N_{\rm S}/N_{\rm B}\approx
0.06$ and $S\approx 34$, thus the option of a polarised positron beam
at the ILC doubles the significance for radiative neutralino
production, but is not needed or essential to observe this process at
$\sqrt{s}=500\GeV$ and ${\mathcal{L}}=500\fb^{-1}$ for the SPS~1a
scenario.

The conclusion of Ref.~\cite{Baer:2001ia} is, however, that an almost
pure level of beam polarisations is needed at the ILC to observe this
process at all.  The authors have used a scenario with $M_1 = M_2$,
leading to a lightest neutralino, which is mostly a wino. Thus larger
couplings to the left selectron than to the right selectron are
obtained.  In such a scenario, one cannot simultaneously enhance the
signal and reduce the background.  Moreover their large selectron
masses $m_{\tilde e_{L,R}}=500\GeV$ lead to an additional suppression
of the signal, see also Sec.~\ref{selectronmassdependence}.

Finally we note that positive electron beam polarisation and negative
positron beam polarisation also suppress the cross section of
radiative sneutrino production, see Fig.~\ref{varBeamPol}(d).  Since
it is the corresponding SUSY process to radiative neutrino production,
we expect such a similar quantitative behaviour.

\subsection{$\mu$ \& $M_2$ Dependence}

%
\begin{figure}
\setlength{\unitlength}{1cm}
 \begin{picture}(20,20)(0,-2)
        \put(2.2,16.5){\fbox{$\sigma(e^+e^- \to \tilde\chi^0_1\tilde
 \chi^0_1\gamma)$ in fb}}
\put(-4.5,-4.5){\includegraphics{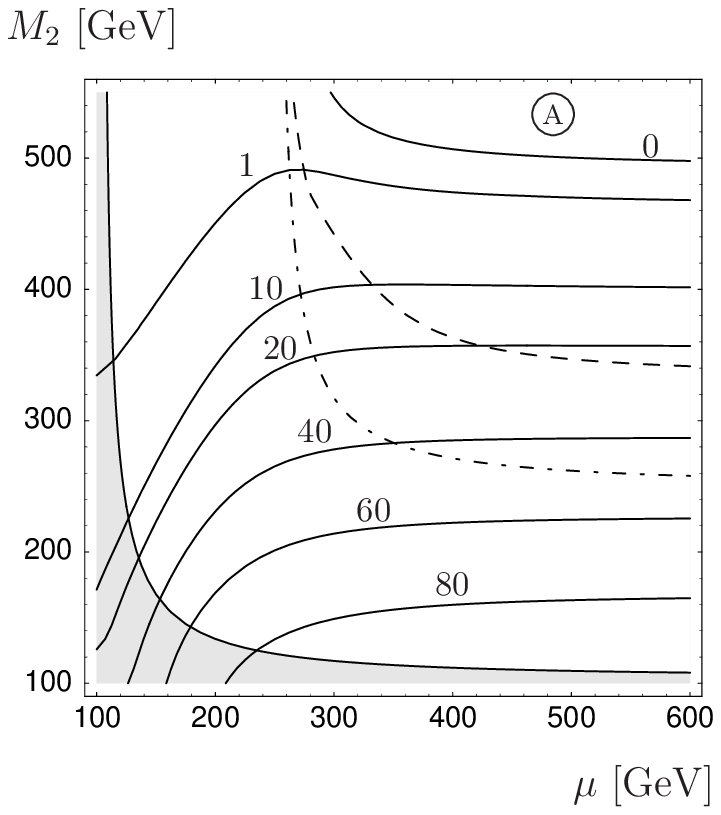}}
\put(1.,8.8){(a)}
   \put(3.5,-4.5){\includegraphics{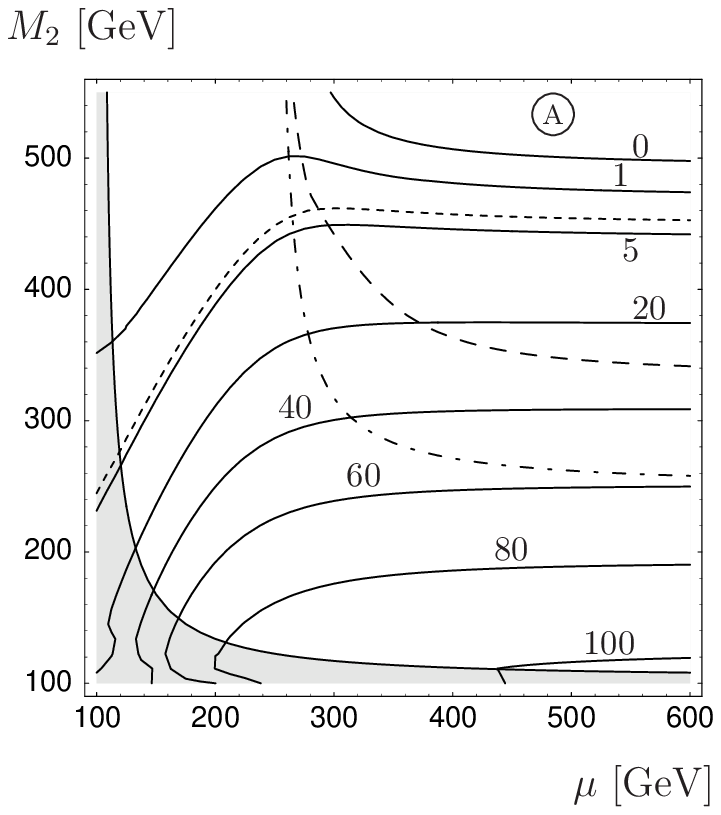}}
        \put(11.,16.5){\fbox{$S=\frac{\sigma}{\sqrt{\sigma+
\sigma_{\rm B}}}\sqrt{\mathcal{L}} $} }
\put(9.,8.8){(b)}
        \put(-4.5,-13.5){\includegraphics{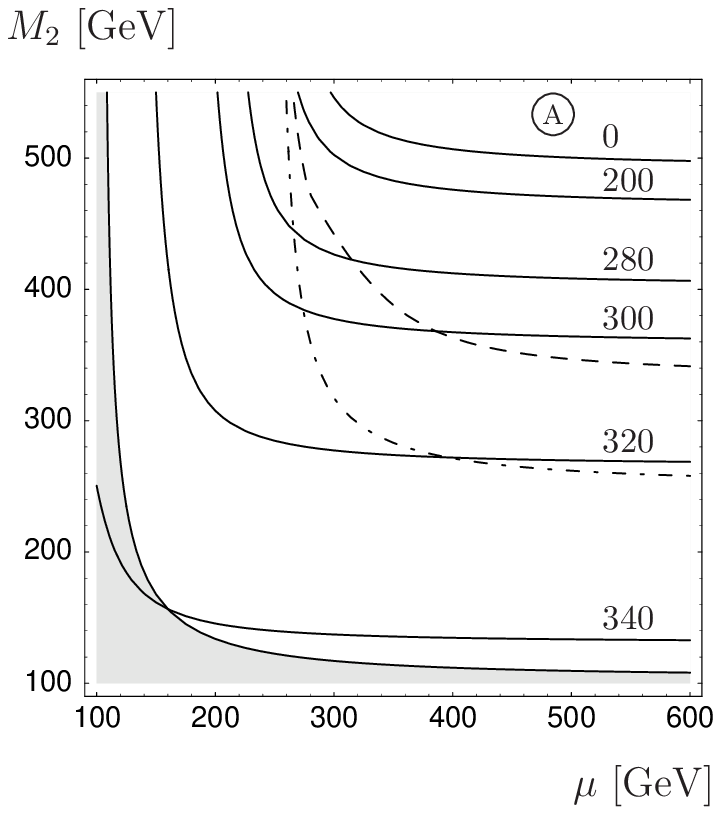}}
        \put(2.2,7.5){\fbox{$\sigma_{\rm B}(e^+e^-\to\nu\bar\nu
                      \gamma)$ in fb }}
\put(1.0,-0.3){(c)}
   \put(3.5,-13.5){\includegraphics{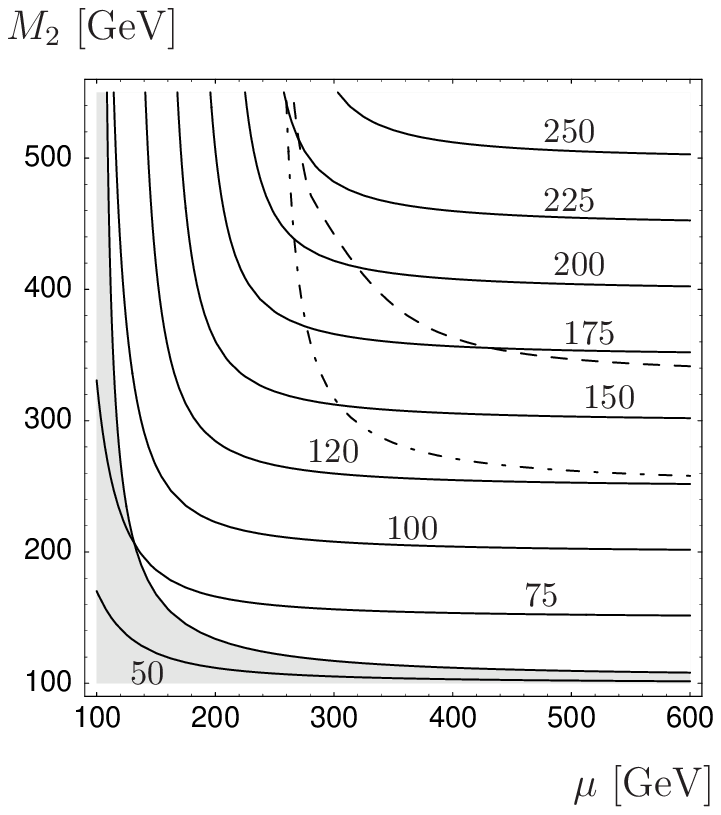}}
        \put(11.2,7.5){\fbox{$m_{\chi_1^0}$ in GeV}}
\put(9.0,-0.3){(d)}
\end{picture}
\vspace*{-1.5cm}
\caption{%
        Contour lines (solid) of (a) the cross section 
        $\sigma(e^+e^- \to \tilde\chi^0_1\tilde\chi^0_1\gamma)$, 
        (b) the significance $S$, (c) the neutrino background
        $\sigma_{\rm B}(e^+e^-\to\nu\bar\nu\gamma)$, and (d) the 
        neutralino mass $m_{\chi_1^0}$ in the $\mu$-$M_2$ plane
        for $\sqrt{s}=500$~GeV, $(P_{e^-},P_{e^+})=(0.8,-0.6)$,
        ${\mathcal{L}}=500~{\rm fb}^{-1}$, with $\tan\beta = 10$, 
        $m_0=100$~GeV, and RGEs for the selectron masses, see 
        Eqs.~(\ref{sleptonR}), (\ref{sleptonL}). The grey area is 
        excluded by $m_{\chi_1^\pm}<104$~GeV. The dashed line
        indicates the kinematical limit $m_{\chi_1^0}+m_{\chi_2^0} 
        =\sqrt{s}$, and the dot-dashed line the kinematical limit 
        $2 m_{\chi_1^\pm} =\sqrt{s}$. Along the dotted line in (b) 
        the signal to background ratio is $\sigma/\sigma_{\rm B}= 
        0.01$. The area A is kinematically forbidden by the cut on 
        the photon energy $E_\gamma$, see Eq.~(\ref{cuts}). 
        \label{CrossSectionMuM2}}
\end{figure}

In Fig.~\ref{CrossSectionMuM2}(a) we show contour lines of the cross
section $\sigma(e^+e^- \to \tilde\chi^0_1\tilde\chi^0_1\gamma)$ in fb
in the $(\mu,\,M_2)$-plane. For $\mu \gsim 300\GeV$ the signal and the
background cross sections are nearly independent of $\mu$, and
consequently also the significance, which is shown in
Fig.~\ref{CrossSectionMuM2}(b). In addition, the dependence of the
neutralino mass on $\mu$ is fairly weak for $\mu \gsim 300\GeV$, as
can be seen in Fig.~\ref{CrossSectionMuM2}(d). Also the couplings have a
rather mild $\mu$-dependence in this parameter region.
 
The cross section $\sigma_{\rm B}(e^+e^- \to \nu\bar\nu\gamma)$ of the
SM background process due to radiative neutrino production, shown in
Fig.~\ref{CrossSectionMuM2}(c), can reach more than $340\fb$ and is
considerably reduced due to the upper cut on the photon energy $x^{\rm
  max}$, see Eq.~(\ref{cuts}).  Without this cut we would have
$\sigma_{\rm B}= 825\fb$.  Thus the signal can be observed with high
statistical significance $S$, see Fig.~\ref{CrossSectionMuM2}(b).  Due
to the large integrated luminosity ${\mathcal{L}}=500~{\rm fb}^{-1}$
of the ILC, we have $S\gsim 25$ with $N_{\rm S}/N_{\rm B}\gsim 1/4$
for $M_2\lsim 350\GeV$.  For $\mu<0$ we get similar results for the
cross sections in shape and size, since the dependence of $N_{11}$
on the sign of $\mu$, see Eq.~(\ref{eq:coefficients}), is weak due to
the large value of $\tan\beta=10$.

In Fig.~\ref{CrossSectionMuM2}, we also indicate the kinematical
limits of the lightest observable associated neutralino production
process, $e^+e^- \to \tilde\chi^0_1\tilde\chi^0_2$ (dashed), and those
of the lightest chargino production process,
$e^+e^-\to\tilde\chi^+_1\tilde \chi^-_1$ (dot-dashed). In the region
above these lines $\mu,M_2\gsim 300\GeV$, heavier neutralinos and charginos
are too heavy to be pair-produced at the first stage of the ILC with
$\sqrt{s}=500\GeV$. In this case radiative neutralino production
$e^+e^-\to\tilde\chi^0_1\tilde\chi^0 _1\gamma$ will be the only
channel to study the gaugino sector. Here significances of $5<S\lsim
25$ can be obtained for $350~{\rm GeV}\lsim M_2 \lsim 450\GeV$, see
Fig.~\ref{CrossSectionMuM2}(b). Note that the production of right
sleptons $e^+e^- \to\tilde\ell^+_R\tilde\ell^-_R$, $\tilde\ell=\tilde
e,\tilde\mu$, and in particular the production of the lighter staus
$e^+e^ - \to \tilde\tau^+_1\tilde\tau^-_1$, due to mixing in the stau
sector~\cite{Bartl:2002bh}, are still open channels to study the
direct production of SUSY particles for $M_2\lsim 500\GeV$ in our GUT
scenario with $m_0=100\GeV$.

\subsection{Dependence on the Selectron Masses
       \label{selectronmassdependence}}

%
\begin{figure}
\setlength{\unitlength}{1cm}
 \begin{picture}(20,20)(0,-2)
        \put(2.2,16.5){\fbox{$\sigma(e^+e^- \to \tilde\chi^0_1\tilde
\chi^0_1\gamma)$ in fb}}
\put(-4.5,-4.5){\includegraphics{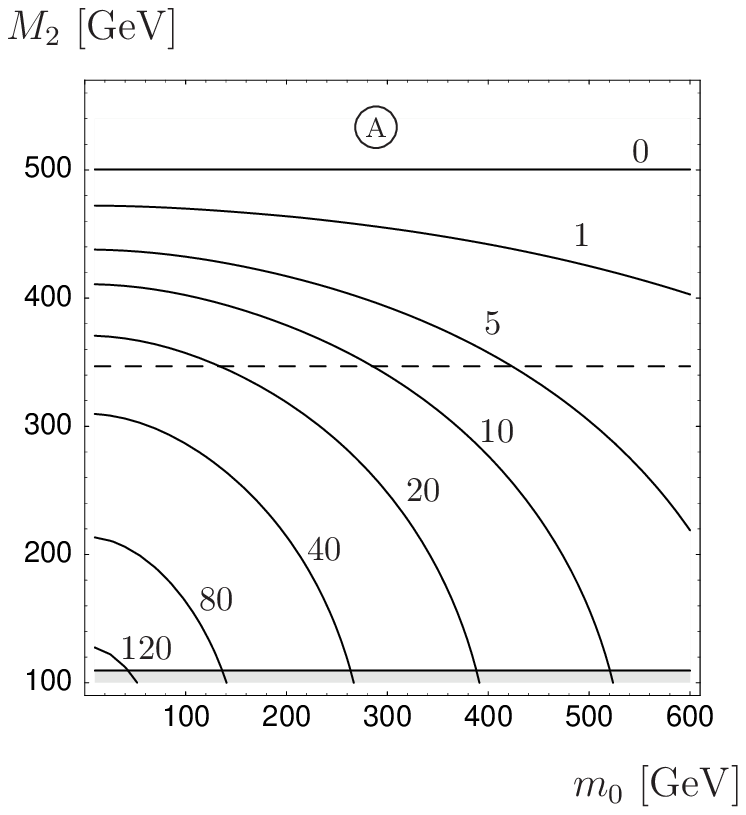}}
\put(1.,8.8){(a)}
   \put(3.5,-4.5){\includegraphics{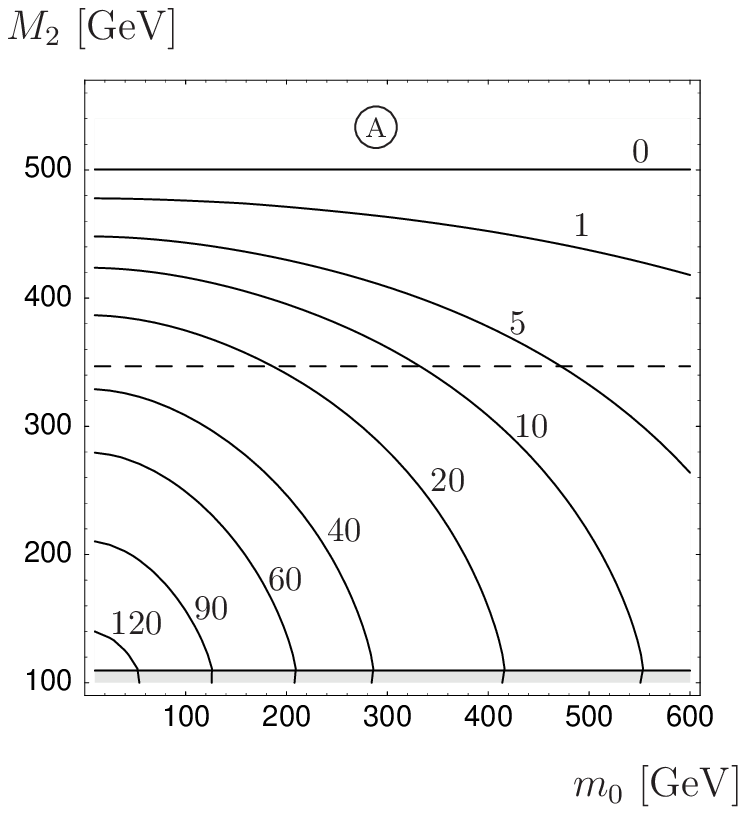}}
        \put(11.,16.5){\fbox{$S=\frac{\sigma}{\sqrt{\sigma+\sigma_{\rm B}}}\sqrt{\mathcal{L}} $} }
\put(9.,8.8){(b)}
        \put(-4.5,-13.5){\includegraphics{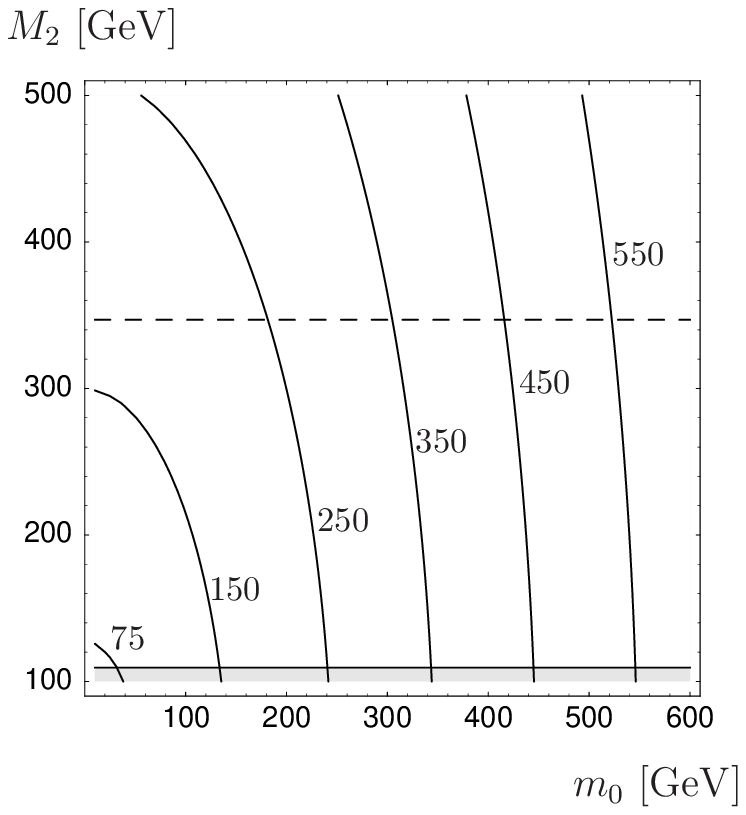}}
        \put(3.2,7.5){\fbox{$ m_{\tilde e_R}$ in GeV }}
\put(1.0,-0.3){(c)}
   \put(3.5,-13.5){\includegraphics{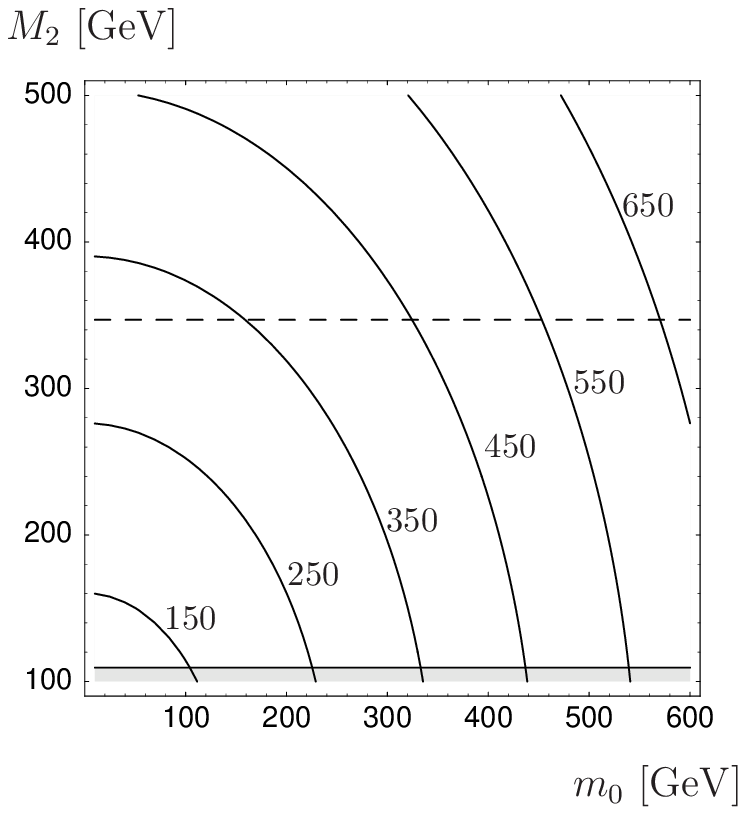}}
        \put(11.2,7.5){\fbox{$m_{\tilde e_L}$ in GeV}}
 \put(9.0,-0.3){(d)}
\end{picture}
\vspace*{-1.5cm}
\caption{%
        (a) Contour lines of the cross section $\sigma(e^+e^- \to 
        \tilde\chi^0_1\tilde\chi^0_1\gamma)$, (b) the significance $S$,
        and (c), (d) the selectron masses $ m_{\tilde e_R}$, $ m_{\tilde e_L}$,
        respectively, in the $m_0$-$M_2$ plane for $\sqrt{s}=500$~GeV, 
        $(P_{e^-},P_{e^+})=(0.8,-0.6)$, ${\mathcal{L}}=500~{\rm fb}^{-1}$, 
        with $\mu=500$~GeV, $\tan\beta = 10$, and RGEs for the selectron 
        masses, see Eqs.~(\ref{sleptonR}), (\ref{sleptonL}). The dashed line 
        indicates the kinematical limit $m_{\chi_1^0}+m_{\chi_2^0} =\sqrt{s}$.
        The grey area is excluded by $m_{\chi_1^\pm}<104$~GeV, the area A is 
        kinematically forbidden. 
\label{CrossSectionM0M2}}
\end{figure}

The cross section for radiative neutralino production $\sigma(e^+e^-
\to\tilde\chi^0_1\tilde\chi^0_1\gamma)$ proceeds mainly via selectron
$\tilde e_{R,L}$ exchange in the $t$ and $u$-channels. Besides the
beam polarisations, which enhance $\tilde e_{R}$ or $\tilde e_{L}$
exchange, the cross section is also very sensitive to the selectron
masses. In the mSUGRA universal supersymmetry breaking 
scenario~\cite{weldon}, the masses are parametrised by $m_0$ and $M_2$, 
besides $\tan\beta$, which enter the RGEs, see Eqs.~(\ref{sleptonR}) and
(\ref{sleptonL}).  We show the contour lines of the selectron masses
$\tilde e_{R,L}$ in the $m_0$-$ M_2$ plane in
Fig.~\ref{CrossSectionM0M2}(c) and \ref{CrossSectionM0M2}(d),
respectively. 
The selectron masses increase with increasing $m_0$ and $M_2$.

For the polarisations $(P_{e^-},P_{e^+})=(0.8,-0.6)$, 
the cross section is dominated by
$\tilde{e}_R$ exchange, as discussed in Sec.~\ref{beampoldep}. 
In Fig.~\ref{CrossSectionM0M2}(a) and~\ref{CrossSectionM0M2}(b) we
show the $m_0$ and $M_2$ dependence of the cross section and the
significance $S$, Eq.~(\ref{significance}). With increasing $m_0$ and
$M_2$ the cross
section and the significance decrease, due to  
the increasing mass of $\tilde{e}_R$.
In Fig.~\ref{CrossSectionMuM2}(d) we see that for $\mu\gsim 7/10\, M_2$, 
the neutralino mass $m_{\chi^0_1}$ is practically independent 
of $\mu$ and rises with $M_2$.
Thus for increasing $M_2$, and thereby increasing neutralino mass,
the cross section $\sigma(e^+e^- \to\tilde\chi^0_1\tilde\chi^0_1 \gamma)$
reaches the kinematical limit at $M_2\approx500$~GeV for $\sqrt s =500$~GeV.
A potential background from radiative sneutrino production is only
relevant for $M_2\lsim200\GeV$, $m_0\lsim 200\GeV$.  For larger values
the production is kinematically forbidden.

\begin{figure}[t]
\setlength{\unitlength}{1cm}
\begin{center}
\begin{picture}(10,8)(0,0)
   \put(-6.,-12.5){\includegraphics{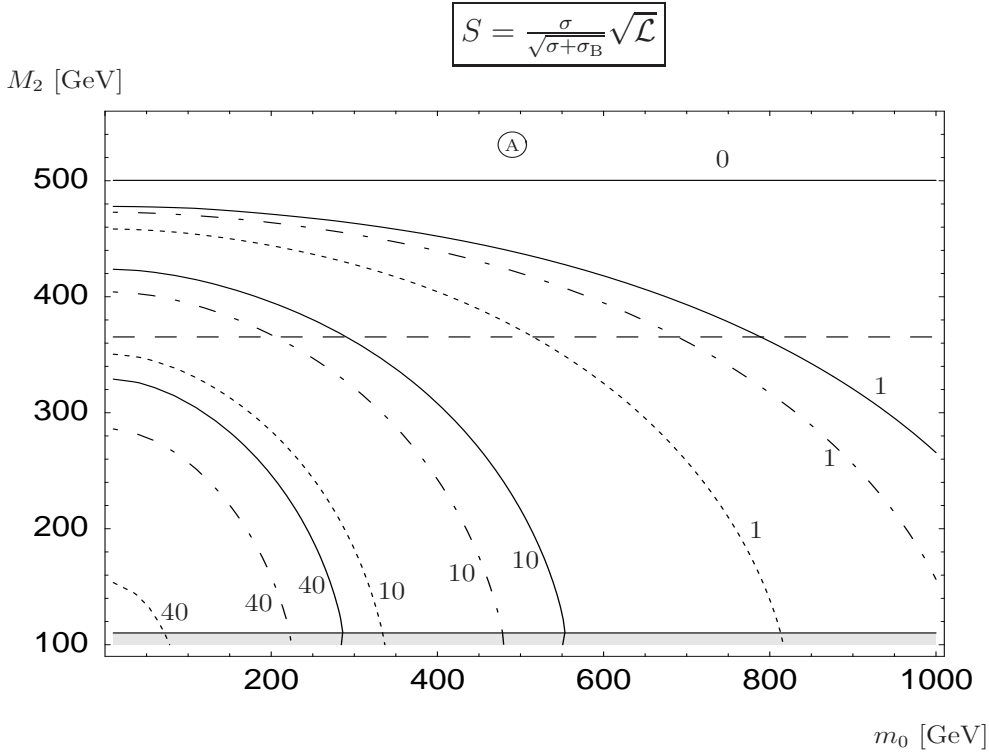}}
        \put(4.,7.4){\fbox{$S=\frac{\sigma}{\sqrt{\sigma+
        \sigma_{\rm B}}}\sqrt{\mathcal{L}} $} }
\end{picture}
\end{center}
\vspace*{1.5cm}
\caption{Contour lines of the significance $S$ 
        in the $m_0$-$M_2$ plane
        for different beam polarisations
        $(P_{e^-},P_{e^+})=(0.8,-0.6)$ (solid),
        $(P_{e^-},P_{e^+})=(0.8,0)$ (dot-dashed),
        and $(P_{e^-},P_{e^+})=(0,0)$ (dotted),
        for $\sqrt{s}=500$~GeV, ${\mathcal{L}}=500~{\rm fb}^{-1}$,
        $\mu=500$~GeV, $\tan\beta = 10$,
        and RGEs for the selectron masses, see Eqs.~(\ref{sleptonR}), (\ref{sleptonL}).
        The dashed line indicates the kinematical limit 
        $m_{\chi_1^0}+m_{\chi_2^0} =\sqrt{s}$.
        The grey area is excluded by $m_{\chi_1^\pm}<104$~GeV,
        the area A is kinematically forbidden.
        \label{plotDiffSig}}
\end{figure}

In Fig.~\ref{CrossSectionM0M2} we also indicate the kinematical limit
of associated neutralino pair production $m_{\chi_1^0}+m_{\chi_2^0}=
\sqrt{s}=500 $~GeV, reached for $M_2\approx350$~GeV. If in addition $
m_0>200$~GeV, also selectron and smuon pairs cannot be produced at $
\sqrt{s}=500 \GeV$ due to $m_{\tilde \ell_R}>250\GeV$. Thus, in this
parameter range $M_2>350$~GeV and $m_0> 200\GeV$, radiative
production of neutralinos will be the \textit{only} possible production
process of SUSY particles, if we neglect stau mixing.  
A statistical significance of $S>1$ can
be obtained for selectron masses not larger than $m_{\tilde e_R}
\approx 50 0$~GeV, corresponding to $m_0\lsim500$~GeV and $M_2\lsim450
$~GeV. Thus radiative neutralino production extends the discovery
potential of the ILC in the parameter range $m_0\in[200,500]$~GeV and
$M_2\in[350,450]$~GeV. Here, the beam polarisations will be essential,
see Fig.~\ref{plotDiffSig}. We show contour lines of the statistical
significance $S$ for three different sets of $(P_{e^-},P_{e^+})$.  The
first set has both beams polarised, $(P_{e^-},P_{e^+})=(0.8,-0.6)$,
the second one has only electron beam polarisation, $(P_{e^-},P_{e^+})
=(0.8,0)$, and the third has zero beam polarisations $(P_{e^-},P_{e^+}
)=(0,0)$.  We observe that beam polarisations significantly enhance
the discovery potential of the ILC. At least electron polarisation
$P_{e^-}=0.8$ is needed to extend an exploration of the $m_0$-$M_2$
parameter space.

%

%

\subsection{Note on LEP2 \label{sec:collider}}

We have also calculated the unpolarised cross sections and the
significances for radiative neutralino production at LEP2 energies $
\sqrt{s} = 200\enspace \mathrm{GeV}$, for a luminosity of $\mathcal{L
}=100\pb^{-1}$. We have used the cuts $|\cos\theta_\gamma|\le0.95$ and
$0.2 \le x\le1-m_{\chi_1^0}^2/E_{\rm beam}^2$, cf.  Eq.~(\ref{cuts}).
Even for rather small selectron masses $m_{\tilde e_{R,L}}=80\GeV$,
the cross sections are not larger than $100\fb$. If we alter the GUT
relation, Eq.~(\ref{eq:gutrelation}), to $M_1 = r\,M_2$, and vary $r$,
we only obtain statistical significances of $S<0.2$. These values
have also been reported by other theoretical studies at LEP2 energies,
see for example Ref.~\cite{Ambrosanio:1995it}.

If we drop the GUT relation, $M_1$ is a free parameter. For
\begin{eqnarray}
M_1 = \frac{M_2 m_Z^2 \sin(2\beta)\sw[2]}{\mu M_2 - m_Z^2\sin(2\beta)\cw[2]}
\end{eqnarray}
the neutralino is massless~\cite{Gogoladze:2002xp} at tree-level and
is apparently experimentally allowed~\cite{lightneutralino}. A
massless neutralino should enlarge the cross section for radiative
neutralino production due to the larger phase space, although the
coupling is also modified to almost pure bino. However, we still find
$S=\mathcal {O}( 10^{-1})$ at most. This is in accordance with the
experimental SUSY searches in photon events with missing energy at
LEP~\cite{Abbiendi:2002vz,Heister:2002ut,Abdallah:2003np,Achard:2003tx,Abbiendi:2000hh},
where no evidence of SUSY particles was found.

\section{Summary and Conclusions \label{sec:conclusion}}

We have studied radiative neutralino production $e^+e^- \to
\tilde\chi^0_1 \tilde\chi^0_1\gamma$ at the linear collider with
polarised beams.  We have considered the Standard Model background
process $e^+e^- \to \nu \bar\nu \gamma$ and the SUSY background
$e^+e^- \to \tilde\nu \tilde\nu^\ast \gamma$, which also has the
signature of a high energetic photon and missing energy, if the
sneutrinos decay invisibly. For these processes we have given the
complete tree-level amplitudes and the full squared matrix elements
including longitudinal polarisations from the electron and positron
beam.  In the MSSM, we have studied the dependence of the cross
sections on the beam polarisations, on the gaugino and higgsino mass
parameters $M_2$ and $\mu$, as well as the dependence on the selectron
masses.  Finally, in order to quantify whether an excess of signal
photons, $N_{\mathrm{S}}$, can be measured over the background
photons, $N_{\rm B}$, from radiative neutrino production, we have
analysed the theoretical statistical significance $S=N_{\rm S}/\sqrt{
N_{\rm S} + N_{\rm B}}$ and the signal to background ratio $N_{\rm S}
/ N_{\rm B}$.  Our results can be summarised as follows.
 
\begin{itemize}
\item The cross section for $e^+e^- \to \tilde\chi^0_1 \tilde\chi^0_1
\gamma$ reaches up to $100\fb$ in the $\mu$-$M_2$ and the $m_0$-$M_2$ 
plane at $\sqrt{s} = 500\GeV$. The significance can be as large as
$120$, for a luminosity of $\mathcal{L} = 500\fb^{-1}$, such that
radiative neutralino production should be well accessible at the ILC.

\item At the ILC, electron and positron beam polarisations can be used
to significantly enhance the signal and suppress the background
simultaneously.  We have shown that the significance can then be
increased almost by an order of magnitude, e.g., with $(P_{e^-},P_{e^+}
)=(0.8,-0.6)$ compared to $(P_{e^-},P_{e^+})=(0,0)$.  In the SPS~1a 
scenario the cross section $\sigma(\signal)$ increases from $25\fb$ to
$70\fb$ with polarised beams, whereas the background $\sigma(e^+e^-\to
\nu \bar\nu\gamma)$ is reduced from $3600\fb$ to $330\fb$.  Although a
polarised positron beam is not essential to study radiative neutralino 
production at the ILC, it will help to increase statistics.
\item We note that charginos and heavier neutralinos could be too heavy to be
pair-produced at the ILC in the first stage at $\sqrt{s} = 500$~GeV.
If only slepton pairs are accessible, the radiative production of the lightest
neutralino might be the only SUSY process to study the neutralino
sector.  Even in the regions of the parameter space near the
kinematical limits of $\tilde{\chi}^0_1$ - $\tilde{\chi}^0_2$ pair production 
we find a cross section of about $20\fb$ and
corresponding significances up to $20$.
\item Finally we want to remark that our given values for the 
statistical significance $S$ can only be seen as rough estimates,
since we do not include a detector simulation. However, since we have
obtained large values up to $S\approx 120$, we hope that our results
encourage further experimental studies, including detailed Monte Carlo
simulations.

\end{itemize}

\section*{Acknowledgments}
\noindent
We thank I.~Brock, G.~Moortgat-Pick, M.~Schumacher, X. Tata and G.~Weiglein for
enlightening discussions. We thank H.~Fraas for reading the manuscript and
valuable comments.  One of us (H.K.D.) would like to thank the Aspen Center for
Physics for hospitality while part of this work was completed. A
preliminary version of this work was presented there with valuable
ensuing discussions.

\begin{appendix}
\section{Radiative Neutralino Production}
\label{sec:app:chifore}
\subsection{Neutralino Mixing Matrix}
\label{subsec:mixing}
In the bino, wino, higgsino basis $(\tilde{B},\tilde{W}^0_3,\tilde{H}
_u, \tilde{H}_d)$, the neutralino mass matrix is given by~\cite{Gunion:1984yn,Bartl:1989ms}
\begin{eqnarray}
\label{eq:chimix}
M = 
\begin{pmatrix}
M_1 & 0   & - m_Z \sw \cos\beta & \phantom{-}m_Z\sw \sin\beta \\
0   & M_2 & \phantom{-} m_Z \cw \cos\beta  & -m_Z \cw\sin\beta \\
 - m_Z \sw \cos\beta &\phantom{-} m_Z \cw \cos\beta  & 0 & -\mu\\
\phantom{-}m_Z\sw \sin\beta& -m_Z \cw\sin\beta & -\mu & 0   
\end{pmatrix}, 
\end{eqnarray}
with $M_1$ and $M_2$ the $U(1)_Y$ and the $SU(2)_w$ gaugino mass
parameters, respectively, $\mu$ the higgsino mass parameter,
$\tan\beta = \frac{v_2}{v_1}$ the ratio of the two vacuum expectation
values of the Higgs fields, $m_Z$ the $Z$ boson mass, and $\tw$ the
weak mixing angle. In the CP conserving case, $M$ is real symmetric
and can be diagonalised by an orthogonal matrix. Since at least one
eigenvalue of $M$ is negative, we can use a unitary matrix $N$ to get
a positive semidefinite diagonal matrix with the neutralino masses
$m_{\chi_i^0}$~\cite{Haber:1984rc}
\begin{eqnarray}
\label{eq:chidiag}
\mathrm{diag}\begin{pmatrix}m_{\chi_1^0}, & m_{\chi_2^0}, & m_{\chi_3^0}, 
& m_{\chi_4^0} \end{pmatrix} = N^\ast M N^{-1}.
\end{eqnarray}
Note that the transformation Eq.~(\ref{eq:chidiag}) is only a
similarity transformation if $N$ is real.
\subsection{Lagrangian and Couplings}
\begin{table}[t!]
\begin{center}
\caption{Vertex factors with parameters $a$, $b$, $c$, $d$, $f$, and
$g$ defined in Eqs.~(\ref{eq:coefficients}), (\ref{eq:Zcoefficients}), with $e>0.$}
\vspace*{5mm}
\begin{tabular}{cl}
\toprule
\vspace{2mm}
Vertex & Factor
\vspace*{1mm}\\
\midrule
\hspace{-10mm}
{%
\unitlength=1.0 pt
\SetScale{1.0}
\SetWidth{0.7}      
\scriptsize    
{} \qquad\allowbreak
\begin{picture}(95,79)(0,0)
\Text(15.0,60.0)[r]{$\tilde e_R$}
\DashArrowLine(16.0,60.0)(58.0,60.0){1.0} 
\Text(80.0,70.0)[l]{$\tilde\chi^0_1$}
\Line(58.0,60.0)(79.0,70.0) 
\Text(80.0,50.0)[l]{$e$}
\ArrowLine(58.0,60.0)(79.0,50.0) 
\end{picture} \ 
}
& \raisebox{2cm}{$\ie e \sqrt{2} a P_L$}\\[-10mm]
\hspace{-10mm}
{%
\unitlength=1.0 pt
\SetScale{1.0}
\SetWidth{0.7}      
\scriptsize    
{} \qquad\allowbreak
\begin{picture}(95,79)(0,0)
\Text(15.0,60.0)[r]{$\tilde{e}_L$}
\DashArrowLine(16.0,60.0)(58.0,60.0){1.0} 
\Text(80.0,70.0)[l]{$\tilde{\chi}^0_1$}
\Line(58.0,60.0)(79.0,70.0) 
\Text(80.0,50.0)[l]{$e$}
\ArrowLine(58.0,60.0)(79.0,50.0) 
\end{picture} \ 
}
&\raisebox{2cm}{$\ie e \sqrt{2} b P_R$}\\[-10mm]
\hspace{-10mm}
{%
\unitlength=1.0 pt
\SetScale{1.0}
\SetWidth{0.7}      
\scriptsize    
{} \qquad\allowbreak
\begin{picture}(95,79)(0,0)
\Text(15.0,60.0)[r]{$\gamma$}
\Photon(16.0,60.0)(58.0,60.0){1.0}{5} 
\Text(80.0,70.0)[l]{$e$}
\ArrowLine(58.0,60.0)(79.0,70.0) 
\Text(80.0,50.0)[l]{$e$}
\ArrowLine(79.0,50.0)(58.0,60.0) 
\end{picture} \ 
}
&\raisebox{2cm}{$ \ie e \gamma^\mu$}\\[-10mm]
\hspace{-10mm}
{%
\unitlength=1.0 pt
\SetScale{1.0}
\SetWidth{0.7}      
\scriptsize    
{} \qquad\allowbreak
\begin{picture}(95,79)(0,0)
\Text(15.0,60.0)[r]{$\gamma$}
\Photon(16.0,60.0)(58.0,60.0){1.0}{5} 
\Text(80.0,70.0)[l]{$\tilde{e}_{L,R}$}
\DashArrowLine(58.0,60.0)(79.0,70.0){1.0} 
\Text(80.0,50.0)[l]{$\tilde{e}_{L,R}^\ast$}
\DashArrowLine(79.0,50.0)(58.0,60.0){1.0} 
\Text(68,71)[c]{\rotatebox{30}{$\rightarrow p_1$}}
\Text(68,49)[c]{\rotatebox{-30}{$\leftarrow p_2$}}
\end{picture} \ 
}
&\raisebox{2cm}{$ \ie e (p_1 + p_2)^\mu$}\\[-10mm]
\hspace{-10mm}
{%
\unitlength=1.0 pt
\SetScale{1.0}
\SetWidth{0.7}      
\scriptsize    
{} \qquad\allowbreak
\begin{picture}(95,79)(0,0)
\Text(15.0,60.0)[r]{$Z$}
\DashLine(16.0,60.0)(58.0,60.0){3.0} 
\Text(80.0,70.0)[l]{$e$}
\ArrowLine(58.0,60.0)(79.0,70.0) 
\Text(80.0,50.0)[l]{$e$}
\ArrowLine(79.0,50.0)(58.0,60.0) 
\end{picture} \ 
}
&\raisebox{2cm}{$ \ie {e}\gamma^\mu\left(c P_L + d P_R\right)$}\\[-10mm]
\hspace{-10mm}
{%
\unitlength=1.0 pt
\SetScale{1.0}
\SetWidth{0.7}      
\scriptsize    
{} \qquad\allowbreak
\begin{picture}(95,79)(0,0)
\Text(15.0,60.0)[r]{$Z$}
\DashLine(16.0,60.0)(58.0,60.0){3.0} 
\Text(80.0,70.0)[l]{$\tilde{\chi}^0_1$}
\Line(58.0,60.0)(79.0,70.0) 
\Text(80.0,50.0)[l]{$\tilde{\chi}^0_1$}
\Line(58.0,60.0)(79.0,50.0) 
\end{picture} \ 
}
&\raisebox{2cm}{$\displaystyle{\frac{\ie {e}}{2}} \gamma^\mu \left(g P_L + fP_R
\right)$}\\[-15mm]
\bottomrule
\end{tabular}
\label{tab:vertex}
\end{center}
\end{table}

For radiative neutralino production  
\begin{eqnarray}
e^-(p_1) + e^+(p_2) \rightarrow \tilde{\chi}_1^0(k_1) + \tilde{\chi}_1^0(k_2) 
+ \gamma(q), 
\end{eqnarray} 
the SUSY Lagrangian is given by~\cite{Haber:1984rc}
\begin{eqnarray}
\label{eq:lagrangian}
{\mathcal L}    &=& \sqrt{2}e a \bar{f}_eP_L\tilde{\chi}^0_1\tilde{e}_R
                +{\sqrt{2}}e b \bar{f}_e P_R\tilde{\chi}^0_1\tilde{e}_L
                +\half e Z_\mu \bar{\tilde{\chi}}_1^0\gamma^\mu\big[g P_L + f P_R\big]\tilde{\chi}_1^0
                + \mathrm{h.~c.},
\end{eqnarray} 
with the electron, selectron, neutralino and $Z$ boson fields $f_e$, $\tilde{e}_{L,R}$, 
$\tilde{\chi}_1^0$, and $Z_\mu$, respectively, 
and $P_{L,R} = \left(1 \mp \gamma^5\right)/2$.
The couplings are
\begin{equation}
\label{eq:coefficients}
\begin{array}{llllll} 
a &= & -\frac{1}{\cw}N_{11}^\ast,\qquad&
\;\; b&=& \phantom{-}\frac{1}{2\sw} (N_{12} + \tw N_{11}),\\[3mm] 
g &=&  -\frac{1}{2\sw\cw}\left(|N_{13}|^2 - |N_{14}|^2\right),\;\;\qquad&
\;\;f &=& -g,
\end{array}
\end{equation}
see the Feynman rules in Tab.~\ref{tab:vertex}. 
The $Z$-$e$-$e$ couplings are
\begin{equation}
\label{eq:Zcoefficients}
\begin{array}{llllll}
c &=& \phantom{-}\frac{1}{\sw\cw}\left(\frac{1}{2} - \sw[2]\right),\qquad&
\;\; d &=& - \tw.
\end{array}
\end{equation}
\subsection{Amplitudes for Radiative Neutralino Production}
We define the selectron and $Z$ boson propagators as
\begin{eqnarray}
\Delta_{\tilde{e}_{L,R}}(p_i,k_j) & \equiv & 
\frac{1}{m_{\tilde{e}_{\scriptscriptstyle{L,R}}}^2 - m_{\chi_1^0}^2 + 
2\mink{p_i}{k_j}},\\ 
\Delta_Z(k_1,k_2)& \equiv & \frac{1}{m_Z^2 -  2 m_{\chi_1^0}^2 -
  2\mink{k_1}{k_2} - \ie \Gamma_Z m_Z}\,.
\end{eqnarray}
The tree-level amplitudes for right selectron exchange in the $t$-channel, see the
diagrams 1-3 in Fig.~\ref{fig:diagrams}, are
\begin{eqnarray}
\M_1 &\!=\!& 2 \ie e^3 |a|^2 \, \Big[\uu(k_1) P_R \,\frac{(\ssl{p}_1 -
\ssl{q})}{2 \mink{p_1}{q}}\,\ssl{\epsilon}^\ast u(p_1)\Big]\, 
         \Big[\vv(p_2) P_L v(k_2)\Big]\Delta_{\tilde{e}_{R}}(p_2,k_2)\,,
\label{eq:m1}\\[2mm]
\M_2 &\!=\!& 2 \ie e^3 |a|^2 \, \Big[\uu(k_1) P_R  u(p_1) \Big]
          \Big[\vv(p_2)\ssl{\epsilon}^\ast\,\frac{(\ssl{q} -\ssl{p}_2)}{2 
\mink{p_2}{q}}\, P_L v(k_2)\Big]\,
          \Delta_{\tilde{e}_{R}}(p_1,k_1)\label{eq:m2}\,,\\[2mm] 
\M_3 &\!=\!& 2 \ie e^3 |a|^2 \, \Big[\uu(k_1) P_R u(p_1)\Big]\,\Big[\vv(p_2) 
P_L v(k_2)\Big]
              {\mink{(2 p_1 - 2 k_1 -q)}{\epsilon^\ast}}
              \Delta_{\tilde{e}_{R}}(p_1,k_1)\Delta_{\tilde{e}_{R}}(p_2,k_2)
\notag\,.\\[-1mm]
&&\label{eq:m3}
\end{eqnarray}
The amplitudes for $u$-channel $\tilde e_R$ exchange, 
see the diagrams 4-6 in Fig.~\ref{fig:diagrams}, are
\begin{eqnarray}
\M_4 &\!=\!& -2 \ie e^3 |a|^2 \, \Big[\uu(k_2) P_R \,\frac{(\ssl{p}_1 -
\ssl{q})}{2 \mink{p_1}{q}}\,\ssl{\epsilon}^\ast u(p_1)\Big]\,
         \Big[\vv(p_2) P_L v(k_1) \Big]\Delta_{\tilde{e}_{R}}(p_2,k_1)
\label{eq:m4}\,,\\[2mm] 
\M_5 &\!=\!& -2 \ie e^3 |a|^2 \, \Big[\uu(k_2) P_R  u(p_1)\Big]
         \Big[\vv(p_2)\ssl{\epsilon}^\ast\,\frac{(\ssl{q} -\ssl{p}_2)}{2 
\mink{p_2}{q}}\, P_L v(k_1)\Big]\, 
           \Delta_{\tilde{e}_{R}}(p_1,k_2)\label{eq:m5}\,,\\[2mm]
\M_6 &\!=\!& -2 \ie e^3 |a|^2 \, \Big[\uu(k_2) P_R u(p_1)\Big]\,
\Big[\vv(p_2) P_L v(k_1)\Big] 
            {\mink{(2 p_1 - 2 k_2 -q)}{\epsilon^\ast}}\Delta_{\tilde{e}_{R}}
(p_1,k_2)\Delta_{\tilde{e}_{R}}(p_2,k_1)\notag\,. \\[-1mm]
&&\label{eq:m6}
\end{eqnarray}
In the photino limit, our amplitudes $\M_1$-$\M_6$, Eqs.~\eqref{eq:m1}-\eqref{eq:m6}, 
agree with those given in Ref.~\cite{Grassie:1983kq}.

The amplitudes for $Z$ boson exchange, see the diagrams 7 and 8 in
Fig.~\ref{fig:diagrams}, are
\begin{eqnarray}
\M_7 &\!=\!& {\ie e^3}\Big[\vv(p_2) \gamma^\mu \left(c P_L + d P_R \right)  
             \frac{(\ssl{p}_1 -\ssl{q})}{2 \mink{p_1}{q}}\,\ssl{\epsilon}^
\ast u(p_1)\Big]\,
           \Big[\uu(k_1) \gamma_\mu  \left(g P_L + f P_R \right) \,v(k_2)
\Big] \Delta_Z(k_1,k_2)\,,\notag\\[-1mm]
&&\label{eq:m7}\\
\M_8 &\!=\!& {\ie e^3}\Big[ \vv(p_2)\ssl{\epsilon}^\ast\frac{(\ssl{q} -
\ssl{p}_2)}{2 \mink{p_2}{q}}
               \gamma^\mu \left(c P_L + d P_R \right)  u(p_1)\Big]\, 
             \Big[\uu(k_1) \gamma_\mu  \left(g P_L + f P_R \right) \,
v(k_2)\Big] \Delta_Z(k_1,k_2)\,. \notag\\[-1mm]
&&\label{eq:m8}
\end{eqnarray}
Note that additional sign factors appear in the amplitudes
$\M_4$-$\M_6$ and $\M_7$-$\M_8$, compared to $\M_1$-$\M_3$.
They stem from the reordering of fermionic operators in the Wick 
expansion of the $S$ matrix. For radiative neutralino production
$e^+e^- \to \tilde\chi_1^0\tilde\chi_1^0\gamma$, such sign factors
appear since the two external neutralinos are 
fermions.\footnote{ 
Note that in Ref.~\cite{Bayer} the relative
sign in the amplitudes for $Z$ boson and $t$-channel $\tilde e_R$ exchange is missing.}
For details 
see Refs.~\cite{Bartl:1986hp,Fraas:1991ky}. 
In addition an extra factor 2 is obtained in the Wick 
expansion of the $S$ matrix, since the Majorana field $\tilde\chi_1^0$ contains both
creation and annihilation operators. 
In our conventions we follow here closely  Ref.~\cite{Bartl:1986hp}.
Other authors take care of this
factor by multiplying the $Z\tilde\chi_1^0\tilde\chi_1^0$ vertex 
by a factor 2 already~\cite{Haber:1984rc}.
For more details of this subtlety, see Ref.~\cite{Haber:1984rc}.

The amplitudes $\M_9-\M_{14}$ for left selectron exchange, see the diagrams 
9-14 in Fig.~\ref{fig:diagrams},  are obtained from the 
$\tilde e_R $ amplitudes by substituting
\begin{eqnarray}
a \rightarrow b, \qquad P_L \rightarrow P_R,\qquad  P_R \rightarrow P_L, \qquad
\Delta_{\tilde{e}_{R}} \rightarrow \Delta_{\tilde{e}_{L}}.
\end{eqnarray}
For $\tilde e_L$ exchange in the $t$-channel they read
\begin{eqnarray}
\M_9 &\!=\!& 2 \ie e^3 |b|^2 \, \Big[\uu(k_1) P_L \,\frac{(\ssl{p}_1 -
\ssl{q})}{2 \mink{p_1}{q}}\,\ssl{\epsilon}^\ast u(p_1)\Big]\, 
         \Big[\vv(p_2) P_R v(k_2)\Big]\Delta_{\tilde{e}_{L}}(p_2,k_2)
\,,\label{eq:m9}\\[2mm]
\M_{10} &\!=\!& 2 \ie e^3 |b|^2 \,\Big[\uu(k_1) P_L  u(p_1)\Big] 
          \Big[\vv(p_2)\ssl{\epsilon}^\ast\,\frac{(\ssl{q} -\ssl{p}_2)}
{2 \mink{p_2}{q}}\, P_R v(k_2)\Big]\,
          \Delta_{\tilde{e}_{L}}(p_1,k_1)\,,\label{eq:m10}\\[2mm] 
\M_{11} &\!=\!& 2 \ie e^3 |b|^2 \, \Big[\uu(k_1) P_L u(p_1)\Big]\,
\Big[\vv(p_2) P_R v(k_2)\Big]
            {\mink{(2 p_1 - 2 k_1 -q)}{\epsilon^\ast}}\Delta_{\tilde{e}_{L}}
(p_1,k_1)\Delta_{\tilde{e}_{L}}(p_2,k_2)\,.\notag\\[-1mm]
&&\label{eq:m11} 
\end{eqnarray}
The $u$-channel $\tilde e_L$ exchange amplitudes are
\begin{eqnarray}
\M_{12} &\!=\!& -2 \ie e^3 |b|^2 \,  \Big[\uu(k_2) P_L \,\frac{(\ssl{p}_1 -
\ssl{q})}{2 \mink{p_1}{q}}\,\ssl{\epsilon}^\ast u(p_1)\Big]\,
         \Big[\vv(p_2) P_R v(k_1)\Big]\Delta_{\tilde{e}_L}(p_2,k_1)\,, 
\label{eq:m12}\\[2mm] 
\M_{13} &\!=\!& -2 \ie e^3 |b|^2 \, \big[\uu(k_2) P_L  u(p_1)\Big]
         \Big[\vv(p_2)\ssl{\epsilon}^\ast\,\frac{(\ssl{q} -\ssl{p}_2)}
{2 \mink{p_2}{q}}\, P_R v(k_1)\Big]\, 
           \Delta_{\tilde{e}_{L}}(p_1,k_2)\,,\label{eq:m13}\\[2mm]
\M_{14} &\!=\!& -2 \ie e^3 |b|^2 \,\Big[ \uu(k_2) P_L u(p_1)\Big]\,
\Big[\vv(p_2) P_R v(k_1)\Big] 
            \mink{(2 p_1 - 2 k_2 -q)}{\epsilon^\ast}\Delta_{\tilde{e}_{L}}
(p_1,k_1)\Delta_{\tilde{e}_{L}}(p_2,k_2)\,.\notag\\[-1mm]
&&\label{eq:m14}
\end{eqnarray} 
Our amplitudes $\M_1$-$\M_{14}$ agree with those given in  
Ref.~\cite{Weidner,Fraas:1991ky}, however there is an obvious 
misprint in amplitude $T_5$ of Ref.~\cite{Weidner}.
In addition we have checked that the amplitudes $\M_i=
\epsilon_\mu\M^\mu_i$ for $i=1,\dots,14$ fulfill the Ward identity $q_\mu
(\sum_i\M^\mu_i)=0$, as done in Refs.~\cite{Bayer, Weidner}.
We find 
$q_\mu(\M^\mu_1+\M^\mu_2+\M^\mu_3)=0$ for $t$-channel $\tilde e_R$ exchange,
$q_\mu(\M^\mu_4+\M^\mu_5+\M^\mu_6)=0$ for $u$-channel $\tilde e_R$
exchange, $q_\mu(\M^\mu_7+\M^\mu_8)=0$ for $Z$ boson exchange,
and analog relations for the $\tilde e_L$ exchange amplitudes.

\subsection{Spin Formalism and Squared Matrix Elements}
We include the longitudinal beam polarisations of electron, $P_{e^-}$, and positron, 
$P_{e^+}$, with $ -1 \le P_{e^\pm}\le +1$ in their density matrices
\begin{eqnarray}
\label{eq:density1}
\rho^{-}_{\lambda_{-} \lambda_{-}^\prime}  &=& 
     \half\left(\delta_{\lambda_{-} \lambda_{-}^\prime} + 
      P_{e^-}\sigma^3_{\lambda_{-} \lambda_{-}^\prime}\right),\\[2mm]
\label{eq:density2}
\rho^{+}_{\lambda_{+} \lambda_{+}^\prime}  &=& 
     \half\left(\delta_{\lambda_{+} \lambda_{+}^\prime} + 
   P_{e^+}\sigma^3_{\lambda_{+} \lambda_{+}^\prime}\right),
\end{eqnarray}
where  $\sigma^3$ is the third Pauli matrix and  $\lambda_{-},\lambda^\prime_{-}$ 
and $\lambda_{+},\lambda^\prime_{+}$
are the helicity indices of electron and positron, respectively.
The squared matrix elements are then obtained by
\begin{eqnarray}
T_{ii} &=& \sum_{\lambda_{-},\lambda_{-}^\prime,\lambda_{+},\lambda_{+}^\prime}
             \rho^{-}_{\lambda_{-} \lambda_{-}^\prime}  \rho^{+}_{\lambda_{+} \lambda_{+}^\prime} 
               \M_i^{\lambda_{-} \lambda_{+}}{\M_i^{\ast}}^{\lambda_{-}^\prime \lambda_{+}^\prime},
\label{Tii}\\[2mm]
T_{ij} &=& 2\, \real\left( 
                \sum_{\lambda_{-},\lambda_{-}^\prime,\lambda_{+},\lambda_{+}^\prime}
               \rho^{-}_{\lambda_{-} \lambda_{-}^\prime}  \rho^{+}_{\lambda_{+} \lambda_{+}^\prime} 
               \M_i^{\lambda_{-} \lambda_{+}} {\M_j^{\ast}}^{\lambda_{-}^\prime \lambda_{+}^\prime} \right),
	      	\enspace i \neq j , 
\label{Tij}\\[2mm]
|\M|^2 & = & \sum_{i \leq j} T_{ij} ,
\end{eqnarray}
where an internal sum over the helicities of the outgoing neutralinos, as well as a sum over 
the polarisations of the photon is included.
Note that we suppress the electron and positron helicity indices of the amplitudes 
$\M_i^{\lambda_{-} \lambda_{+}}$ in the formulas~(\ref{eq:m1})-(\ref{eq:m14}).
The product of the amplitudes in Eqs.~(\ref{Tii}) and (\ref{Tij}) contains
the projectors
\begin{eqnarray}
\label{eq:beam1}
u(p,\lambda_{-})\uu(p,\lambda_{-}^\prime) &=& 
                   \half\left(\delta_{\lambda_{-} \lambda_{-}^\prime} + 
                   \gamma^5\sigma^3_{\lambda_{-}\lambda_{-}^\prime} \right)\ssl{p},\\[2mm]
\label{eq:beam2}
v(p,\lambda_{+}^\prime)\vv(p,\lambda_{+}) &=& 
                   \half\left(\delta_{\lambda_{+} \lambda_{+}^\prime} + 
                   \gamma^5\sigma^3_{\lambda_{+}\lambda_{+}^\prime} \right)\ssl{p}.
\end{eqnarray} 
The contraction with the density matrices of the electron and positron beams leads finally to
\begin{eqnarray}
\sum_{\lambda_{-}, \lambda_{-}^\prime}
\rho^{-}_{\lambda_{-} \lambda_{-}^\prime} u(p,\lambda_{-})\uu(p,\lambda_{-}^\prime) &=& 
\left(\frac{1-P_{e^-}}{2}P_L + \frac{1+P_{e^-}}{2} P_R\right)\ssl{p},\\[2mm]
\sum_{\lambda_{+}, \lambda_{+}^\prime}
\rho^{+}_{\lambda_{+} \lambda_{+}^\prime} v(p,\lambda_{+}^\prime)\vv(p,\lambda_{+}) &=& 
\left(\frac{1+P_{e^+}}{2}P_L + \frac{1-P_{e^+}}{2} P_R\right)\ssl{p}.
\end{eqnarray}
In the squared amplitudes, we include the electron and positron 
beam polarisations in the coefficients
\begin{eqnarray}
C_R = (1 + P_{e^-})(1-P_{e^+}),\qquad C_L = (1 - P_{e^-})(1 + P_{e^+}).
\end{eqnarray}
In the following we give the squared amplitudes $T_{ij}$, 
as defined in Eqs.~(\ref{Tii}) and (\ref{Tij}),
for $\tilde e_R$ and $Z$ exchange.
To obtain the corresponding squared amplitudes for $\tilde e_L$ exchange,
one has to substitute
\begin{eqnarray}
a \rightarrow b, \qquad  d \leftrightarrow c, \qquad  f\leftrightarrow g,
\qquad C_R \rightarrow C_L,\qquad \Delta_{\tilde{e}_R}\rightarrow \Delta_{\tilde{e}_L}\,. 
\end{eqnarray}
There is no interference between diagrams with $\tilde e_R$ and
$\tilde e_L$ exchange, since we assume the high energy limit where
ingoing particles are considered massless.  

\begin{eqnarray}
T_{11} &=& 4 e^6 C_R |a|^4 \,\Delta^2_{\tilde{e}_{R}}(p_2,k_2)\,
\frac{\mink{p_2}{k_2} \,\,q \cdot k_1}{\mink{q}{p_1}}\\[4mm] 
T_{22} &=& 4 e^6 C_R |a|^4 \,\Delta^2_{\tilde{e}_{R}}(p_1,k_1)\,\frac{\mink{p_1}{k_1}
\,\,\mink{q}{k_2}}{\mink{q}{p_2}}\\[4mm]
T_{33} &=& 4 e^6 C_R |a|^4 \,\Delta^2_{\tilde{e}_{R}}(p_1,k_1)\Delta^2_{\tilde{e}_{R}}
(p_2,k_2)\,{\mink{p_1}{k_1}\,\,\mink{p_2}{k_2}} 
       \Bigl[-4 m_{\chi_1^0}^2 + 8 \mink{p_1}{k_1} + 4 \mink{q}{p_1} - 4\mink{q}{k_1}
\Bigr]\notag\\[1mm]
&&\\[0mm]
T_{44} &=& 4 e^6 C_R |a|^4 \,\Delta^2_{\tilde{e}_{R}}(p_2,k_1)\,\frac{\mink{p_2}{k_1} 
\,\,\mink{q}{k_2}}{\mink{q}{p_1}}\\[4mm]
T_{55} &=& 4 e^6 C_R |a|^4 \,\Delta^2_{\tilde{e}_{R}}(p_1,k_2)\,\frac{\mink{p_1}{k_2}
\,\,\mink{q}{k_1}}{\mink{q}{p_2}}\\[4mm]
T_{66} &=& 4 e^6 C_R |a|^4 \,\Delta^2_{\tilde{e}_{R}}(p_1,k_2)\Delta^2_{\tilde{e}_{R}}
(p_2,k_1){\mink{p_1}{k_2}\,\,\mink{p_2}{k_1}}
       \Bigl[-4 m_{\chi_1^0}^2 + 8 \mink{p_1}{k_2} + 4 \mink{q}{p_1} - 4
       \mink{q}{k_2}\Bigr]\notag\\[1mm]
&&\\
T_{77} &=& \frac{4e^6}{\mink{q}{p_1}} |\Delta_Z(k_1,k_2)|^2 
               \Bigl[(C_R d^2 f^2 + C_L c^2 g^2)\mink{p_2}{k_1}\mink{q}{k_2}+
                (C_R d^2 g^2 + C_Lc^2 f^2)\mink{p_2}{k_2}\mink{q}{k_1} \Bigr.
\notag\\[2mm]
               &&  \hspace*{30mm}+ \Bigl.f g (C_L c^2 + C_R d^2) m_{\chi_1^0}^2 \,\mink{q}{p_2}\Bigr] \\[4mm] 
T_{88} &=& \frac{4e^6}{\mink{q}{p_2}} |\Delta_Z(k_1,k_2)|^2 
               \Bigl[(C_R d^2 f^2 + C_L c^2 g^2)\mink{p_1}{k_2}\mink{q}{k_1} + 
                 (C_R d^2 g^2 + C_L c^2 f^2)\mink{p_1}{k_1}\mink{q}{k_2} \Bigr.
\notag\\[2mm]
               &&  \hspace*{30mm}\Bigl.+ f g (C_L c^2 + C_R d^2) m_{\chi^0_1}^2 
\,\mink{q}{p_1}\Bigr] \\[4mm]
T_{12} &=& - 4 e^6 C_R |a|^4 \Delta_{\tilde{e}_{R}}(p_1,k_1)\Delta_{\tilde{e}_{R}}(p_2,k_2)
             \frac{1}{\mink{q}{p_1}\mink{q}{p_2}}\notag \\[2mm]
        &&\Bigl[\mink{q}{k_2}\mink{p_1}{k_1}\mink{p_1}{p_2}-\mink{p_1}{k_1}
\mink{q}{p_2}\mink{p_1}{k_2}
         +\mink{p_1}{k_1}\mink{p_2}{k_2}\mink{q}{p_1}\Bigr.
        + \mink{p_1}{p_2}\mink{q}{k_1}\mink{p_2}{k_2}\notag\\[2mm]
        &&\hspace*{5mm}-\mink{q}{p_1}\mink{p_2}{k_2}\mink{p_2}{k_1} 
         + \mink{p_1}{k_1}\mink{p_2}{k_2}\mink{q}{p_2} 
        \Bigl. -2\mink{p_1}{p_2}\mink{p_1}{k_1}\mink{p_2}{k_2}  \Bigr]\\[4mm]
T_{13} &=& 8 e^6  C_R |a|^4  \Delta_{\tilde{e}_{R}}(p_1,k_1)\Delta^2_{\tilde{e}_{R}}(p_2,k_2)
                             \frac{\mink{p_2}{k_2}}{\mink{q}{p_1}}\notag\\[2mm]
        &&\Bigl[m_{\chi^0_1}^2 \mink{q}{p_1} + 2 (\mink{p_1}{k_1})^2 + 
		\mink{p_1}{k_1}\mink{q}{p_1}  
             - 2\mink{p_1}{k_1} \mink{q}{k_1}\Bigr]\\[4mm]
T_{14} &=& - 4 e^6 C_R |a|^4  \Delta_{\tilde{e}_{R}}(p_2,k_1)\Delta_{\tilde{e}_{R}}(p_2,k_2)
              \frac{m_{\chi^0_1}^2 \mink{q}{p_2}}{\mink{q}{p_1}} \\[4mm]
T_{15} &=&   4 e^6 C_R |a|^4  \Delta_{\tilde{e}_{R}}(p_1,k_2)\Delta_{\tilde{e}_{R}}(p_2,k_2) 
             \frac{m_{\chi^0_1}^2\mink{p_1}{p_2}}{\mink{q}{p_1} \mink{q}{p_2}}
    \Bigl[ \mink{q}{p_1} -\mink{p_1}{p_2}+ \mink{q}{p_2} \Bigr]\\[4mm]
T_{16} &=&  4 e^6 C_R |a|^4  \Delta_{\tilde{e}_{R}}(p_1,k_2)\Delta_{\tilde{e}_{R}}(p_2,k_1) 
	\Delta_{\tilde{e}_{R}}(p_2,k_2)
               \frac{m_{\chi^0_1}^2}{\mink{q}{p_1}} \notag\\[2mm]
              &&\Bigl[-2 \mink{p_1}{k_2}\mink{p_1}{p_2}- \mink{q}{p_1}\mink{p_1}{p_2} +
              \mink{q}{k_2}\mink{p_1}{p_2} \Bigr.
              -\mink{q}{p_1}\mink{p_2}{k_2}+\mink{q}{p_2}\mink{p_1}{k_2}\Bigr]\\[4mm]
T_{17} &=& 4 e^6 |a|^2 C_R d \frac{1}{\mink{q}{p_1}}\Delta_{\tilde{e}_{R}}(p_2,k_2)
            \real\{\Delta_Z(k_1,k_2)\} \bigl[2 g  \mink{p_2}{k_2}\mink{q}{k_1}
             + fm_{\chi^0_1}^2 \mink{q}{p_2}\bigr]\\[4mm]
T_{18} &=& - 4 e^6 C_R |a|^2 d\frac{1}{\mink{q}{p_1}\mink{q}{p_2}}
            \Delta_{\tilde{e}_{R}}(p_2,k_2)\real\{\Delta_Z(k_1,k_2)\}\notag\\[2mm]
          &&\Bigl[g \Bigl(-2 \mink{p_1}{p_2}\mink{p_2}{k_2}\mink{p_1}{k_1} 
	+\mink{p_2}{k_2}(\mink{q}{k_1}\mink{p_1}{p_2}
          - \mink{p_2}{k_1}\mink{q}{p_1}
          + \mink{p_1}{k_1}\mink{q}{p_2}) \bigr.\notag\\[2mm]
          &&\bigl.+\mink{p_1}{k_1}(\mink{q}{p_1}\mink{p_2}{k_2} 
           +\bigl.\mink{q}{k_2}\mink{p_1}{p_2} -\mink{q}{p_2}\mink{p_1}{k_2})\Bigr) \notag\\[2mm] 
          && - f m_{\chi^0_1}^2 \mink{p_1}{p_2} 
          \bigl(\mink{p_1}{p_2}-\mink{q}{p_2}-\mink{q}{p_1}
          \bigr)\Bigl]
\\[4mm]
T_{23} &=& 8 e^6  C_R |a|^4 \frac{\mink{p_1}{k_1}}{\mink{q}{p_2}}
         \Delta^2_{\tilde{e}_{R}}(p_1,k_1)\Delta_{\tilde{e}_{R}}(p_2,k_2)\notag \\[2mm]
      && \hspace*{0mm}\Bigl[m_{\chi^0_1}^2 \mink{q}{p_2} + 2 (\mink{p_2}{k_2})^2 + 
	\mink{p_2}{k_2}\mink{q}{p_2}  
         - 2\mink{p_2}{k_2} \mink{q}{k_2}\Bigr]\\[4mm] 
T_{24} &=& 4 e^6  C_R |a|^4 \Delta_{\tilde{e}_{R}}(p_1,k_1)\Delta_{\tilde{e}_{R}}(p_2,k_1) 
         \frac{m_{\chi^0_1}^2\mink{p_1}{p_2}}{\mink{q}{p_1} \mink{q}{p_2}}
    \Bigl( \mink{q}{p_1} -\mink{p_1}{p_2}+ \mink{q}{p_2} \Bigr)\\[4mm]
T_{25} &=& 4 e^6  C_R |a|^4 \Delta_{\tilde{e}_{R}}(p_1,k_1)\Delta_{\tilde{e}_{R}}
(p_1,k_2)
             \frac{m_{\chi^0_1}^2  \mink{q}{p_1}}{\mink{q}{p_2}}\\[4mm]
T_{26} &=&  4 e^6 C_R |a|^4 \Delta_{\tilde{e}_{R}}(p_1,k_2)\Delta_{\tilde{e}_{R}}(p_2,k_1)
	\Delta_{\tilde{e}_{R}}(p_1,k_1)
              \frac{m_{\chi^0_1}^2}{\mink{q}{p_2}}\notag\\[2mm]
              &&\Bigl[-2 \mink{p_2}{k_1}\mink{p_1}{p_2}- \mink{q}{p_2}\mink{p_1}{p_2} +
              \mink{q}{k_1}\mink{p_1}{p_2} 
              - \mink{q}{p_2}\mink{p_1}{k_1}+\mink{q}{p_1}\mink{p_2}{k_1}\Bigr]\\[4mm] 
T_{27} &=&  \frac{ 4 e^6 C_R |a|^2 d}{\mink{q}{p_1}\mink{q}{p_2}}
               \Delta_{\tilde{e}_{R}}(p_1,k_1)\real\{\Delta_Z(k_1,k_2)\}\notag\\[2mm] 
          &&\Bigl[g \Bigl(2 \mink{p_1}{p_2}\mink{p_2}{k_2}\mink{p_1}{k_1} +\mink{p_2}
{k_2}(-\mink{q}{k_1}\mink{p_1}{p_2}
          + \mink{p_2}{k_1}\mink{q}{p_1}- \mink{p_1}{k_1}\mink{q}{p_2}) \bigr.\notag\\[2mm] 
          &&+ \mink{p_1}{k_1}(-\mink{q}{p_1}\mink{p_2}{k_2}  
          \bigl. - \mink{q}{k_2}\mink{p_1}{p_2} + \mink{q}{p_2}\mink{p_1}{k_2})  
          \Bigr) \Bigr.\notag\\[2mm] 
           &&+ f m_{\chi^0_1}^2 \mink{p_1}{p_2} 
          \bigl(\mink{p_1}{p_2}-\mink{q}{p_2}-\mink{q}{p_1}
          \bigr)\Bigl]
\\[4mm]
T_{28} &=&  \frac{4 e^6 C_R |a|^2 d}{\mink{q}{p_2}}\Delta_{\tilde{e}_{R}}(p_1,k_1)\real\{\Delta_Z(k_1,k_2)\}
                \Bigl[2 g  \mink{p_1}{k_1}\mink{q}{k_2}
                           + f m_{\chi^0_1}^2 \mink{q}{p_1}\Bigr]\\[4mm]
T_{34} &=& - 4 e^6  C_R |a|^4 \frac{m_{\chi^0_1}^2}{\mink{q}{p_1}}
               \Delta_{\tilde{e}_{R}}(p_1,k_1)\Delta_{\tilde{e}_{R}}(p_2,k_1)\Delta_{\tilde{e}_{R}}(p_2,k_2)\notag\\[2mm]
           &&\Bigl[2\mink{p_1}{p_2}\mink{p_1}{k_1} + \mink{p_1}{p_2}\mink{q}{p_1}-\mink{p_1}{k_1}\mink{q}{p_2}
            +\mink{p_2}{k_1}\mink{q}{p_1} -\mink{p_1}{p_2}\mink{q}{k_1}\Bigr]\\[4mm]
T_{35} &=&  -4 e^6  C_R |a|^4 \frac{m_{\chi^0_1}^2}{\mink{q}{p_2}}
             \Delta_{\tilde{e}_{R}}(p_1,k_1)\Delta_{\tilde{e}_{R}}(p_1,k_2)\Delta_{\tilde{e}_{R}}(p_2,k_2)\notag\\[2mm]
          &&\Bigl[2\mink{p_1}{p_2}\mink{p_2}{k_2}- \mink{p_1}{p_2}\mink{q}{k_2}+
          \mink{p_1}{p_2}\mink{q}{p_2}- \mink{p_2}{k_2} \mink{q}{p_1} +\mink{p_1}{k_2}\mink{q}{p_2}\Bigr]\\[4mm]
T_{36} &=& 8 e^6  C_R |a|^4 \Delta_{\tilde{e}_{R}}(p_1,k_1)\Delta_{\tilde{e}_{R}}(p_1,k_2)
                          \Delta_{\tilde{e}_{R}}(p_2,k_1)\Delta_{\tilde{e}_{R}}(p_2,k_2) \notag\\[2mm]
                &&        m_{\chi^0_1}^2 \mink{p_1}{p_2}
         \Bigl[- 2\mink{p_1}{k_1} - 2\mink{q}{p_1} -2\mink{p_1}{k_2} + 2 \mink{k_1}{k_2} +  \mink{q}{k_2}
         + \mink{q}{k_1}\Bigr]\\[4mm] 
T_{37} &=& \frac{ 4e^6 C_R |a|^2 d}{\mink{q}{p_1}}
                   \Delta_{\tilde{e}_{R}}(p_1,k_1)\Delta_{\tilde{e}_{R}}(p_2,k_2) \real\{\Delta_Z(k_1,k_2)\}\notag\\[2mm]
              && \Bigl[ 2 g\mink{p_2}{k_2} \bigl( 
                  \mink{q}{p_1}\mink{p_1}{k_1}-2\mink{p_1}{k_1}\mink{q}{k_1} + 2(\!\mink{p_1}{k_1}\!)^2
                  + m_{\chi^0_1}^2\mink{q}{p_1}\bigr) \Bigr.\notag\\[2mm] 
          &&\Bigl.+ f m_{\chi^0_1}^2\bigl(
                  2 \mink{p_1}{p_2}\mink{p_1}{k_1}+\mink{p_1}{p_2}\mink{q}{p_1}-\mink{p_1}{p_2}\mink{q}{k_1}  
                  -\mink{p_1}{k_1}\mink{q}{p_2}+\mink{q}{p_1}\mink{p_2}{k_1}
                  \bigr) \Bigr]
                   \\[4mm]
T_{38} &=& \frac{4 e^6 C_R |a|^2 d}{\mink{q}{p_2}}
           \Delta_{\tilde{e}_{R}}(p_1,k_1)\Delta_{\tilde{e}_{R}}(p_2,k_2)\real\{\Delta_Z(k_1,k_2)\}\notag\\[2mm]
           &&\Bigl[ 2 g\mink{p_1}{k_1}
             \bigl(2 (\!\mink{p_2}{k_2}\!)^2 + \mink{p_2}{k_2} \mink{q}{p_2} - 2\mink{p_2}{k_2} \mink{q}{k_2}
             + m_{\chi^0_1}^2  \mink{q}{p_2}  \bigr)\Bigr.\notag\\[2mm] 
             &&\bigl.+ f m_{\chi^0_1}^2  \bigl(2 \mink{p_1}{p_2}\mink{p_2}{k_2} 
             +\mink{p_1}{p_2}\mink{q}{p_2} - \mink{p_1}{p_2}\mink{q}{k_2} + \mink{q}{p_2}\mink{p_1}{k_2} - 
             \mink{q}{p_1}\mink{p_2}{k_2} \bigr) \Bigr]
           \\[4mm]
T_{45} &=& - \frac{4 e^6 C_R |a|^4} {\mink{q}{p_1} \mink{q}{p_2}} 
          \Delta_{\tilde{e}_{R}}(p_1,k_2) \Delta_{\tilde{e}_{R}}(p_2,k_1)\notag \\[2mm] 
        &&\Bigl[\mink{q}{k_1}\mink{p_1}{k_2}\mink{p_1}{p_2}-\mink{p_1}{k_2}\mink{q}{p_2}\mink{p_1}{k_1}
        +\mink{p_1}{k_2}\mink{p_2}{k_1}\mink{q}{p_1}
        + \mink{p_1}{p_2}\mink{q}{k_2}\mink{p_2}{k_1}\Bigr.\notag \\[2mm]        
       && -\mink{q}{p_1}\mink{p_2}{k_1}\mink{p_2}{k_2}         
        + \mink{p_1}{k_2}\mink{p_2}{k_1}\mink{q}{p_2} 
        -2\mink{p_1}{p_2}\mink{p_1}{k_2}\mink{p_2}{k_1}  \Bigr]\\[4mm] 
T_{46} &=&  8 e^6  C_R |a|^4 \frac{\mink{p_2}{k_1}}{\mink{q}{p_1}}
                    \Delta_{\tilde{e}_{R}}(p_1,k_2) \Delta^2_{\tilde{e}_{R}}(p_2,k_1)\notag \\[2mm]
           &&\hspace*{0mm} \left[m_{\chi^0_1}^2 \mink{q}{p_1} + 2 (\mink{p_1}{k_2})^2 + \mink{p_1}{k_2}\mink{q}{p_1}  
            - 2\mink{p_1}{k_2} \mink{q}{k_2}\right]\\[4mm] 
T_{47} &=&  - \frac{4 e^6 C_R |a|^2 d}{\mink{q}{p_1}}
              \Delta_{\tilde{e}_{R}}(p_2,k_1)\real\{\Delta_Z(k_1,k_2)\}
        \Bigl[2 g  \mink{p_2}{k_1}\mink{q}{k_2} + f m_{\chi^0_1}^2 \mink{q}{p_2}\Bigr]\\[4mm]
T_{48} &=&  \frac{ -4 e^6 C_R |a|^2 d}{\mink{q}{p_1}\mink{q}{p_2}}
             \Delta_{\tilde{e}_{R}}(p_2,k_1)\real\{\Delta_Z(k_1,k_2)\}\notag\\[2mm]
          &&\Bigl[g \Bigl(2 \mink{p_1}{p_2}\mink{p_1}{k_2}\mink{p_2}{k_1} 
            +\mink{p_2}{k_1} \big( -\mink{q}{k_2}\mink{p_1}{p_2} - \mink{p_1}{k_2}\mink{q}{p_2}
            + \mink{p_2}{k_2}\mink{q}{p_1}\big)  \notag\\[2mm]
           && +\mink{p_1}{k_2}(-\mink{q}{p_1}\mink{p_2}{k_1} 
           +\mink{q}{p_2}\mink{p_1}{k_1} - \mink{q}{k_1}\mink{p_1}{p_2}) \Bigr)  \notag\\[2mm]
          && + f m_{\chi^0_1}^2 \mink{p_1}{p_2}
          \bigl(\mink{p_1}{p_2}-\mink{q}{p_2} - \mink{q}{p_1} 
          \bigr)\Bigl]
             \\[4mm]
T_{56} &=&  8 e^6 C_R |a|^4 \frac{\mink{p_1}{k_2}}{\mink{q}{p_2}} 
                \Delta^2_{\tilde{e}_{R}}(p_1,k_2) \Delta_{\tilde{e}_{R}}(p_2,k_1) \notag\\[2mm] 
          &&\hspace*{0mm} \Bigl[\mink{p_2}{k_1}\mink{q}{p_2} - 2 \mink{p_2}{k_1}\mink{q}{k_1}
           + 2 (\!\mink{p_2}{k_1} \!)^2 + m_{\chi^0_1}^2 \mink{q}{p_2}\Bigr]\\[4mm] 
T_{57} &=& - \frac{4 e^6 C_R |a|^2 d}{\mink{q}{p_2}\mink{q}{p_1}}
                \Delta_{\tilde{e}_{R}}(p_1,k_2)\real\{\Delta_Z(k_1,k_2)\}\notag\\[2mm]
                     &&\Bigl[ g \Bigl(
                     2\mink{p_1}{p_2}\mink{p_1}{k_2}\mink{p_2}{k_1} 
                     + \mink{p_1}{k_2} \bigl(-\mink{p_1}{p_2}\mink{q}{k_1}+\mink{p_1}{k_1}\mink{q}{p_2} 
                     -\mink{q}{p_1}\mink{p_2}{k_1}\!\bigr) \notag\\[2mm] 
                     &&+ \mink{p_2}{k_1} \bigl(-\mink{p_1}{p_2}\mink{q}{k_2} 
                     -\mink{p_1}{k_2}\mink{q}{p_2} + \mink{q}{p_1}\mink{p_2}{k_2} \bigr)\Bigr) \Bigl. \notag\\[2mm]
                     &&+ f m_{\chi^0_1}^2 \mink{p_1}{p_2} 
                     \bigl(\mink{p_1}{p_2}-\mink{q}{p_2} - \mink{q}{p_1}\bigr) \Bigr]
                \\[4mm]             
T_{58} &=& - \frac{ 4 e^6 C_R |a|^2 d}{\mink{q}{p_2}}\Delta_{\tilde{e}_{R}}(p_1,k_2)\real\{\Delta_Z(k_1,k_2)\}
                    \Bigl[2 g \mink{p_1}{k_2}\mink{q}{k_1} + f m_{\chi^0_1}^2 \mink{q}{p_1}\Bigr] \\[4mm]
T_{67} &=&   - \frac{ 4 e^6 C_R |a|^2 d }{\mink{q}{p_1}}
               \Delta_{\tilde{e}_{R}}(p_1,k_2)\Delta_{\tilde{e}_{R}}(p_2,k_1)
               \real\{\Delta_Z(k_1,k_2)\}\notag\\[2mm]
               &&\Bigl[ 2 g \mink{p_2}{k_1}\bigl(\mink{p_1}{k_2}\mink{q}{p_1} -
                 2\mink{q}{k_2}\mink{p_1}{k_2} 
                 +2(\!\mink{p_1}{k_2}\!)^2 + m_{\chi^0_1}^2 
                 \mink{q}{p_1}\bigr)\Bigr.\notag \\[2mm] 
                 && \Bigl.\bigl. + f m_{\chi^0_1}^2 \bigl(2\mink{p_1}{k_2}\mink{p_1}{p_2} +
                 \mink{q}{p_1}\mink{p_1}{p_2} -\mink{q}{k_2}\mink{p_1}{p_2} - 
                 \mink{q}{p_2}\mink{p_1}{k_2}  +\mink{q}{p_1}\mink{p_2}{k_2} \bigr)\Bigr]
               \\[4mm]
T_{68} &=&  - \frac{4 e^6 C_R |a|^2 d}{\mink{q}{p_2}}
                  \Delta_{\tilde{e}_{R}}(p_1,k_2)\Delta_{\tilde{e}_{R}}(p_2,k_1)\real\{\Delta_Z(k_1,k_2)\}\notag\\[2mm]
             &&\Bigl[2g\mink{p_1}{k_2}\bigl(2 (\!\mink{p_2}{k_1}\!)^2 + \mink{q}{p_2}\mink{p_2}{k_1}-
                    2 \mink{p_2}{k_1} \mink{q}{k_1} + m_{\chi^0_1}^2\mink{q}{p_2}\bigr) \bigr.\Bigr.\notag \\[2mm] 
       && + f m_{\chi^0_1}^2\bigl(
                      2\mink{p_1}{p_2} \mink{p_2}{k_1} + \mink{p_1}{p_2} \mink{q}{p_2}- \mink{p_2}{k_1} \mink{q}{p_1} 
                      +\mink{p_1}{k_1} \mink{q}{p_2} -\mink{p_1}{p_2}
                       \mink{q}{k_1}\bigr) \Bigr] 
                  \\[4mm]
T_{78} &=&  \frac{ 4e^6}{\mink{q}{p_2}\mink{q}{p_1}}|\Delta_Z(k_1,k_2)|^2 
            \Bigl[(C_R g^2 d^2 + C_L f^2 c^2)\Bigl(2\mink{p_1}{p_2}\mink{p_1}{k_1} \mink{p_2}{k_2}\notag \\[2mm] 
              &&\hspace*{40mm} +\mink{p_1}{k_1}\bigl( \mink{p_1}{k_2}\mink{q}{p_2} - \mink{p_1}{p_2}\mink{q}{k_2} - 
                               \mink{p_2}{k_2} \mink{q}{p_2} \bigr)\Bigr) \notag \\[2mm]     
&&\hspace*{40mm}+\mink{p_2}{k_2}\bigl( \mink{p_2}{k_1}\mink{q}{p_1} - \mink{p_1}{p_2}\mink{q}{k_1} - 
                               \mink{p_1}{k_1} \mink{q}{p_1} \bigr)\Bigr) \notag \\[2mm]
     &&  + (C_L g^2 c^2 + C_R f^2 d^2)\Bigl(2\mink{p_1}{p_2}\mink{p_1}{k_2} \mink{p_2}{k_1} \notag\\[2mm] 
               &&\hspace*{40mm} +\mink{p_1}{k_2}\bigl( \mink{p_1}{k_1}\mink{q}{p_2} - \mink{p_1}{p_2}\mink{q}{k_1} - 
                               \mink{p_2}{k_1} \mink{q}{p_2} \bigr)\Bigr) \notag \\[2mm]
     &&   \hspace*{40mm}+\mink{p_2}{k_1}\bigl( \mink{p_2}{k_2}\mink{q}{p_1} - \mink{p_1}{p_2}\mink{q}{k_2} - 
                               \mink{p_1}{k_2} \mink{q}{p_1} \bigr)\Bigr) \notag \\[2mm]                               
     &&  + \Bigl. 2 g f (C_L c^2 + C_R d^2) m_{\chi_1^0}^2\mink{p_1}{p_2}  
                               \bigl(\mink{p_1}{p_2}-\mink{q}{p_2} - \mink{q}{p_1}\bigr) \Bigr]
\end{eqnarray}
We have calculated the squared amplitudes with \texttt{FeynCalc}~\cite{Kublbeck:1992mt}.
When integrating the squared amplitude over the phase space, see
Appendix~\ref{sec:phasespace}, the $s$-$t$-interference terms cancel
the $s$-$u$-interference terms due to a symmetry in these channels,
caused by the Majorana properties of the neutralinos~\cite{
  Choi:1999bs}.  
Note that in principle also terms proportional to 
$\eps \imag \{\Delta_Z\}$
would appear in the squared amplitudes $T_{ij}$,
due to the inclusion of the $Z$ width 
to regularise the pole of the propagator $\Delta_Z$. 
However, since this is a higher order effect
which is small far off the $Z$ resonance,
we neglect such terms.
In addition they would
vanish after performing a complete phase
space integration.

\section{Amplitudes for Radiative Neutrino 
Production \label{sec:app:nuback}}

\begin{figure}[t]
{%
\unitlength=0.8 pt
\SetScale{0.8}
\SetWidth{0.7}      
\scriptsize    
\allowbreak
\begin{picture}(95,79)(0,0)
\Text(15.0,70.0)[r]{$e$}
\ArrowLine(16.0,70.0)(58.0,70.0) 
\Text(80.0,70.0)[l]{$\gamma$}
\Photon(58.0,70.0)(79.0,70.0){1.0}{4} 
\Text(54.0,60.0)[r]{$e$}
\ArrowLine(58.0,70.0)(58.0,50.0) 
\Text(80.0,50.0)[l]{$\nu_e$}
\ArrowLine(58.0,50.0)(79.0,50.0) 
\Text(54.0,40.0)[r]{$W^+$}
\DashArrowLine(58.0,30.0)(58.0,50.0){3.0} 
\Text(15.0,30.0)[r]{$\bar{e}$}
\ArrowLine(58.0,30.0)(16.0,30.0) 
\Text(80.0,30.0)[l]{$\bar{\nu}_e$}
\ArrowLine(79.0,30.0)(58.0,30.0) 
\Text(47,0)[b] {diagr. 1}
\end{picture} \ 
{} \quad
\begin{picture}(95,79)(0,0)
\Text(15.0,70.0)[r]{$e$}
\ArrowLine(16.0,70.0)(58.0,70.0) 
\Text(80.0,70.0)[l]{$\nu_e$}
\ArrowLine(58.0,70.0)(79.0,70.0) 
\Text(54.0,60.0)[r]{$W^+$}
\DashArrowLine(58.0,50.0)(58.0,70.0){3.0} 
\Text(80.0,50.0)[l]{$\bar{\nu_e}$}
\ArrowLine(79.0,50.0)(58.0,50.0) 
\Text(54.0,40.0)[r]{$e$}
\ArrowLine(58.0,50.0)(58.0,30.0) 
\Text(15.0,30.0)[r]{$\bar{e}$}
\ArrowLine(58.0,30.0)(16.0,30.0) 
\Text(80.0,30.0)[l]{$\gamma$}
\Photon(58.0,30.0)(79.0,30.0){1.0}{4} 
\Text(47,0)[b] {diagr. 2}
\end{picture} \ 
{} \quad
\begin{picture}(95,79)(0,0)
\Text(15.0,70.0)[r]{$e$}
\ArrowLine(16.0,70.0)(58.0,70.0) 
\Text(80.0,70.0)[l]{$\nu_e$}
\ArrowLine(58.0,70.0)(79.0,70.0) 
\Text(54.0,60.0)[r]{$W^+$}
\DashArrowLine(58.0,50.0)(58.0,70.0){3.0} 
\Text(80.0,50.0)[l]{$\gamma$}
\Photon(58.0,50.0)(79.0,50.0){1.0}{4} 
\Text(54.0,40.0)[r]{$W^+$}
\DashArrowLine(58.0,30.0)(58.0,50.0){3.0} 
\Text(15.0,30.0)[r]{$\bar{e}$}
\ArrowLine(58.0,30.0)(16.0,30.0) 
\Text(80.0,30.0)[l]{$\bar{\nu}_e$}
\ArrowLine(79.0,30.0)(58.0,30.0) 
\Text(47,0)[b] {diagr. 3}
\end{picture} \ 
{} \quad
\begin{picture}(95,79)(0,0)
\Text(15.0,60.0)[r]{$e$}
\ArrowLine(16.0,60.0)(37.0,60.0) 
\Photon(37.0,60.0)(58.0,60.0){1.0}{4} 
\Text(80.0,70.0)[l]{$\gamma$}
\Photon(58.0,60.0)(79.0,70.0){1.0}{4} 
\Text(33.0,50.0)[r]{$e$}
\ArrowLine(37.0,60.0)(37.0,40.0) 
\Text(15.0,40.0)[r]{$\bar{e}$}
\ArrowLine(37.0,40.0)(16.0,40.0) 
\Text(47.0,41.0)[b]{$Z$}
\DashLine(37.0,40.0)(58.0,40.0){3.0} 
\Text(80.0,50.0)[l]{$\nu$}
\ArrowLine(58.0,40.0)(79.0,50.0) 
\Text(80.0,30.0)[l]{$\bar{\nu}$}
\ArrowLine(79.0,30.0)(58.0,40.0) 
\Text(47,0)[b] {diagr. 4}
\end{picture} \ 
{} \quad
\begin{picture}(95,79)(0,0)
\Text(15.0,60.0)[r]{$e$}
\ArrowLine(16.0,60.0)(37.0,60.0) 
\Text(47.0,61.0)[b]{$Z$}
\DashLine(37.0,60.0)(58.0,60.0){3.0} 
\Text(80.0,70.0)[l]{$\nu$}
\ArrowLine(58.0,60.0)(79.0,70.0) 
\Text(80.0,50.0)[l]{$\bar{\nu}$}
\ArrowLine(79.0,50.0)(58.0,60.0) 
\Text(33.0,50.0)[r]{$e$}
\ArrowLine(37.0,60.0)(37.0,40.0) 
\Text(15.0,40.0)[r]{$\bar{e}$}
\ArrowLine(37.0,40.0)(16.0,40.0) 
\Photon(37.0,40.0)(58.0,40.0){1.0}{4} 
\Text(80.0,30.0)[l]{$\gamma$}
\Photon(58.0,40.0)(79.0,30.0){1.0}{4} 
\Text(47,0)[b] {diagr. 5}
\end{picture} \ 
}
\caption{Contributing diagrams to 
$e^+e^- \rightarrow {\nu}{\bar{\nu}}\gamma$~\cite{Boos:2004kh}.}
\label{fig:neutrino}
\end{figure}
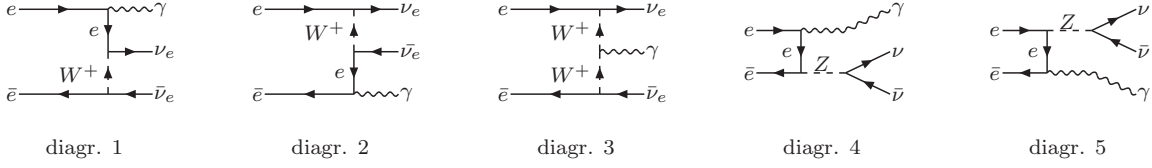
\noindent

For radiative neutrino production 
\begin{eqnarray}
e^-(p_1) + e^+(p_2) \rightarrow \nu(k_1) + \bar{\nu}(k_2) + \gamma(q),
\end{eqnarray}
we define the $W$ and $Z$ boson propagators as
\begin{eqnarray}
\Delta_W(p_i,k_j) & \equiv & 
\frac{1}{m_W^2 + 2\mink{p_i}{k_j}}\,,\\ 
\Delta_Z(k_1,k_2)& \equiv & \frac{1}{m_Z^2 -
  2\mink{k_1}{k_2} - \ie \Gamma_Z m_Z}\,.
\end{eqnarray}
\begin{table}[t]
\begin{center}
\caption{Vertex factors with the parameters $a$, $c$, $d$, and $f$ 
defined in Eq.~(\ref{eq:ncouplinga}) 
  and (\ref{eq:ncoupling}).}
\vspace*{5mm}
\begin{tabular}{cl}
\toprule
\vspace{2mm}
Vertex & Factor
\vspace*{1mm}\\
\midrule
{
\unitlength=1.25 pt
\SetScale{1.25}
\SetWidth{0.7}      
\scriptsize    
{} \qquad\allowbreak
\begin{picture}(95,79)(0,0)
\Text(15.0,60.0)[r]{$Z$}
\DashLine(16.0,60.0)(58.0,60.0){3.0} 
\Text(80.0,70.0)[l]{$\nu_\ell$}
\ArrowLine(58.0,60.0)(79.0,70.0) 
\Text(80.0,50.0)[l]{$\nu_\ell$}
\ArrowLine(79.0,50.0)(58.0,60.0) 
\end{picture} \ 
}
&\raisebox{2.5cm}{$ - \displaystyle{\frac{\ie {e}}{2}}f \gamma^\mu  
P_L, \quad \ell=e,\mu,\tau$}\\[-15mm]
{
\unitlength=1.25 pt
\SetScale{1.25}
\SetWidth{0.7}      
\scriptsize    
{} \qquad\allowbreak
\begin{picture}(95,79)(0,0)
\Text(15.0,63.0)[r]{$\beta \atop{\displaystyle \gamma}$}
\Photon(16.0,60.0)(58.0,60.0){1.0}{4}
\Text(37,65)[c]{$k_1\to$}
\Text(68,71)[c]{\rotatebox{30}{$\leftarrow k_3$}}
\Text(68,49)[c]{\rotatebox{-30}{$\leftarrow k_2$}}
\Text(80.0,70.0)[l]{$\alpha \atop {W^+}$}
\DashArrowLine(79.0,70.0)(58.0,60.0){3.0} 
\Text(80.0,50.0)[l]{$\mu \atop {W^-}$}
\DashArrowLine(79.0,50.0)(58.0,60.0){3.0} 
\end{picture} \ 
}
&\raisebox{2.5cm}{$- \ie e[(k_1 - k_2)_\alpha g_{\beta\mu} + (k_2 - k_3)_\beta g_{\mu\alpha}+ 
    (k_3 - k_1)_\mu g_{\alpha\beta}]$}\\[-15mm]
{
\unitlength=1.25 pt
\SetScale{1.25}
\SetWidth{0.7}      
\scriptsize    
{} \qquad\allowbreak
\begin{picture}(95,79)(0,0)
\Text(15.0,60.0)[r]{$W^+$}
\DashArrowLine(16.0,60.0)(58.0,60.0){3.0} 
\Text(80.0,70.0)[l]{$\nu_e$}
\ArrowLine(58.0,60.0)(79.0,70.0) 
\Text(80.0,50.0)[l]{$e$}
\ArrowLine(79.0,50.0)(58.0,60.0) 
\end{picture} \ 
}
&\raisebox{2.5cm}{$\displaystyle - \frac{1}{\sqrt{2}}\ie e a \gamma_\mu P_L$}\\[-15mm]
\bottomrule
\end{tabular}
\label{tab:vertexnu}
\end{center} 
\end{table}
The tree-level amplitudes for $W$ boson exchange, see the diagrams 1-3
in Fig.~\ref{fig:neutrino}, are then
\begin{eqnarray}
\M_1 &\!=\!& \frac{\ie e^3 a^2}{4\mink{q}{p_1}}\Delta_W(p_2,k_2)\Big[\vv(p_2)
\gamma^\mu P_L v(k_2)\Big]\,
         \Big[\uu(k_1)\gamma_\mu P_L (\ssl{q} - \ssl{p}_1)\ssl{\epsilon}^\ast 
u(p_1)\Big],\\[2mm]
\M_2 &\!=\!& \frac{\ie e^3 a^2}{4\mink{q}{p_2}}\Delta_W(p_1,k_1)\Big[\uu(k_1)
\gamma^\mu P_L u(p_1)\Big]
         \Big[\vv(p_2)\ssl{\epsilon}^\ast (\ssl{p}_2 - \ssl{q})\gamma_\mu P_L 
v(k_2)\Big],\\[2mm]         
\M_3 &\!=\!& \frac{1}{2}\ie e^3 a^2\Delta_W(p_1,k_1) \Delta_W(p_2,k_2) 
\Big[\uu(k_1)\gamma^\beta P_L u(p_1)\Big]\,
        \Big[\vv(p_2)\gamma^\alpha   P_L v(k_2)\Big] \notag\\[1mm]
    && \big((2 k_1 - 2 p_1 + q)_\mu g_{\alpha \beta}
        +(p_1 - k_1 -2 q)_\beta g_{\mu\alpha}+ (p_1-k_1+q)_\alpha g_{\beta\mu}\big) (\epsilon^\mu)^\ast,
\end{eqnarray}
with the parameter
\begin{eqnarray}
\label{eq:ncouplinga}
a  = \frac{1}{\sw}.
\end{eqnarray}
The amplitudes for $Z$ boson exchange, see diagrams 4 and 5 in
Fig.~\ref{fig:neutrino}, are
\begin{eqnarray}
\M_4 &\!=\!& \frac{\ie e^3 f}{4 \mink{q}{p_1}}\Delta_Z(k_1,k_2)\Big[\uu(k_1)
\gamma^\nu P_L v(k_2)\Big]\,
          \Big[\vv(p_2)\gamma_\nu(c P_L + d P_R)(\ssl{q} - \ssl{p}_1)\ssl{\epsilon}^
\ast u(p_1)\Big],\hspace*{10mm}\\[2mm]
\M_5 &\!=\!&  \frac{\ie e^3 f}{4 \mink{q}{p_2}}\Delta_Z(k_1,k_2)\Big[\uu(k_1)\gamma^
\nu P_L v(k_2)\Big]\,
          \Big[\vv(p_2)\ssl{\epsilon}^\ast(\ssl{p}_2 - \ssl{q})\gamma_\nu
(c P_L + d P_R)u(p_1)\Big], 
\end{eqnarray}
with the parameters 
\begin{eqnarray}
\label{eq:ncoupling}
c = \frac{1}{\sw\cw}\left(\frac{1}{2}-\sw[2]\right),\qquad
d = -\tw,\qquad        f =\frac{1}{\sw\cw}.
\end{eqnarray}
We have checked that the amplitudes $\M_i=
\epsilon_\mu\M^\mu_i$ for $i=1,\dots,5$ fulfill the Ward identity 
$q_\mu(\sum_i\M^\mu_i)=0$.
We find 
$q_\mu(\M^\mu_1+\M^\mu_2+\M^\mu_3)=0$ for  $W$ exchange and
$q_\mu(\M^\mu_4+\M^\mu_5)=0$ for $Z$ exchange.

We obtain the squared amplitudes $T_{ii}$ and $T_{ij}$ 
as defined in Eqs.~(\ref{Tii}) and (\ref{Tij}):
\begin{eqnarray}    
T_{11}& =& \frac{e^6 C_L a^4}{\mink{q}{p_1}}\Delta_W^2(p_2,k_2) 
        \mink{p_2}{k_1}  \mink{q}{k_2}\\[2mm]
T_{22} &=& \frac{e^6 C_L a^4}{\mink{q}{p_2}} \Delta_W^2(p_1,k_1) 
        \mink{p_1}{k_2}  \mink{q}{k_1}\\[2mm]   
T_{33} &=& e^6 C_L a^4  \Delta_W^2(p_2,k_2) \Delta_W^2(p_1,k_1) 
        \Big[\mink{p_2}{k_2}\mink{p_1}{k_1}\mink{p_1}{k_1} + 
        (\mink{p_2}{k_1}(7\mink{p_1}{k_2} -6\mink{q}{k_2}) + \notag\\[1mm] 
         &&\mink{p_2}{k_2}(\mink{q}{k_1} - \mink{q}{p_1}) - 
        \mink{q}{k_2}(\mink{p_1}{p_2} + 2\mink{q}{p_2})+ 
        \mink{p_1}{k_2}(\mink{p_1}{p_2} + 
        6\mink{q}{p_2}))\mink{p_1}{k_1} + \notag\\[1mm] 
         &&\mink{p_2}{k_1}\mink{q}{k_1}\mink{p_1}{k_2} - 
        3\mink{q}{k_1}\mink{p_1}{k_2}\mink{p_1}{p_2} +  
        \mink{q}{k_1}\mink{q}{k_2}\mink{p_1}{p_2} -
        \mink{p_2}{k_1}\mink{p_1}{k_2}\mink{q}{p_1} + \notag\\[1mm] 
        &&\mink{q}{k_1}\mink{p_2}{k_2}\mink{q}{p_1}+ 
        2\mink{p_2}{k_1}\mink{q}{k_2}\mink{q}{p_1} +   
        2\mink{q}{k_1}\mink{p_1}{k_2}\mink{q}{p_2} + 
        \mink{k_1}{k_2}\big(-2\mink{q}{k_1}\mink{p_1}{p_2} + \notag\\[1mm]  
        &&\mink{p_1}{k_1}(\mink{p_2}{k_1} - \mink{p_1}{p_2}+
        \mink{q}{p_2}) + \mink{q}{p_1}(3\mink{p_2}{k_1} +  
        2 \mink{p_1}{p_2} + \mink{q}{p_2})\big)\Big]\\[2mm]    
T_{44} &=& 3\frac{e^6 f^2}{\mink{q}{p_1}} |\Delta_Z(k_1,k_2)|^2  
        (C_L c^2 \mink{p_2}{k_1}\mink{q}{k_2} + 
        C_R d^2 \mink{p_2}{k_2}  \mink{q}{k_1})\\[2mm] 
T_{55} &=& 3\frac{e^6 f^2}{\mink{q}{p_2}}|\Delta_Z(k_1,k_2)|^2 
        (C_L c^2 \mink{p_1}{k_2}\mink{q}{k_1} + 
        C_R d^2 \mink{p_1}{k_1}\mink{q}{k_2})\\[2mm] 
T_{12} &=& \frac{e^6 C_L a^4}{\mink{q}{p_1}\mink{q}{p_2}}\Delta_W(p_1,k_1)\Delta_W(p_2,k_2) \notag\\[1mm]  
        &&\Big[2\mink{p_2}{k_1}\mink{p_1}{k_2}\mink{p_1}{p_2} - 
        \mink{q}{k_1}\mink{p_1}{k_2}\mink{p_1}{p_2} - 
        \mink{p_2}{k_1}\mink{q}{k_2}\mink{p_1}{p_2} -
        \mink{p_2}{k_1}\mink{p_1}{k_2}\mink{q}{p_1}  + \notag\\[1mm]   
        &&\mink{p_2}{k_1}\mink{p_2}{k_2}\mink{q}{p_1} + 
        \mink{p_1}{k_1}\mink{p_1}{k_2}\mink{q}{p_2}- 
       \mink{p_2}{k_1}\mink{p_1}{k_2}\mink{q}{p_2}\Big] \\[2mm] 
T_{13} &=& \frac{e^6 C_L a^4}{\mink{q}{p_1}} \Delta_W^2(p_2,k_2) \Delta_W(p_1,k_1) \notag\\[1mm]  
     &&\Big[4\mink{p_1}{k_1}\mink{p_2}{k_1}\mink{p_1}{k_2} -
        \mink{p_2}{k_1}\mink{q}{k_1}\mink{p_1}{k_2}-
        3\mink{q}{k_1}\mink{p_1}{p_2}\mink{p_1}{k_2}+
        3\mink{p_1}{k_1}\mink{q}{p_2}\mink{p_1}{k_2}-\notag\\[1mm]
        &&3\mink{p_1}{k_1}\mink{p_2}{k_1}\mink{q}{k_2}+
        \mink{q}{k_1}\mink{q}{k_2}\mink{p_1}{p_2}+
        \mink{k_1}{k_2}\mink{p_2}{k_1}\mink{q}{p_1}-
        \mink{p_1}{k_1}\mink{p_2}{k_2}\mink{q}{p_1}+\notag\\[1mm] 
        &&3 \mink{p_2}{k_1}\mink{q}{k_2}\mink{q}{p_1}+ 
        \mink{k_1}{k_2}\mink{p_1}{p_2}\mink{q}{p_1}-
        \mink{p_1}{k_1}\mink{q}{k_2}\mink{q}{p_2}\Big]\\[2mm]
T_{14} &=& - \frac{2e^6 C_L c f a^2}{\mink{q}{p_1}} \Delta_W(p_2,k_2)\real \{\Delta_Z(k_1,k_2)\}  
        \mink{p_2}{k_1} \mink{q}{k_2}\\[2mm]    
T_{15} &=& -\frac{e^6 C_L c f a^2}{\mink{q}{p_1}\mink{q}{p_2}}\Delta_W(p_2,k_2)\real \{\Delta_Z(k_1,k_2)\} \notag\\[1mm]  
        &&\Big[2\mink{p_2}{k_1}\mink{p_1}{k_2}\mink{p_1}{p_2} - 
        \mink{q}{k_1}\mink{p_1}{k_2}\mink{p_1}{p_2} -  
        \mink{p_2}{k_1}\mink{q}{k_2}\mink{p_1}{p_2} - \notag\\[1mm] 
       && \mink{p_2}{k_1}\mink{p_1}{k_2}\mink{q}{p_1} + 
        \mink{p_2}{k_1}\mink{p_2}{k_2}\mink{q}{p_1} + 
        \mink{p_1}{k_1}\mink{p_1}{k_2}\mink{q}{p_2} -  
        \mink{p_2}{k_1}\mink{p_1}{k_2}\mink{q}{p_2}\Big] 
       \\[2mm]
T_{23} &=& \frac{e^6 C_L a^4}{\mink{q}{p_2}}\Delta_W^2(p_1,k_1) \Delta_W(p_2,k_2)\notag\\[1mm]  
         && \Big[-3\mink{p_1}{k_2}\mink{p_2}{k_1}\mink{p_2}{k_1} +
             3\mink{q}{k_1}\mink{p_1}{k_2}\mink{p_2}{k_1} -
              \mink{p_1}{k_1}\mink{p_2}{k_1}\mink{p_2}{k_2}  + 
          \mink{k_1}{k_2}\mink{p_1}{p_2}\mink{p_2}{k_1} +\notag\\[1mm]
            &&  2\mink{p_1}{k_2}\mink{p_1}{p_2}\mink{p_2}{k_1} -
             \mink{q}{k_2}\mink{p_1}{p_2}\mink{p_2}{k_1}  - 
           2\mink{p_1}{k_2}\mink{q}{p_1}\mink{p_2}{k_1}+
             \mink{p_2}{k_2}\mink{q}{p_1}\mink{p_2}{k_1} -\notag\\[1mm]
           &&  3\mink{p_1}{k_2}\mink{q}{p_2}\mink{p_2}{k_1}+ 
          \mink{p_1}{k_1}\mink{q}{k_1}\mink{p_2}{k_2}-
            \mink{k_1}{k_2}\mink{q}{k_1}\mink{p_1}{p_2} -
            \mink{q}{k_1}\mink{q}{k_2}\mink{p_1}{p_2}+ \notag\\[1mm]
          && \mink{q}{k_1}\mink{p_2}{k_2}\mink{q}{p_1}+
             2\mink{p_1}{k_1}\mink{p_1}{k_2}\mink{q}{p_2}+
             3 \mink{q}{k_1}\mink{p_1}{k_2}\mink{q}{p_2}\Big] \\[2mm]
T_{24} &=& -\frac{e^6 C_L c f a^2}{\mink{q}{p_1}\mink{q}{p_2}}\Delta_W(p_1,k_1)
	\real \{\Delta_Z(k_1,k_2)\}\notag\\[1mm]  
        &&\Big[2\mink{p_2}{k_1}\mink{p_1}{k_2}\mink{p_1}{p_2} - 
        \mink{q}{k_1}\mink{p_1}{k_2}\mink{p_1}{p_2} -  
        \mink{p_2}{k_1}\mink{q}{k_2}\mink{p_1}{p_2} - 
        \mink{p_2}{k_1}\mink{p_1}{k_2}\mink{q}{p_1}  + \notag\\[1mm]  
        &&\mink{p_2}{k_1}\mink{p_2}{k_2}\mink{q}{p_1} + 
        \mink{p_1}{k_1}\mink{p_1}{k_2}\mink{q}{p_2} -  
        \mink{p_2}{k_1}\mink{p_1}{k_2}\mink{q}{p_2}\Big] 
          \\[2mm]
T_{25} &=& - \frac{2e^6 C_L c f a^2}{\mink{q}{p_2}}\Delta_W(p_1,k_1)\real \{\Delta_Z(k_1,k_2)\} 
        \mink{p_1}{k_2} \mink{q}{k_1} \\[2mm]
T_{34} &=& - \frac{e^6 C_L c f a^2}{\mink{q}{p_1}} \Delta_W(p_1,k_1)\Delta_W(p_2,k_2)\real \{\Delta_Z(k_1,k_2)\} \notag\\[1mm]  
         &&\Big[4\mink{p_1}{k_1}\mink{p_2}{k_1}\mink{p_1}{k_2} -  
             \mink{p_2}{k_1}\mink{q}{k_1}\mink{p_1}{k_2}-
             3\mink{q}{k_1}\mink{p_1}{p_2}\mink{p_1}{k_2}+ 
          3\mink{p_1}{k_1}\mink{q}{p_2}\mink{p_1}{k_2} - \notag\\[1mm]
            && 3\mink{p_1}{k_1}\mink{p_2}{k_1}\mink{q}{k_2}+
             \mink{q}{k_1} \mink{q}{k_2}\mink{p_1}{p_2}+
          \mink{k_1}{k_2}\mink{p_2}{k_1}\mink{q}{p_1}-
             \mink{p_1}{k_1}\mink{p_2}{k_2}\mink{q}{p_1}+ \notag\\[1mm]
          &&    3\mink{p_2}{k_1}\mink{q}{k_2}\mink{q}{p_1}+
          \mink{k_1}{k_2}\mink{p_1}{p_2}\mink{q}{p_1}-
             \mink{p_1}{k_1}\mink{q}{k_2}\mink{q}{p_2}\Big]
               \\[2mm] 
T_{35} &=& -\frac{e^6 C_L c f a^2}{\mink{q}{p_2}}\Delta_W(p_1,k_1)\Delta_W(p_2,k_2) 
		\real \{\Delta_Z(k_1,k_2)\}\notag\\[1mm]  
    && \Big[-3\mink{p_1}{k_2}\mink{p_2}{k_1}\mink{p_2}{k_1}+
        3\mink{q}{k_1}\mink{p_1}{k_2}\mink{p_2}{k_1}-
        \mink{p_1}{k_1}\mink{p_2}{k_2}\mink{p_2}{k_1}+
     \mink{k_1}{k_2}\mink{p_1}{p_2}\mink{p_2}{k_1}+ \notag\\[1mm]   
     && 2\mink{p_1}{k_2}\mink{p_1}{p_2}\mink{p_2}{k_1} -
        \mink{q}{k_2}\mink{p_1}{p_2}\mink{p_2}{k_1}-
      2\mink{p_1}{k_2}\mink{q}{p_1}\mink{p_2}{k_1}+
        \mink{p_2}{k_2}\mink{q}{p_1}\mink{p_2}{k_1}-\notag\\[1mm]   
        && 3\mink{p_1}{k_2}\mink{q}{p_2}\mink{p_2}{k_1}+
       \mink{p_1}{k_1}\mink{q}{k_1}\mink{p_2}{k_2}-
        \mink{k_1}{k_2}\mink{p_1}{p_2}\mink{q}{k_1}-
        \mink{q}{k_1}\mink{q}{k_2}\mink{p_1}{p_2}+\notag\\[1mm]   
      &&\mink{q}{k_1}\mink{p_2}{k_2}\mink{q}{p_1}+
        2\mink{p_1}{k_1}\mink{p_1}{k_2}\mink{q}{p_2}+
        3\mink{q}{k_1}\mink{p_1}{k_2}\mink{q}{p_2} \Big]
     \\[2mm]
T_{45} &=& \frac{3 e^6f^2}{\mink{q}{p_1}\mink{q}{p_2}} |\Delta_Z(k_1,k_2)|^2  \notag\\[1mm]  
        &&\Big[C_L c^2 \big(2\mink{p_2}{k_1}\mink{p_1}{k_2}\mink{p_1}{p_2} - 
        \mink{q}{k_1}\mink{p_1}{k_2}\mink{p_1}{p_2} -  
        \mink{p_2}{k_1}\mink{q}{k_2}\mink{p_1}{p_2} -\notag\\[1mm] 
        &&\mink{p_2}{k_1}\mink{p_1}{k_2}\mink{q}{p_1}  + 
        \mink{p_2}{k_1}\mink{p_2}{k_2}\mink{q}{p_1} + 
        \mink{p_1}{k_1}\mink{p_1}{k_2}\mink{q}{p_2} - 
        \mink{p_2}{k_1}\mink{p_1}{k_2}\mink{q}{p_2}\big) + \notag\\[1mm] 
        &&C_R d^2  \big(2\mink{p_1}{k_1}\mink{p_2}{k_2}\mink{p_1}{p_2} -  
        \mink{q}{k_1}\mink{p_2}{k_2}\mink{p_1}{p_2} - 
        \mink{p_1}{k_1}\mink{q}{k_2}\mink{p_1}{p_2} -  \notag\\[1mm] 
        &&\mink{p_1}{k_1}\mink{p_2}{k_2}\mink{q}{p_1}  + 
        \mink{p_2}{k_1}\mink{p_2}{k_2}\mink{q}{p_1} +   
        \mink{p_1}{k_1}\mink{p_1}{k_2}\mink{q}{p_2} -
        \mink{p_1}{k_1}\mink{p_2}{k_2}\mink{q}{p_2}\big)\Big]\notag\\
&&    
\end{eqnarray}   
We have calculated the squared amplitudes with \texttt{FeynCalc}~\cite{Kublbeck:1992mt}.
We neglect terms proportional to 
$\eps \imag \{\Delta_Z\}$,
see the discussion at the end of Appendix~\ref{sec:app:chifore}.

\section{Amplitudes for Radiative Sneutrino Production}
\label{sec:app:snuback}

\begin{figure}[t]
{%
\unitlength=1.0pt
\SetScale{1.0}
\SetWidth{0.7}      
\scriptsize    
\allowbreak
\begin{picture}(95,79)(0,0)
\Text(15.0,70.0)[r]{$e$}
\ArrowLine(16.0,70.0)(58.0,70.0) 
\Text(80.0,70.0)[l]{$\gamma$}
\Photon(58.0,70.0)(79.0,70.0){1.0}{4} 
\Text(54.0,60.0)[r]{$e$}
\ArrowLine(58.0,70.0)(58.0,50.0) 
\Text(80.0,50.0)[l]{$\widetilde{\nu}_e$}
\DashArrowLine(58.0,50.0)(79.0,50.0){1.0} 
\Text(54.0,40.0)[r]{$\widetilde{\chi}^+_1$}
\ArrowLine(58.0,30.0)(58.0,50.0) 
\Text(15.0,30.0)[r]{$\bar{e}$}
\ArrowLine(58.0,30.0)(16.0,30.0) 
\Text(80.0,30.0)[l]{${\widetilde{\nu}^\ast_e}$}
\DashArrowLine(79.0,30.0)(58.0,30.0){1.0} 
\Text(47,0)[b] {diagr. 1}
\end{picture} \ 
{} \qquad\allowbreak
\begin{picture}(95,79)(0,0)
\Text(15.0,70.0)[r]{$e$}
\ArrowLine(16.0,70.0)(58.0,70.0) 
\Text(80.0,70.0)[l]{$\widetilde{\nu}_e$}
\DashArrowLine(58.0,70.0)(79.0,70.0){1.0} 
\Text(54.0,60.0)[r]{$\widetilde{\chi}^+_1$}
\ArrowLine(58.0,50.0)(58.0,70.0) 
\Text(80.0,50.0)[l]{${\widetilde{\nu}^\ast_e}$}
\DashArrowLine(79.0,50.0)(58.0,50.0){1.0} 
\Text(54.0,40.0)[r]{$e$}
\ArrowLine(58.0,50.0)(58.0,30.0) 
\Text(15.0,30.0)[r]{$\bar{e}$}
\ArrowLine(58.0,30.0)(16.0,30.0) 
\Text(80.0,30.0)[l]{$\gamma$}
\Photon(58.0,30.0)(79.0,30.0){1.0}{4} 
\Text(47,0)[b] {diagr. 2}
\end{picture} \ 
{} \qquad\allowbreak
\begin{picture}(95,79)(0,0)
\Text(15.0,70.0)[r]{$e$}
\ArrowLine(16.0,70.0)(58.0,70.0) 
\Text(80.0,70.0)[l]{$\widetilde{\nu}_e$}
\DashArrowLine(58.0,70.0)(79.0,70.0){1.0} 
\Text(54.0,60.0)[r]{$\widetilde{\chi}^+_1$}
\ArrowLine(58.0,50.0)(58.0,70.0) 
\Text(80.0,50.0)[l]{$\gamma$}
\Photon(58.0,50.0)(79.0,50.0){1.0}{4} 
\Text(54.0,40.0)[r]{$\widetilde{\chi}^+_1$}
\ArrowLine(58.0,30.0)(58.0,50.0) 
\Text(15.0,30.0)[r]{$\bar{e}$}
\ArrowLine(58.0,30.0)(16.0,30.0) 
\Text(80.0,30.0)[l]{$\widetilde{\nu}^\ast_e$}
\DashArrowLine(79.0,30.0)(58.0,30.0){1.0} 
\Text(47,0)[b] {diagr. 3}
\end{picture} \ 
{} \qquad\allowbreak
\begin{picture}(95,79)(0,0)
\Text(15.0,70.0)[r]{$e$}
\ArrowLine(16.0,70.0)(58.0,70.0) 
\Text(80.0,70.0)[l]{$\gamma$}
\Photon(58.0,70.0)(79.0,70.0){1.0}{4} 
\Text(54.0,60.0)[r]{$e$}
\ArrowLine(58.0,70.0)(58.0,50.0) 
\Text(80.0,50.0)[l]{$\widetilde{\nu}_e$}
\DashArrowLine(58.0,50.0)(79.0,50.0){1.0} 
\Text(54.0,40.0)[r]{$\widetilde{\chi}^+_2$}
\ArrowLine(58.0,30.0)(58.0,50.0) 
\Text(15.0,30.0)[r]{$\bar{e}$}
\ArrowLine(58.0,30.0)(16.0,30.0) 
\Text(80.0,30.0)[l]{${\widetilde{\nu}^\ast_e}$}
\DashArrowLine(79.0,30.0)(58.0,30.0){1.0} 
\Text(47,0)[b] {diagr. 4}
\end{picture} \ 
{} \qquad\allowbreak
\begin{picture}(95,79)(0,0)
\Text(15.0,70.0)[r]{$e$}
\ArrowLine(16.0,70.0)(58.0,70.0) 
\Text(80.0,70.0)[l]{$\widetilde{\nu}_e$}
\DashArrowLine(58.0,70.0)(79.0,70.0){1.0} 
\Text(54.0,60.0)[r]{$\widetilde{\chi}^+_2$}
\ArrowLine(58.0,50.0)(58.0,70.0) 
\Text(80.0,50.0)[l]{${\widetilde{\nu}^\ast_e}$}
\DashArrowLine(79.0,50.0)(58.0,50.0){1.0} 
\Text(54.0,40.0)[r]{$e$}
\ArrowLine(58.0,50.0)(58.0,30.0) 
\Text(15.0,30.0)[r]{$\bar{e}$}
\ArrowLine(58.0,30.0)(16.0,30.0) 
\Text(80.0,30.0)[l]{$\gamma$}
\Photon(58.0,30.0)(79.0,30.0){1.0}{5} 
\Text(47,0)[b] {diagr. 5}
\end{picture} \ 
{} \qquad\allowbreak
\begin{picture}(95,79)(0,0)
\Text(15.0,70.0)[r]{$e$}
\ArrowLine(16.0,70.0)(58.0,70.0) 
\Text(80.0,70.0)[l]{$\widetilde{\nu}_e$}
\DashArrowLine(58.0,70.0)(79.0,70.0){1.0} 
\Text(54.0,60.0)[r]{$\widetilde{\chi}^+_2$}
\ArrowLine(58.0,50.0)(58.0,70.0) 
\Text(80.0,50.0)[l]{$\gamma$}
\Photon(58.0,50.0)(79.0,50.0){1.0}{4} 
\Text(54.0,40.0)[r]{$\widetilde{\chi}^+_2$}
\ArrowLine(58.0,30.0)(58.0,50.0) 
\Text(15.0,30.0)[r]{$\bar{e}$}
\ArrowLine(58.0,30.0)(16.0,30.0) 
\Text(80.0,30.0)[l]{${\widetilde{\nu}^\ast}_e$}
\DashArrowLine(79.0,30.0)(58.0,30.0){1.0} 
\Text(47,0)[b] {diagr. 6}
\end{picture} \ 
{} \qquad\allowbreak
\begin{picture}(95,79)(0,0)
\Text(15.0,60.0)[r]{$e$}
\ArrowLine(16.0,60.0)(37.0,60.0) 
\Photon(37.0,60.0)(58.0,60.0){1.0}{4} 
\Text(80.0,70.0)[l]{$\gamma$}
\Photon(58.0,60.0)(79.0,70.0){1.0}{4} 
\Text(33.0,50.0)[r]{$e$}
\ArrowLine(37.0,60.0)(37.0,40.0) 
\Text(15.0,40.0)[r]{$\bar{e}$}
\ArrowLine(37.0,40.0)(16.0,40.0) 
\Text(47.0,41.0)[b]{$Z$}
\DashLine(37.0,40.0)(58.0,40.0){3.0} 
\Text(80.0,50.0)[l]{$\widetilde{\nu}$}
\DashArrowLine(58.0,40.0)(79.0,50.0){1.0} 
\Text(80.0,30.0)[l]{${\widetilde{\nu}^\ast}$}
\DashArrowLine(79.0,30.0)(58.0,40.0){1.0} 
\Text(47,0)[b] {diagr. 7}
\end{picture} \ 
{} \qquad\allowbreak
\begin{picture}(95,79)(0,0)
\Text(15.0,60.0)[r]{$e$}
\ArrowLine(16.0,60.0)(37.0,60.0) 
\Text(47.0,61.0)[b]{$Z$}
\DashLine(37.0,60.0)(58.0,60.0){3.0} 
\Text(80.0,70.0)[l]{$\widetilde{\nu}$}
\DashArrowLine(58.0,60.0)(79.0,70.0){1.0} 
\Text(80.0,50.0)[l]{${\widetilde{\nu}^\ast}$}
\DashArrowLine(79.0,50.0)(58.0,60.0){1.0} 
\Text(33.0,50.0)[r]{$e$}
\ArrowLine(37.0,60.0)(37.0,40.0) 
\Text(15.0,40.0)[r]{$\bar{e}$}
\ArrowLine(37.0,40.0)(16.0,40.0) 
\Photon(37.0,40.0)(58.0,40.0){1.0}{4} 
\Text(80.0,30.0)[l]{$\gamma$}
\Photon(58.0,40.0)(79.0,30.0){1.0}{4} 
\Text(47,0)[b] {diagr. 8}
\end{picture} \
}
\caption{Contributing diagrams to 
$e^+e^- \rightarrow \tilde{\nu}\tilde{\nu}^\ast\gamma$~\cite{Boos:2004kh}.}
\label{fig:sneutrino}
\end{figure}
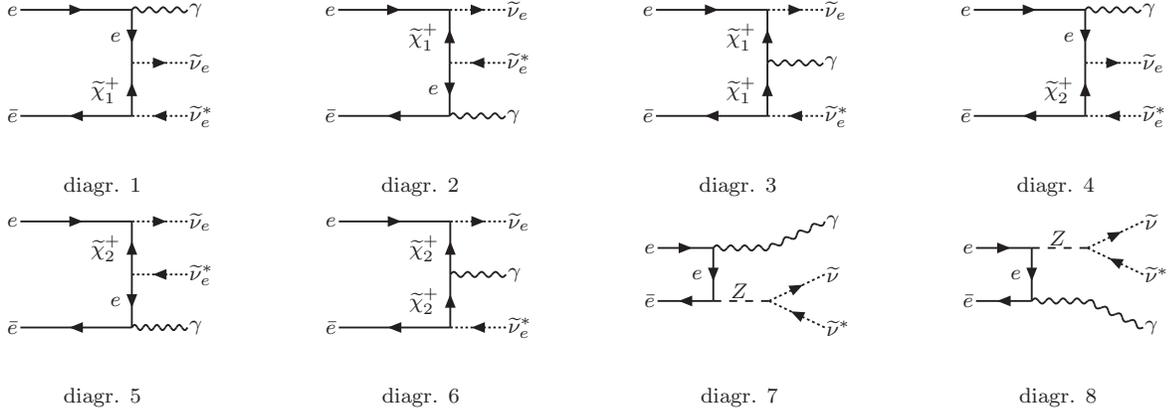
\noindent
For radiative sneutrino production 
\begin{eqnarray}
e^-(p_1) + e^+(p_2) \rightarrow \tilde{\nu}(k_1) + \tilde{\nu}^\ast(k_2) + \gamma(q)
\end{eqnarray}
we define the chargino and $Z$ boson propagators as
\begin{eqnarray}
\Delta_{{\chi}_{1,2}^+}(p_i,k_j) & \equiv & 
\frac{1}{m_{{\chi}_{1,2}^+}^2 - m_{\tilde\nu}^2 + 2\mink{p_i}{k_j}},\\ 
\Delta_Z(k_1,k_2)& \equiv & \frac{1}{m_Z^2 -  2 m_{\tilde\nu}^2 - 2\mink{k_1}{k_2} - \ie \Gamma_Z m_Z}.
\end{eqnarray}

\begin{table}[t]
\begin{center}
\caption{Vertex factors with parameters $a$, $f$ defined in Eqs.~(\ref{eq:ncouplinga}) 
and (\ref{eq:ncoupling}), and $C$ the charge conjugation operator.}
\vspace*{5mm}
\begin{tabular}{cl}
\toprule
\vspace{2mm}
Vertex & Factor
\vspace*{1mm}\\
\midrule
{
\unitlength=1.25 pt
\SetScale{1.25}
\SetWidth{0.7}      
\scriptsize    
{} \qquad\allowbreak
\begin{picture}(95,79)(0,0)
\Text(15.0,60.0)[r]{$Z$}
\DashLine(16.0,60.0)(58.0,60.0){3.0} 
\Text(80.0,70.0)[l]{$\widetilde\nu_\ell$}
\DashArrowLine(58.0,60.0)(79.0,70.0){1.0} 
\Text(80.0,50.0)[l]{$\widetilde\nu^\ast_\ell$}
\DashArrowLine(79.0,50.0)(58.0,60.0){1.0} 
\Text(68,71)[c]{\rotatebox{30}{$\rightarrow p_{\tilde\nu}$}}
\Text(68,49)[c]{\rotatebox{-30}{$\leftarrow p_{\tilde\nu^\ast}$}}
\end{picture} \ 
}
&\raisebox{2.5cm}{$-\half \ie e f (p_{\tilde\nu} + p_{\tilde\nu^\ast})_\mu, 
                   \quad \ell = e,\mu,\tau$}\\[-15mm]
{
\unitlength=1.25 pt
\SetScale{1.25}
\SetWidth{0.7}      
\scriptsize    
{} \qquad\allowbreak
\begin{picture}(95,79)(0,0)
\Text(15.0,60.0)[r]{$\widetilde{\chi}^+_j$}
\ArrowLine(16.0,60.0)(58.0,60.0) 
\Text(80.0,70.0)[l]{$\widetilde\nu_e$}
\DashArrowLine(58.0,60.0)(79.0,70.0){1.0} 
\Text(80.0,50.0)[l]{$ e$}
\ArrowLine(79.0,50.0)(58.0,60.0) 
\end{picture} \ 
}
&\raisebox{2.5cm}{$-\ie e a  V_{j1} P_R C, \quad \tilde\chi^+_j\,{\rm transposed} $}\\[-15mm]
{
\unitlength=1.25 pt
\SetScale{1.25}
\SetWidth{0.7}      
\scriptsize    
{} \qquad\allowbreak
\begin{picture}(95,79)(0,0)
\Text(15.0,60.0)[r]{$\gamma$}
\Photon(16.0,60.0)(58.0,60.0){1.0}{4} 
\Text(80.0,70.0)[l]{$\widetilde\chi^-_j$}
\ArrowLine(58.0,60.0)(79.0,70.0) 
\Text(80.0,50.0)[l]{$\widetilde\chi^+_j$}
\ArrowLine(79.0,50.0)(58.0,60.0) 
\end{picture} \ 
}
&\raisebox{2.5cm}{$ -\ie e \gamma_\mu$}\\[-15mm]
\bottomrule
\end{tabular}
\label{tab:vertexsnu}
\end{center}
\end{table}
The tree-level amplitudes for chargino $\tilde{\chi}_1^\pm$ exchange,
see the contributing diagrams 1-3 in Fig.~\ref{fig:sneutrino}, are
\begin{eqnarray}
\M_1 &=& \frac{\ie e^3 a^2|V_{11}|^2}{2 \mink{q}{p_1}}\Delta_{\chi_1^+}(p_2,k_2) 
    \Big[\vv(p_2) P_R (\ssl{p}_2 - \ssl{k}_2 -  m_{{\chi}_1^+})P_L(\ssl{p}_1 - 
\ssl{q})\ssl{\epsilon}^\ast u(p_1)\Big],
   \hspace*{0mm}\\[2mm]
\M_2 &=&-\frac{\ie e^3 a^2|V_{11}|^2}{2 \mink{q}{p_2}}\Delta_{\chi_1^+}(p_1,k_1)
\Big[\vv(p_2)\ssl{\epsilon}^\ast (\ssl{p}_2 - \ssl{q}) 
           P_R(\ssl{k}_1 - \ssl{p}_1 -  m_{{\chi}_1^+})P_L u(p_1)\Big],\\[3mm]
\M_3 &=& -{\ie e^3 a^2|V_{11}|^2}\Delta_{\chi_1^+}(p_1,k_1)\Delta_{\chi_1^+}(p_2,k_2) 
\notag \\[2mm]
       &&   \Big[\vv(p_2) P_R (\ssl{p}_2 - \ssl{k}_2 -  m_{{\chi}_1^+})
\ssl{\epsilon}^\ast
            (\ssl{k}_1 - \ssl{p}_1 -  m_{{\chi}_1^+})P_L u(p_1)\Big],
\end{eqnarray}
with the parameter $a$ defined in Eq.~(\ref{eq:ncouplinga}).  The
$2\times2$ matrices $U$ and $V$ diagonalise the chargino mass matrix
$X$~\cite{Haber:1984rc}
\begin{eqnarray}
U^\ast X V^{-1} = \mathrm{diag}\begin{pmatrix}m_{\chi_1^+},& m_{\chi_2^+}\end{pmatrix}.
\end{eqnarray}
The amplitudes for  chargino $\tilde{\chi}_2^\pm$ exchange, 
see the contributing diagrams 4-6 in Fig.~\ref{fig:sneutrino}, are
\begin{eqnarray}
\M_4 &=& \frac{\ie e^3 a^2|V_{21}|^2}{2 \mink{q}{p_1}}\Delta_{\chi_2^+}(p_2,k_2) 
\Big[\vv(p_2) P_R
       (\ssl{p}_2 - \ssl{k}_2 -  m_{{\chi}_2^+})P_L(\ssl{p}_1 - \ssl{q})
\ssl{\epsilon}^\ast u(p_1)\Big],\\[2mm] 
\M_5 &=&-\frac{\ie e^3 a^2|V_{21}|^2}{2 \mink{q}{p_1}}\Delta_{\chi_2^+}(p_1,k_1)
\Big[\vv(p_2)\ssl{\epsilon}^\ast (\ssl{p}_2 - \ssl{q})
           P_R(\ssl{k}_1 - \ssl{p}_1 -  m_{{\chi}_2^+})P_L u(p_1)\Big],\\[3mm] 
\M_6 &=& -{\ie e^3 a^2|V_{21}|^2}\Delta_{\chi_2^+}(p_1,k_1)\Delta_{\chi_2^+}
(p_2,k_2) \notag\\[2mm]
       &&   \Big[\vv(p_2) P_R (\ssl{p}_2 - \ssl{k}_2 -  m_{{\chi}_2^+})
\ssl{\epsilon}^\ast 
            (\ssl{k}_1 - \ssl{p}_1 -  m_{{\chi}_2^+})P_L u(p_1)\Big].
\end{eqnarray}
The amplitudes for $Z$ boson exchange, see the diagrams 7 and 8 in 
Fig.~\ref{fig:sneutrino}, read
\begin{eqnarray}
\M_7 &=&  \frac{\ie e^3 f}{4 \mink{q}{p_1}}\Delta_Z(k_1,k_2)\Big[\vv(p_2)
(\ssl{k}_1-\ssl{k}_2) 
          (c P_L + d P_R) (\ssl{p}_1 - \ssl{q})\ssl{\epsilon}^\ast u(p_1)\Big],\\[2mm]
\M_8 &=&  \frac{\ie e^3 f}{4 \mink{q}{p_2}}\Delta_Z(k_1,k_2)\Big[\vv(p_2)
\ssl{\epsilon}^\ast 
          (\ssl{q} - \ssl{p}_2)(\ssl{k}_1-\ssl{k}_2)(c P_L + d P_R)u(p_1)\Big],
\end{eqnarray}
with the parameters $c$, $d$, and $f$ defined in
Eq.~(\ref{eq:ncoupling}).  We have checked that the amplitudes $\M_i=
\epsilon_\mu\M^\mu_i$, $i=1,\dots,8$, fulfill the Ward identity $q_\mu
(\sum_i\M^\mu_i)=0$, as done in Ref.~\cite{Franke:thesis}. We find $q_
\mu(\M^\mu_1+\M^\mu_2+\M^\mu_3)=0$ for $\tilde{\chi}_1^\pm$ exchange,
$q_\mu(\M^\mu_4+\M^\mu_5+\M^\mu_6)=0$ for $\tilde{\chi}_2^\pm$
exchange, and $q_\mu(\M^\mu_7+\M^\mu_8)=0$ for $Z$ boson exchange.
Our amplitudes for chargino and $Z$ boson exchange agree with those
given in Refs.~\cite{Franke:thesis,Franke:1994ph}, and in the limit of
vanishing chargino mixing with those of Ref.~\cite{Chen:1987ux}.
However, there are obvious misprints in the amplitudes $M_2$ and $M_4$
of Ref.~\cite{Franke:1994ph}, see their Eqs.~(7) and (9),
respectively, and in the amplitude $T_5$ of Ref.~\cite{Chen:1987ux},
see their Eq.~(F.3).

We then obtain the squared amplitudes 
$T_{ii}$ and $T_{ij}$ 
as defined in Eqs.~(\ref{Tii}) and (\ref{Tij}):
\begin{eqnarray}
  T_{1 1} &=& \frac{e^6 C_L a^4 |V_{11}|^4}{2 \mink{q}{p_1} } \Delta_{\chi_1^+}^2(p_2,k_2)
  (2\mink{p_2}{k_2} \mink{q}{k_2} - m_{\tilde{\nu}}^2 \mink{q}{p_2})  \\[2mm]
  T_{2 2} &=& \frac{e^6 C_L a^4 |V_{11}|^4}{2 \mink{q}{p_2} } \Delta_{\chi_1^+}^2(p_1,k_1)
  (2\mink{p_1}{k_1} \mink{q}{k_1} - m_{\tilde{\nu}}^2 \mink{q}{p_1})   \\[2mm]
  T_{3 3} &=&{e^6 C_L a^4 |V_{11}|^4}\Delta_{\chi_1^+}^2(p_1,k_1) \Delta_{\chi_1^+}^2(p_2,k_2)
  \big[m_{\chi_1^+}^4 \mink{p_1}{p_2} + 
  4 m_{{\chi_1^+}}^2 \mink{p_1}{k_1} \mink{p_2}{k_2} - \notag\\[1mm]
  && 2m_{\tilde{\nu}}^2 \mink{p_1}{k_1} \mink{p_2}{k_1} +
  4\mink{k_1}{k_2} \mink{p_1}{k_1} \mink{p_2}{k_2} -  
  2m_{\tilde{\nu}}^2 \mink{p_1}{k_2} \mink{p_2}{k_2} + 
  m_{\tilde{\nu}}^4 \mink{p_1}{p_2} \big] \\[2mm] 
  T_{4 4} &=& \frac{e^6 C_L a^4 |V_{21}|^4}{2\mink{q}{p_1}} \Delta_{\chi_2^+}^2(p_2,k_2)
  (2\mink{p_2}{k_2} \mink{q}{k_2} - m_{\tilde{\nu}}^2 \mink{q}{p_2})  \\[2mm] 
  T_{5 5} &=& \frac{e^6 C_L a^4 |V_{21}|^4}{2\mink{q}{p_2} } \Delta_{\chi_2^+}^2(p_1,k_1) 
  (2\mink{p_1}{k_1} \mink{q}{k_1} - m_{\tilde{\nu}}^2 \mink{q}{p_1})  \\[2mm] 
  T_{6 6} &=& {e^6 C_L a^4 |V_{21}|^4}\Delta_{\chi_2^+}^2(p_1,k_1) \Delta_{\chi_2^+}^2(p_2,k_2)
  \big[m_{\chi^+_2}^4 \mink{p_1}{p_2} + 
  4  m_{{\chi_2^+}}^2\mink{p_1}{k_1} \mink{p_2}{k_2} -  \notag\\[1mm]
  &&2m_{\tilde{\nu}}^2 \mink{p_1}{k_1} \mink{p_2}{k_1} +  
  4\mink{k_1}{k_2} \mink{p_1}{k_1} \mink{p_2}{k_2} -  
  2m_{\tilde{\nu}}^2 \mink{p_1}{k_2} \mink{p_2}{k_2} + 
  m_{\tilde{\nu}}^4 \mink{p_1}{p_2}\big]  \\[2mm] 
  T_{7 7} &=& 3\frac{e^6 f^2 (C_L c^2 + C_R d^2)}{4 \mink{q}{p_1}}  |\Delta_Z(k_1,k_2)|^2 \notag\\[1mm]
  &&\big[\mink{p_2}{k_1} \mink{q}{k_1} - \mink{p_2}{k_2} \mink{q}{k_1} - 
  \mink{p_2}{k_1} \mink{q}{k_2} + \mink{p_2}{k_2} \mink{q}{k_2} -
  m_{\tilde{\nu}}^2 \mink{q}{p_2} + \mink{k_1}{k_2} \mink{q}{p_2}\big]  \\[2mm]
  T_{8 8} &=& 3\frac{e^6 f^2 (C_L c^2 + C_R d^2)}{4 \mink{q}{p_2}}  |\Delta_Z(k_1,k_2)|^2 \notag\\[1mm]
  &&\big[\mink{p_1}{k_1} \mink{q}{k_1} - \mink{p_1}{k_2} \mink{q}{k_1} - 
  \mink{p_1}{k_1} \mink{q}{k_2} + \mink{p_1}{k_2} \mink{q}{k_2} -
   m_{\tilde{\nu}}^2 \mink{q}{p_1} + \mink{k_1}{k_2} \mink{q}{p_1}\big] \\[2mm]
  T_{1 2} &=& -\frac{e^6 C_L a^4 |V_{11}|^4}{\mink{q}{p_1} \mink{q}{p_2}}  
  \Delta_{\chi_1^+}(p_1,k_1) \Delta_{\chi_1^+}(p_2,k_2)\notag\\[1mm]
  && \big[-\mink{k_1}{k_2} \mink{p_1}{p_2} \mink{p_1}{p_2} + 
  \mink{p_2}{k_1} \mink{p_1}{k_2} \mink{p_1}{p_2} +  
  \mink{p_1}{k_1} \mink{p_2}{k_2} \mink{p_1}{p_2} -\notag\\[1mm]
 && \mink{q}{k_1} \mink{p_2}{k_2} \mink{p_1}{p_2} -  
  \mink{p_1}{k_1} \mink{q}{k_2} \mink{p_1}{p_2} + 
  \mink{k_1}{k_2} \mink{q}{p_1} \mink{p_1}{p_2} + 
  \mink{k_1}{k_2} \mink{q}{p_2} \mink{p_1}{p_2} - \notag\\[1mm]
 && \mink{p_2}{k_1} \mink{p_1}{k_2} \mink{q}{p_1} +  
  \mink{p_2}{k_1} \mink{p_2}{k_2} \mink{q}{p_1} +
  \mink{p_1}{k_1} \mink{p_1}{k_2} \mink{q}{p_2} -  
  \mink{p_2}{k_1} \mink{p_1}{k_2} \mink{q}{p_2}\big]  \\[2mm]
  T_{1 3} &=& -\frac{e^2 C_L a^4 |V_{11}|^4}{\mink{q}{p_1}}   
  \Delta_{\chi_1^+}(p_1,k_1) \Delta_{\chi_1^+}^2(p_2,k_2) \notag\\[1mm]
  &&\big[m_{\chi^+_1}^2 \mink{q}{k_2} \mink{p_1}{p_2} +  
  m_{\chi^+_1}^2 \mink{q}{p_1} \mink{p_2}{k_2} - 
  m_{\chi^+_1}^2 \mink{q}{p_2} \mink{p_1}{k_2} -
   4\mink{p_1}{k_1} \mink{p_1}{k_2} \mink{p_2}{k_2} + \notag\\[1mm] 
   &&4\mink{p_1}{k_1} \mink{p_2}{k_2} \mink{q}{k_2} +  
   2m_{\tilde{\nu}}^2 \mink{p_1}{k_1} \mink{p_1}{p_2} - \ 
   2m_{\tilde{\nu}}^2 \mink{p_1}{k_1} \mink{q}{p_2}\big]  \\[2mm]
  T_{1 4} &=& \frac{e^6 C_L a^4 |V_{11}|^2 |V_{21}|^2}{\mink{q}{p_1}}
  	\Delta_{\chi_1^+}(p_2,k_2) \Delta_{\chi_2^+}(p_2,k_2) 
  \big[2\mink{p_2}{k_2} \mink{q}{k_2} -  
  m_{\tilde{\nu}}^2 \mink{q}{p_2}\big] \\[2mm]
  T_{1 5} &=& -\frac{e^6 C_L a^4 |V_{11}|^2 |V_{21}|^2}{\mink{q}{p_1} \mink{q}{p_2}}
  \Delta_{\chi_2^+}(p_1,k_1) \Delta_{\chi_1^+}(p_2,k_2)\notag\\[1mm]
  &&\big[-\mink{k_1}{k_2} \mink{p_1}{p_2} \mink{p_1}{p_2} + 
  \mink{p_2}{k_1} \mink{p_1}{k_2} \mink{p_1}{p_2} +  
  \mink{p_1}{k_1} \mink{p_2}{k_2} \mink{p_1}{p_2} -
  \mink{q}{k_1} \mink{p_2}{k_2} \mink{p_1}{p_2} - \notag\\[1mm] 
  &&\mink{p_1}{k_1} \mink{q}{k_2} \mink{p_1}{p_2} + 
  \mink{k_1}{k_2} \mink{q}{p_1} \mink{p_1}{p_2} + 
  \mink{k_1}{k_2} \mink{q}{p_2} \mink{p_1}{p_2} - 
  \mink{p_2}{k_1} \mink{p_1}{k_2} \mink{q}{p_1} +  \notag\\[1mm]
  &&\mink{p_2}{k_1} \mink{p_2}{k_2} \mink{q}{p_1} +
  \mink{p_1}{k_1} \mink{p_1}{k_2} \mink{q}{p_2} -  
  \mink{p_2}{k_1} \mink{p_1}{k_2} \mink{q}{p_2}\big] \\[2mm]
  T_{1 6} &=& -\frac{e^6 C_L a^4 |V_{11}|^2 |V_{21}|^2}{\mink{q}{p_1}} 
  \Delta_{\chi_2^+}(p_1,k_1) \Delta_{\chi_1^+}(p_2,k_2) \Delta_{\chi_2^+}(p_2,k_2)\notag\\[1mm]
  &&\big[m_{\chi^+_2}^2 \mink{q}{k_2} \mink{p_1}{p_2} +  
  m_{\chi^+_2}^2 \mink{q}{p_1} \mink{p_2}{k_2} - 
  m_{\chi^+_2}^2 \mink{q}{p_2} \mink{p_1}{k_2} - 
  4\mink{p_1}{k_1} \mink{p_1}{k_2} \mink{p_2}{k_2} + \notag\\[1mm] 
  &&4\mink{p_1}{k_1} \mink{p_2}{k_2} \mink{q}{k_2} + 
  2m_{\tilde{\nu}}^2 \mink{p_1}{k_1} \mink{p_1}{p_2} - \
  2m_{\tilde{\nu}}^2 \mink{p_1}{k_1} \mink{q}{p_2}\big] \\[2mm] 
  T_{1 7} &=& -\frac{e^6 C_L a^2 c f |V_{11}|^2}{2 \mink{q}{p_1}} 
  \Delta_{\chi_1^+}(p_2,k_2) \real \{\Delta_Z(k_1,k_2)\} \notag\\[1mm]
  &&\big[\mink{q}{k_1} \mink{p_2}{k_2} - 
  2\mink{q}{k_2} \mink{p_2}{k_2} + \mink{p_2}{k_1} \mink{q}{k_2} + m_{\tilde{\nu}}^2 \mink{q}{p_2} 
  - \mink{k_1}{k_2} \mink{q}{p_2}\big] 
  \\[2mm] 
  T_{1 8} &=& -\frac{e^6 C_L a^2 c f |V_{11}|^2}{2\mink{q}{p_1} \mink{q}{p_2}} 
  \Delta_{\chi_1^+}(p_2,k_2) \real \{\Delta_Z(k_1,k_2)\}\notag\\[1mm]
  &&\big[-\mink{q}{p_2} \mink{p_1}{k_2} \mink{p_1}{k_2} + 
  \mink{p_2}{k_1} \mink{p_1}{p_2} \mink{p_1}{k_2} -  
  2\mink{p_2}{k_2} \mink{p_1}{p_2} \mink{p_1}{k_2} +
  \mink{q}{k_2} \mink{p_1}{p_2} \mink{p_1}{k_2} - \notag\\[1mm] 
  &&\mink{p_2}{k_1} \mink{q}{p_1} \mink{p_1}{k_2} + 
  \mink{p_2}{k_2} \mink{q}{p_1} \mink{p_1}{k_2} + 
  \mink{p_1}{k_1} \mink{q}{p_2} \mink{p_1}{k_2} - 
  \mink{p_2}{k_1} \mink{q}{p_2} \mink{p_1}{k_2} + \notag\\[1mm] 
  &&\mink{p_2}{k_2} \mink{q}{p_2} \mink{p_1}{k_2} +
  m_{\tilde{\nu}}^2 \mink{p_1}{p_2} \mink{p_1}{p_2} -  
  \mink{k_1}{k_2} \mink{p_1}{p_2} \mink{p_1}{p_2} + 
  \mink{p_1}{k_1} \mink{p_2}{k_2} \mink{p_1}{p_2} -\notag\\[1mm]
  &&\mink{q}{k_1} \mink{p_2}{k_2} \mink{p_1}{p_2} - 
  \mink{p_1}{k_1} \mink{q}{k_2} \mink{p_1}{p_2} +  
  \mink{p_2}{k_2} \mink{q}{k_2} \mink{p_1}{p_2} -
  \mink{p_2}{k_2} \mink{p_2}{k_2} \mink{q}{p_1} + \notag\\[1mm]
  &&\mink{p_2}{k_1} \mink{p_2}{k_2} \mink{q}{p_1} -  
  m_{\tilde{\nu}}^2 \mink{p_1}{p_2} \mink{q}{p_1} +
  \mink{k_1}{k_2} \mink{p_1}{p_2} \mink{q}{p_1} -  
  m_{\tilde{\nu}}^2 \mink{p_1}{p_2} \mink{q}{p_2} + \notag\\[1mm] 
  &&\mink{k_1}{k_2} \mink{p_1}{p_2} \mink{q}{p_2}\big]
  \\[2mm]
  T_{2 3} &=&-\frac{e^6 C_L a^4 |V_{11}|^4}{\mink{q}{p_2}} \Delta_{\chi_1^+}^2(p_1,k_1) 
  \Delta_{\chi_1^+}(p_2,k_2)\notag\\[1mm]
  &&\big[m_{\chi^+_1}^2 \mink{q}{k_1} \mink{p_1}{p_2} - 
  m_{\chi^+_1}^2 \mink{p_2}{k_1} \mink{q}{p_1} +  
  m_{\chi^+_1}^2 \mink{p_1}{k_1} \mink{q}{p_2} -
  4\mink{p_1}{k_1} \mink{p_2}{k_1} \mink{p_2}{k_2} +  \notag\\[1mm]
  &&4\mink{p_1}{k_1} \mink{q}{k_1} \mink{p_2}{k_2} + 
  2m_{\tilde{\nu}}^2 \mink{p_2}{k_2} \mink{p_1}{p_2} - 
  2m_{\tilde{\nu}}^2 \mink{p_2}{k_2} \mink{q}{p_1}\big] \\[2mm]
  T_{2 4} &=& -\frac{e^6 C_L a^4 |V_{11}|^2 |V_{21}|^2}{\mink{q}{p_1} \mink{q}{p_2}}
  \Delta_{\chi_1^+}(p_1,k_1) \Delta_{\chi_2^+}(p_2,k_2)\notag\\[1mm]
  &&\big[-\mink{k_1}{k_2} \mink{p_1}{p_2} \mink{p_1}{p_2} + 
  \mink{p_2}{k_1} \mink{p_1}{k_2} \mink{p_1}{p_2} +  
  \mink{p_1}{k_1} \mink{p_2}{k_2} \mink{p_1}{p_2} -
  \mink{q}{k_1} \mink{p_2}{k_2} \mink{p_1}{p_2} - \notag\\[1mm] 
  &&\mink{p_1}{k_1} \mink{q}{k_2} \mink{p_1}{p_2} + 
  \mink{k_1}{k_2} \mink{q}{p_1} \mink{p_1}{p_2} + 
  \mink{k_1}{k_2} \mink{q}{p_2} \mink{p_1}{p_2} - 
  \mink{p_2}{k_1} \mink{p_1}{k_2} \mink{q}{p_1} +  \notag\\[1mm]
  &&\mink{p_2}{k_1} \mink{p_2}{k_2} \mink{q}{p_1} +
  \mink{p_1}{k_1} \mink{p_1}{k_2} \mink{q}{p_2} -  
  \mink{p_2}{k_1} \mink{p_1}{k_2} \mink{q}{p_2}\big] \\[2mm]
  T_{2 5} &=& \frac{e^6 C_L a^4 |V_{11}|^2 |V_{21}|^2}{ \mink{q}{p_2}} 
  \Delta_{\chi_1^+}(p_1,k_1) \Delta_{\chi_2^+}(p_1,k_1)
  \big[2\mink{p_1}{k_1} \mink{q}{k_1} - 
  m_{\tilde{\nu}}^2 \mink{q}{p_1}\big] \\[2mm] 
  T_{2 6} &=& - \frac{e^6 C_L a^4 |V_{11}|^2 |V_{21}|^2}{\mink{q}{p_2}}\Delta_{\chi_2^+}(p_1,k_1) 
  \Delta_{\chi_1^+}(p_1,k_1) \Delta_{\chi_2^+}(p_2,k_2)  \notag\\[1mm]
  &&\big[m_{\chi^+_2}^2 \mink{q}{k_1} \mink{p_1}{p_2} - 
  m_{\chi^+_2}^2 \mink{q}{p_1} \mink{p_2}{k_1} +  
  m_{\chi^+_2}^2 \mink{q}{p_2} \mink{p_1}{k_1} -
  4\mink{p_1}{k_1} \mink{p_2}{k_1} \mink{p_2}{k_2} + \notag\\[1mm] 
  &&4\mink{p_1}{k_1} \mink{p_2}{k_2} \mink{q}{k_1} + 
  2m_{\tilde{\nu}}^2 \mink{p_2}{k_2} \mink{p_1}{p_2} -  
  2m_{\tilde{\nu}}^2 \mink{p_2}{k_2} \mink{q}{p_1}\big]\\[2mm]
  T_{2 7} &=& \frac{e^6 C_L a^2 c f |V_{11}|^2}{2 \mink{q}{p_1} \mink{q}{p_2}} 
  \Delta_{\chi_1^+}(p_1,k_1) \real \{\Delta_Z(k_1,k_2)\}\notag\\[1mm]
  &&\big[\mink{q}{p_2} \mink{p_1}{k_1} \mink{p_1}{k_1} + 
  2\mink{p_2}{k_1} \mink{p_1}{p_2} \mink{p_1}{k_1} -  
  \mink{q}{k_1} \mink{p_1}{p_2} \mink{p_1}{k_1} -
  \mink{p_2}{k_2} \mink{p_1}{p_2} \mink{p_1}{k_1} + \notag\\[1mm] 
  &&\mink{q}{k_2} \mink{p_1}{p_2} \mink{p_1}{k_1} - 
  \mink{p_2}{k_1} \mink{q}{p_1} \mink{p_1}{k_1} -  
  \mink{p_2}{k_1} \mink{q}{p_2} \mink{p_1}{k_1} - 
  \mink{p_1}{k_2} \mink{q}{p_2} \mink{p_1}{k_1} - \notag\\[1mm] 
  &&m_{\tilde{\nu}}^2 \mink{p_1}{p_2} \mink{p_1}{p_2} + 
  \mink{k_1}{k_2} \mink{p_1}{p_2} \mink{p_1}{p_2} -  
  \mink{p_2}{k_1} \mink{q}{k_1} \mink{p_1}{p_2} - 
  \mink{p_2}{k_1} \mink{p_1}{k_2} \mink{p_1}{p_2} + \notag\\[1mm] 
  &&\mink{q}{k_1} \mink{p_2}{k_2} \mink{p_1}{p_2} + 
  \mink{p_2}{k_1} \mink{p_2}{k_1} \mink{q}{p_1} +  
  \mink{p_2}{k_1} \mink{p_1}{k_2} \mink{q}{p_1} - 
  \mink{p_2}{k_1} \mink{p_2}{k_2} \mink{q}{p_1} + \notag\\[1mm] 
  &&m_{\tilde{\nu}}^2 \mink{p_1}{p_2} \mink{q}{p_1} - 
  \mink{k_1}{k_2} \mink{p_1}{p_2} \mink{q}{p_1} +  
  \mink{p_2}{k_1} \mink{q}{p_2} \mink{p_1}{k_2} +  
  m_{\tilde{\nu}}^2 \mink{p_1}{p_2} \mink{q}{p_2} - \notag\\[1mm]
  &&\mink{k_1}{k_2} \mink{p_1}{p_2} \mink{q}{p_2}\big]
\\[2mm]
  T_{2 8} &=& \frac{e^6 C_L a^2 c f |V_{11}|^2}{2 \mink{q}{p_2}} \Delta_{\chi_1^+}(p_1,k_1) 
  \real \{\Delta_Z(k_1,k_2)\} \notag\\[1mm]
  &&\big[- \mink{q}{k_1} \mink{p_1}{k_2} + 
  2\mink{q}{k_1} \mink{p_1}{k_1}  - \mink{p_1}{k_1} \mink{q}{k_2} - m_{\tilde{\nu}}^2 \mink{q}{p_1} \
  + \mink{k_1}{k_2} \mink{q}{p_1}\big] 
  \\[2mm] 
  T_{3 4} &=& -\frac{e^6 C_L a^4 |V_{11}|^2 |V_{21}|^2}{ \mink{q}{p_1}} \Delta_{\chi_1^+}(p_1,k_1) 
  \Delta_{\chi_1^+}(p_2,k_2) \Delta_{\chi_2^+}(p_2,k_2)\notag\\[1mm]
  &&\big[m_{\chi^+_1}^2 (\mink{q}{k_2} \mink{p_1}{p_2} + 
                        \mink{p_2}{k_2} \mink{q}{p_1} - 
                       \mink{p_1}{k_2} \mink{q}{p_2}) - 
  4\mink{p_1}{k_1} \mink{p_2}{k_2} (\mink{p_1}{k_2} - \mink{q}{k_2}) + \notag\\[1mm] 
  &&2m_{\tilde{\nu}}^2 \mink{p_1}{k_1} (\mink{p_1}{p_2} - \mink{q}{p_2})\big] \\[2mm]
  T_{3 5} &=&  -\frac{e^6 C_L a^4 |V_{11}|^2 |V_{21}|^2}{\mink{q}{p_2}}\Delta_{\chi_1^+}(p_1,k_1) 
  \Delta_{\chi_1^+}(p_2,k_2) \Delta_{\chi_2^+}(p_1,k_1)\notag\\[1mm]
  &&\big[m_{\chi^+_1}^2 (\mink{q} {k_1} \mink{p_1}{p_2} + 
  \mink{p_1}{k_1} \mink{q}{p_2} - \mink{p_2}{k_1} \mink{q}{p_1}) - 
  4\mink{p_1}{k_1} \mink{p_2}{k_2} (\mink{p_2}{k_1} - \mink{q}{k_1}) + \notag\\[1mm] 
  &&2m_{\tilde{\nu}}^2 \mink{p_2}{k_2} (\mink{p_1}{p_2} - \mink{q}{p_1})\big] \\[2mm]
  T_{3 6} &=& 2{e^6 C_L a^4 |V_{11}|^2 |V_{21}|^2} \Delta_{\chi_1^+}(p_1,k_1) \Delta_{\chi_1^+}(p_2,k_2) 
  \Delta_{\chi_2^+}(p_1,k_1) \Delta_{\chi_2^+}(p_2,k_2)\notag\\[1mm]
  &&\big[2\mink{p_1}{k_1} \mink{p_2}{k_2} (m_{\chi^+_1}^2 + m_{\chi^+_2}^2 + 
  2\mink{k_1}{k_2}) + m_{\chi^+_1}^2 m_{\chi^+_2}^2 \mink{p_1}{p_2} - \notag\\[1mm] 
  &&2m_{\tilde{\nu}}^2 (\mink{p_1}{k_1} \mink{p_2}{k_1} + \mink{p_1}{k_2} \mink{p_2}{k_2}) + 
  m_{\tilde{\nu}}^4 \mink{p_1}{p_2}\big] \\[2mm] 
  T_{3 7} &=& \frac{e^6 C_L a^2 c f |V_{11}|^2}{2\mink{q}{p_1}} \Delta_{\chi_1^+}(p_1,k_1) 
  \Delta_{\chi_1^+}(p_2,k_2) \real \{\Delta_Z(k_1,k_2)\}\notag\\[1mm]
  &&\hspace*{-2mm}\big[m_{\chi^+_1}^2 (\mink{q}{k_1} \mink{p_1}{p_2} - 
  \mink{q}{k_2} \mink{p_1}{p_2} +  
  \mink{p_2}{k_1} \mink{q}{p_1} - 
  \mink{p_2}{k_2} \mink{q}{p_1} - 
  \mink{p_1}{k_1} \mink{q}{p_2} + \notag\\[1mm]
  &&\mink{p_1}{k_2} \mink{q}{p_2}) - 
  2\mink{p_1}{k_1} \mink{p_2}{k_1} \mink{p_1}{k_2} -
  2\mink{p_1}{k_1} \mink{p_1}{k_1} \mink{p_2}{k_2} +  
  2\mink{p_1}{k_1} \mink{p_2}{k_2} \mink{q}{k_1} + \notag\\[1mm]
  &&4\mink{p_1}{k_1} \mink{p_2}{k_2} \mink{p_1}{k_2} + 
  2\mink{p_1}{k_1} \mink{p_2}{k_1} \mink{q}{k_2} - 
  4\mink{p_1}{k_1} \mink{p_2}{k_2} \mink{q}{k_2} -  
  2m_{\tilde{\nu}}^2 \mink{p_1}{p_2} \mink{p_1}{k_1} +\notag\\[1mm]
  &&2\mink{k_1}{k_2} \mink{p_1}{k_1} \mink{p_1}{p_2} +
  2m_{\tilde{\nu}}^2 \mink{p_1}{k_1} \mink{q}{p_2} - 
  2\mink{k_1}{k_2} \mink{p_1}{k_1} \mink{q}{p_2}\big]  
  \\[2mm]
  T_{3 8} &=&  -\frac{e^6 C_L a^2 c f |V_{11}|^2}{2\mink{q}{p_2}}\Delta_{\chi_1^+}(p_1,k_1)
  \Delta_{\chi_1^+}(p_2,k_2) \real\{\Delta_Z(k_1,k_2)\} \notag\\[1mm]
  &&\hspace*{-2mm}\big[m_{\chi^+_1}^2 (\mink{q}{k_1} \mink{p_1}{p_2} - 
  \mink{q}{k_2} \mink{p_1}{p_2} -  
  \mink{p_2}{k_1} \mink{q}{p_1} + 
  \mink{p_2}{k_2} \mink{q}{p_1} + 
  \mink{p_1}{k_1} \mink{q}{p_2} - \notag\\[1mm]
  &&\mink{p_1}{k_2} \mink{q}{p_2}) +   
  2\mink{p_1}{k_1} \mink{p_2}{k_2} \mink{p_2}{k_2} -
  4\mink{p_1}{k_1} \mink{p_2}{k_1} \mink{p_2}{k_2} +  
  4\mink{p_1}{k_1} \mink{p_2}{k_2} \mink{q}{k_1} +\notag\\[1mm] 
  &&2\mink{p_2}{k_1} \mink{p_1}{k_2} \mink{p_2}{k_2} -  
  2\mink{p_1}{k_2} \mink{p_2}{k_2} \mink{q}{k_1} - 
  2\mink{p_1}{k_1} \mink{p_2}{k_2} \mink{q}{k_2} +  
  2m_{\tilde{\nu}}^2 \mink{p_1}{p_2} \mink{p_2}{k_2} -\notag\\[1mm] 
  &&2\mink{k_1}{k_2} \mink{p_2}{k_2} \mink{p_1}{p_2} -  
  2m_{\tilde{\nu}}^2 \mink{p_2}{k_2} \mink{q}{p_1} + 
  2\mink{k_1}{k_2} \mink{p_2}{k_2} \mink{q}{p_1}\big] 
  \\[2mm]
  T_{4 5} &=& -\frac{e^6 C_L a^4 |V_{21}|^4}{\mink{q}{p_1} \mink{q}{p_2}}\
  \Delta_{\chi_2^+}(p_1,k_1) \Delta_{\chi_2^+}(p_2,k_2) \notag\\[1mm]
   &&\big[-\mink{k_1}{k_2} \mink{p_1}{p_2} \mink{p_1}{p_2} + 
  \mink{p_1}{k_2} \mink{p_2}{k_1} \mink{p_1}{p_2} +  
  \mink{p_1}{k_1} \mink{p_2}{k_2} \mink{p_1}{p_2} -
  \mink{q}{k_1} \mink{p_2}{k_2} \mink{p_1}{p_2} -  \notag\\[1mm] 
  &&\mink{p_1}{k_1} \mink{q}{k_2} \mink{p_1}{p_2} + 
  \mink{k_1}{k_2} \mink{q}{p_1} \mink{p_1}{p_2} + 
  \mink{k_1}{k_2} \mink{q}{p_2} \mink{p_1}{p_2} - 
  \mink{p_2}{k_1} \mink{p_1}{k_2} \mink{q}{p_1} + \notag\\[1mm]   
  &&\mink{p_2}{k_1} \mink{p_2}{k_2} \mink{q}{p_1} +
  \mink{p_1}{k_1} \mink{p_1}{k_2} \mink{q}{p_2} -  
  \mink{p_2}{k_1} \mink{p_1}{k_2} \mink{q}{p_2}\big] \\[2mm]
  T_{4 6} &=& -\frac{e^6 C_L a^4 |V_{21}|^4}{\mink{q}{p_1}}
  \Delta_{\chi_2^+}(p_1,k_1) \Delta_{\chi_2^+}^2(p_2,k_2) \notag\\[1mm]
  &&\big[m_{\chi^+_2}^2 \mink{q}{k_2} \mink{p_1}{p_2} + 
  m_{\chi^+_2}^2 \mink{q}{p_1} \mink{p_2}{k_2} -  
  m_{\chi^+_2}^2 \mink{q}{p_2} \mink{p_1}{k_2} -
  4\mink{p_1}{k_1} \mink{p_1}{k_2} \mink{p_2}{k_2} + \notag\\[1mm] 
  &&4\mink{p_1}{k_1} \mink{p_2}{k_2} \mink{q}{k_2} + 
  2m_{\tilde{\nu}}^2 \mink{p_1}{k_1} \mink{p_1}{p_2} - 
  2m_{\tilde{\nu}}^2 \mink{p_1}{k_1} \mink{q}{p_2}\big]  \\[2mm]
  T_{4 7} &=&  -\frac{e^6 C_L a^2 c f |V_{21}|^2}{2 \mink{q}{p_1}}\Delta_{\chi_2^+}(p_2,k_2) 
  \real \{\Delta_Z(k_1,k_2)\}\notag\\[1mm]
  &&[\mink{q}{k_1} \mink{p_2}{k_2} - 
  2\mink{q}{k_2} \mink{p_2}{k_2} + \mink{p_2}{k_1} \mink{q}{k_2} + m_{\tilde{\nu}}^2 \mink{q}{p_2} - 
  \mink{k_1}{k_2} \mink{q}{p_2}] 
  \\[2mm] 
  T_{4 8} &=&  -\frac{e^6 C_L a^2 c f |V_{21}|^2}{2\mink{q}{p_1} \mink{q}{p_2}}\Delta_{\chi_2^+}(p_2,k_2) 
  \real \{\Delta_Z(k_1,k_2)\}\notag\\[1mm]
  &&\big[-\mink{q}{p_2} \mink{p_1}{k_2} \mink{p_1}{k_2} + 
  \mink{p_2}{k_1} \mink{p_1}{p_2} \mink{p_1}{k_2} - 
  2\mink{p_2}{k_2} \mink{p_1}{p_2} \mink{p_1}{k_2} +
  \mink{q}{k_2} \mink{p_1}{p_2} \mink{p_1}{k_2} - \notag\\[1mm]
  &&\mink{p_2}{k_1} \mink{q}{p_1} \mink{p_1}{k_2} + 
  \mink{p_2}{k_2} \mink{q}{p_1} \mink{p_1}{k_2} + 
  \mink{p_1}{k_1} \mink{q}{p_2} \mink{p_1}{k_2} - 
  \mink{p_2}{k_1} \mink{q}{p_2} \mink{p_1}{k_2} + \notag\\[1mm] 
  &&\mink{p_2}{k_2} \mink{q}{p_2} \mink{p_1}{k_2} +
  m_{\tilde{\nu}}^2 \mink{p_1}{p_2} \mink{p_1}{p_2} -  
  \mink{k_1}{k_2} \mink{p_1}{p_2} \mink{p_1}{p_2} + 
  \mink{p_1}{k_1} \mink{p_2}{k_2} \mink{p_1}{p_2} -\notag\\[1mm] 
  &&\mink{q}{k_1} \mink{p_2}{k_2} \mink{p_1}{p_2} - 
  \mink{p_1}{k_1} \mink{q}{k_2} \mink{p_1}{p_2} +  
  \mink{p_2}{k_2} \mink{q}{k_2} \mink{p_1}{p_2} -
  \mink{p_2}{k_2} \mink{p_2}{k_2} \mink{q}{p_1} + \notag\\[1mm] 
  &&\mink{p_2}{k_1} \mink{p_2}{k_2} \mink{q}{p_1} - 
  m_{\tilde{\nu}}^2 \mink{p_1}{p_2} \mink{q}{p_1} + 
  \mink{k_1}{k_2} \mink{p_1}{p_2} \mink{q}{p_1} - 
  m_{\tilde{\nu}}^2 \mink{p_1}{p_2} \mink{q}{p_2} +  \notag\\[1mm]
  &&\mink{k_1}{k_2} \mink{p_1}{p_2} \mink{q}{p_2}\big] 
  \\[2mm]
  T_{5 6} &=& -\frac{e^6 C_L a^4 |V_{21}|^4}{\mink{q}{p_2}}\Delta_{\chi_2^+}^2(p_1,k_1) 
  \Delta_{\chi_2^+}(p_2,k_2)\notag\\[1mm]
  &&\big[m_{\chi^+_2}^2 \mink{q}{k_1} \mink{p_1}{p_2} - 
  m_{\chi^+_2}^2 \mink{p_2}{k_1} \mink{q}{p_1} +   
  m_{\chi^+_2}^2 \mink{p_1}{k_1} \mink{q}{p_2} -
  4\mink{p_1}{k_1} \mink{p_2}{k_1} \mink{p_2}{k_2} +  \notag\\[1mm]
  &&4\mink{p_1}{k_1} \mink{q}{k_1} \mink{p_2}{k_2} + 
  2m_{\tilde{\nu}}^2 \mink{p_2}{k_2} \mink{p_1}{p_2} - 
  2m_{\tilde{\nu}}^2 \mink{p_2}{k_2} \mink{q}{p_1}\big] \\[2mm]
  T_{5 7} &=& \frac{e^6 C_L a^2 c f |V_{21}|^2}{2\mink{q}{p_1} \mink{q}{p_2}}
  \Delta_{\chi_2^+}(p_1,k_1) \real \{\Delta_Z(k_1,k_2)\}\notag\\[1mm] 
  &&\big[\mink{q}{p_2} \mink{p_1}{k_1} \mink{p_1}{k_1} +  
  2\mink{p_2}{k_1} \mink{p_1}{p_2} \mink{p_1}{k_1} - 
  \mink{q}{k_1} \mink{p_1}{p_2} \mink{p_1}{k_1} - 
  \mink{p_2}{k_2} \mink{p_1}{p_2} \mink{p_1}{k_1} +\notag\\[1mm] 
  &&\mink{q}{k_2} \mink{p_1}{p_2} \mink{p_1}{k_1} -  
  \mink{p_2}{k_1} \mink{q}{p_1} \mink{p_1}{k_1} - 
  \mink{p_2}{k_1} \mink{q}{p_2} \mink{p_1}{k_1} -  
  \mink{p_1}{k_2} \mink{q}{p_2} \mink{p_1}{k_1} - \notag\\[1mm]
  &&m_{\tilde{\nu}}^2 \mink{p_1}{p_2} \mink{p_1}{p_2} +  
  \mink{k_1}{k_2} \mink{p_1}{p_2} \mink{p_1}{p_2} - 
  \mink{p_2}{k_1} \mink{q}{k_1} \mink{p_1}{p_2} -  
  \mink{p_2}{k_1} \mink{p_1}{k_2} \mink{p_1}{p_2} + \notag\\[1mm]
  &&\mink{q}{k_1} \mink{p_2}{k_2} \mink{p_1}{p_2} +  
  \mink{p_2}{k_1} \mink{p_2}{k_1} \mink{q}{p_1} + 
  \mink{p_2}{k_1} \mink{p_1}{k_2} \mink{q}{p_1} -  
  \mink{p_2}{k_1} \mink{p_2}{k_2} \mink{q}{p_1} + \notag\\[1mm]
  &&m_{\tilde{\nu}}^2 \mink{p_1}{p_2} \mink{q}{p_1} -  
  \mink{k_1}{k_2} \mink{p_1}{p_2} \mink{q}{p_1} + 
  \mink{p_2}{k_1} \mink{q}{p_2} \mink{p_1}{k_2} +  
  m_{\tilde{\nu}}^2 \mink{p_1}{p_2} \mink{q}{p_2} - \notag\\[1mm]
  &&\mink{k_1}{k_2} \mink{p_1}{p_2} \mink{q}{p_2}\big] 
  \\[2mm]
  T_{5 8} &=& \frac{e^6 C_L a^2 c f |V_{21}|^2}{2 \mink{q}{p_2}} \Delta_{\chi_2^+}(p_1,k_1) 
  \real \{\Delta_Z(k_1,k_2)\}\notag\\[1mm]
  &&\big[2\mink{q}{k_1} \mink{p_1}{k_1} - \mink{q}{k_1} \mink{p_1}{k_2} - 
  \mink{p_1}{k_1} \mink{q}{k_2} - m_{\tilde{\nu}}^2 \mink{q}{p_1} +  
  \mink{k_1}{k_2} \mink{q}{p_1}\big] 
  \\[2mm] 
  T_{6 7} &=&  \frac{e^6 C_L a^2 c f |V_{21}|^2}{2 \mink{q}{p_1}}
  \Delta_{\chi_2^+}(p_1,k_1) \Delta_{\chi_2^+}(p_2,k_2) \real \{\Delta_Z(k_1,k_2)\}\notag\\[1mm]
  &&\big[m_{\chi^+_2}^2 (\mink{q}{k_1} \mink{p_1}{p_2} - 
  \mink{q}{k_2} \mink{p_1}{p_2} +  
  \mink{p_2}{k_1} \mink{q}{p_1} - 
  \mink{p_2}{k_2} \mink{q}{p_1} - 
  \mink{p_1}{k_1} \mink{q}{p_2} + \notag\\[1mm] 
  &&\mink{p_1}{k_2} \mink{q}{p_2}) - 
  2\mink{p_1}{k_1} \mink{p_2}{k_1} \mink{p_1}{k_2} -
  2\mink{p_1}{k_1} \mink{p_1}{k_1} \mink{p_2}{k_2} +  
  2\mink{p_1}{k_1} \mink{p_2}{k_2} \mink{q}{k_1} + \notag\\[1mm] 
  &&4\mink{p_1}{k_1} \mink{p_2}{k_2} \mink{p_1}{k_2} +
  2\mink{p_1}{k_1} \mink{p_2}{k_1} \mink{q}{k_2} - 
  4\mink{p_1}{k_1} \mink{p_2}{k_2} \mink{q}{k_2} -  
  2m_{\tilde{\nu}}^2 \mink{p_1}{p_2} \mink{p_1}{k_1} +\notag\\[1mm] 
  &&2\mink{k_1}{k_2} \mink{p_1}{k_1} \mink{p_1}{p_2} + 
  2m_{\tilde{\nu}}^2 \mink{p_1}{k_1} \mink{q}{p_2} -  
  2\mink{k_1}{k_2} \mink{p_1}{k_1} \mink{q}{p_2}\big] 
  \\[2mm]
  T_{6 8} &=&  -\frac{e^6 C_L a^2 c f |V_{21}|^2}{2 \mink{q}{p_2}} \Delta_{\chi_2^+}(p_1,k_1) 
  \Delta_{\chi_2^+}(p_2,k_2) \real \{\Delta_Z(k_1,k_2)\}\notag\\[1mm]
  &&\big[m_{\chi^+_2}^2 (\mink{q}{k_1} \mink{p_1}{p_2} - 
  \mink{q}{k_2} \mink{p_1}{p_2} -  
  \mink{p_2}{k_1} \mink{q}{p_1} + 
  \mink{p_2}{k_2} \mink{q}{p_1} + 
  \mink{p_1}{k_1} \mink{q}{p_2} - \notag\\[1mm] 
  &&\mink{p_1}{k_2} \mink{q}{p_2}) + 
  2\mink{p_1}{k_1} \mink{p_2}{k_2} \mink{p_2}{k_2} -
  4\mink{p_1}{k_1} \mink{p_2}{k_1} \mink{p_2}{k_2} +  
  4\mink{p_1}{k_1} \mink{p_2}{k_2} \mink{q}{k_1} + \notag\\[1mm]  
  &&2\mink{p_2}{k_1} \mink{p_1}{k_2} \mink{p_2}{k_2} - 
  2\mink{p_1}{k_2} \mink{p_2}{k_2} \mink{q}{k_1} - 
  2\mink{p_1}{k_1} \mink{p_2}{k_2} \mink{q}{k_2} +  
  2m_{\tilde{\nu}}^2 \mink{p_1}{p_2} \mink{p_2}{k_2} -\notag\\[1mm] 
  &&2\mink{k_1}{k_2} \mink{p_2}{k_2} \mink{p_1}{p_2} -
  2m_{\tilde{\nu}}^2 \mink{p_2}{k_2} \mink{q}{p_1} + 
  2\mink{k_1}{k_2} \mink{p_2}{k_2} \mink{q}{p_1}\big]  
  \\[2mm]
  T_{7 8} &=& 3\frac{e^6 f^2 (C_L c^2 + C_R d^2)}{4\mink{q}{p_1} \mink{q}{p_2}} |\Delta_Z(k_1,k_2)|^2\notag\\[1mm]
  &&\big[\mink{p_1}{k_1} \big(\mink{p_1}{k_1} \mink{q}{p_2} + 
                        2\mink{p_2}{k_1} \mink{p_1}{p_2} -  
                         \mink{q}{k_1} \mink{p_1}{p_2} -
                        2\mink{p_2}{k_2} \mink{p_1}{p_2} +  
                        \mink{q}{k_2} \mink{p_1}{p_2} - \notag\\[1mm]
                        &&\hspace*{20mm}\mink{p_2}{k_1} \mink{q}{p_1} +
                        \mink{p_2}{k_2} \mink{q}{p_1} - 
                        \mink{p_2}{k_1} \mink{q}{p_2} - 
                        2\mink{p_1}{k_2} \mink{q}{p_2} +
                        \mink{p_2}{k_2} \mink{q}{p_2}\big) +\notag\\[1mm]
  &&\mink{p_1}{p_2} \big(-2m_{\tilde{\nu}}^2 \mink{p_1}{p_2} + 
                   2\mink{k_1}{k_2} \mink{p_1}{p_2} -  
                   \mink{p_2}{k_1} \mink{q}{k_1} -
                   2\mink{p_2}{k_1} \mink{p_1}{k_2} + 
                   \mink{q}{k_1} \mink{p_1}{k_2} + \notag\\[1mm]
                   &&\hspace*{20mm}\mink{q}{k_1} \mink{p_2}{k_2} +
                   2\mink{p_1}{k_2} \mink{p_2}{k_2} + 
                   \mink{p_2}{k_1} \mink{q}{k_2} - 
                   \mink{p_1}{k_2} \mink{q}{k_2} -
                   \mink{p_2}{k_2} \mink{q}{k_2}\big) +\notag\\[1mm]
  &&\mink{q}{p_1} \big(\mink{p_2}{k_1} \mink{p_2}{k_1} + 
                 \mink{p_2}{k_2} \mink{p_2}{k_2} +  
                 \mink{p_2}{k_1} \mink{p_1}{k_2} -
                 2\mink{p_2}{k_1} \mink{p_2}{k_2} - \notag\\[1mm]
                 &&\hspace*{20mm}\mink{p_1}{k_2} \mink{p_2}{k_2} + 
                 2m_{\tilde{\nu}}^2 \mink{p_1}{p_2} -
                 2\mink{k_1}{k_2} \mink{p_1}{p_2}\big) +\notag\\[1mm]
  &&\mink{q}{p_2} \big(\mink{p_1}{k_2} \mink{p_1}{k_2} + 
                 \mink{p_2}{k_1} \mink{p_1}{k_2} - 
                 \mink{p_1}{k_2} \mink{p_2}{k_2} + 
                 2m_{\tilde{\nu}}^2 \mink{p_1}{p_2} -
                 2\mink{k_1}{k_2} \mink{p_1}{p_2}\big)\big] 
\end{eqnarray}
Formulas for the squared amplitudes for radiative sneutrino production
can also be found in Refs.~\cite{Franke:thesis,Franke:1994ph} for
longitudinal and transverse beam polarisations. Here, we give however
our calculated amplitudes for completeness.
We have calculated the squared amplitudes with \texttt{FeynCalc}~\cite{Kublbeck:1992mt}.
We neglect terms proportional to 
$\eps \imag \{\Delta_Z\}$,
see the discussion at the end of Appendix~\ref{sec:app:chifore}.

\section{Definition of the Differential Cross 
Section and Phase Space}
\label{sec:phasespace}
We present some details of the phase space calculation for radiative neutralino production
\begin{equation}
\label{eq:scatter}
e^-(p_1) + e^+(p_2) \rightarrow \tilde{\chi}^0_1(k_1) +  
\tilde{\chi}^0_1(k_2) + \gamma(q). 
\end{equation} 
The differential cross section for (\ref{eq:scatter}) is given
by~\cite{Eidelman:2004wy}
\begin{eqnarray}
\dif \sigma &=& \half \frac{(2\pi)^4}{2 s}
\prod_f \frac{\dif^3 \mathbf{p}_f}{(2\pi)^3 2E_f}\delta^{(4)}(p_1 + 
p_2 - k_1 - k_2 - q)|\M|^2,
\label{phasespace}
\end{eqnarray}
where $\mathbf{p}_f$ and $E_f$ denote the final three-momenta
and the final energies
of the neutralinos and the photon.  The squared matrix element $|\M|^2$ is given in
Appendix~\ref{sec:app:chifore}.

\subsection{Parametrisation of Momenta and Phase Space 
in the Neutralino System}

We parametrise the four-momenta in the center-of-mass (cms) system of the
incoming particles, which we call the laboratory (lab) system. The beam
momenta  are then parametrised as 
\begin{eqnarray}
p_1 &={\displaystyle \half} 
\begin{pmatrix}
\sqrt{s}, & 0, & 0, & \sqrt{s}
\end{pmatrix},\qquad
p_2 &={\displaystyle \half}
\begin{pmatrix}
\sqrt{s}, & 0, & 0, & -\sqrt{s}
\end{pmatrix}.
\end{eqnarray}
For the outgoing neutralinos and the photon we consider in a first
step the local center-of-mass system of the two neutralinos.  The
photon shall escape along this $x_3$-axis. We start with general
momentum-vectors for the two neutralinos, boost them along their
$x_3$-axis and rotate them around the $x_1$-axis to reach the lab
system.  Note that the three-momenta of the outgoing particles lie in
a plane whose normal vector is inclined by an angle $\theta$ towards
the beam axis.  We parametrise the neutralino momenta in their cms
frame~\cite{Grassie:1983kq}
\begin{eqnarray}
\label{eq:momenta}
k_1^\ast &=&
 \begin{pmatrix}
\frac{1}{2}\sqrt{s^\ast}\\
\phantom{-}k^\ast \sin\theta^\ast\cos\phi^\ast \\
\phantom{-}k^\ast \sin\theta^\ast\sin\phi^\ast \\
\phantom{-}k^\ast \cos\theta^\ast 
\end{pmatrix},\qquad 
k_2^\ast = 
\begin{pmatrix}
\frac{1}{2}\sqrt{s^\ast}\\
-k^\ast \sin\theta^\ast\cos\phi^\ast \\
-k^\ast \sin\theta^\ast\sin\phi^\ast \\
-k^\ast \cos\theta^\ast
\end{pmatrix}, 
\end{eqnarray}
with the local cms energy $s^\ast$ of the two neutralinos
\begin{equation}
s^\ast = (k_1+k_2)^2 = 2 m_{\chi_1^0}^2 + 2\mink{k_1\,}{\,k_2},
\end{equation}
the polar angle $\theta^\ast$, the azimuthal angle $\phi^\ast$ and the
absolute value of the neutralino three-momenta $k^\ast$ in their cms
frame. These momenta are boosted to the lab system with the Lorentz
transformation
\begin{eqnarray}
\label{eq:lorentz}
L(\beta) = 
\begin{pmatrix}
\gamma & 0 & 0 & \gamma \beta \\
0 & 1 & 0 & 0 \\
0 & 0 & 1 & 0 \\
 \gamma \beta & 0 & 0 & \gamma
\end{pmatrix} ,
\end{eqnarray}
with $\gamma = \frac{1}{\sqrt{1-\beta^2}}$ and 
$\beta = \frac{|{\bf k_1 + k_2}|}{(k_1)^0 + (k_2)^0}_{|\mathrm{cms\,\, beam}}$ 
the boost velocity from the cms to the lab system
\begin{eqnarray}
\beta = \frac{|\mathbf{q}|}{\sqrt{s} -E_\gamma} = \frac{s - s^\ast}{s + s^\ast}.
\end{eqnarray}
Boosting the momenta $k_1^\ast$ and $k_2^\ast$, see
Eq.~(\ref{eq:momenta}), at first with the Lorentz transformation
Eq.~(\ref{eq:lorentz}) and then rotating with $\theta$ yields the
neutralino and photon momenta in the lab system~\cite{Grassie:1983kq}
\begin{eqnarray}
k_1 &=& 
\begin{pmatrix}
\gamma E^\ast + \beta\gamma k^\ast \cos\theta^\ast\\[1mm]
k^\ast\sin \theta^\ast \cos \phi^\ast\\[1mm]
k^\ast\sin \theta^\ast \sin \phi^\ast \cos\theta + 
(\beta \gamma E^\ast + \gamma k^\ast \cos\theta^\ast) \sin\theta\\[1mm]
 -k^\ast\sin \theta^\ast \sin \phi^\ast \sin\theta + (\beta \gamma 
E^\ast + \gamma k^\ast \cos\theta^\ast) \cos\theta
\end{pmatrix},\\[3mm]
k_2 &=&
\begin{pmatrix}
\gamma E^\ast - \beta\gamma k^\ast \cos\theta^\ast\\[1mm]
-k^\ast\sin \theta^\ast \cos \phi^\ast\\[1mm]
-k^\ast\sin \theta^\ast \sin \phi^\ast \cos\theta + 
(\beta\gamma E^\ast - \gamma k^\ast \cos\theta^\ast) \sin\theta\\[1mm]
 k^\ast\sin \theta^\ast \sin \phi^\ast \sin\theta + (\beta\gamma E^\ast 
- \gamma k^\ast \cos\theta^\ast) \cos\theta
\end{pmatrix},\\[3mm]
q &=&
\begin{pmatrix}
\frac{s-s^\ast}{2\sqrt{s}}\\[1mm]
0\\[1mm]
-\frac{s-s^\ast}{2\sqrt{s}}\sin\theta\\[1mm]
-\frac{s-s^\ast}{2\sqrt{s}}\cos\theta
\end{pmatrix},
\end{eqnarray} 
with
\begin{eqnarray}
k^\ast &=& \frac{1}{2}\sqrt{s^\ast-4 m_{\chi_1^0}^2},\\
E^\ast &= &\frac{\sqrt{s^\ast}}{2},\\ 
\beta\gamma &=&  \frac{s - s^\ast}{2 \sqrt{ss^\ast}} \enspace.
\end{eqnarray} 
The differential cross section for \signal~now reads~\cite{Grassie:1983kq}
\begin{eqnarray}
\dif\sigma &=& \frac{1}{4096 \pi^4 s}\left(1 - \frac{s^\ast}{s}\right)
\sqrt{1 - \frac{4 m_{\chi_1^0}^2}{s^\ast}}|\M|^2\,\dif\!\cos\theta \, \dif\!\cos\theta^\ast \, 
\dif\phi^\ast \, \dif s^\ast,
\end{eqnarray}
where the integration variables run over
\begin{equation}
\begin{array}{rcccl}
 0 & \le & \phi^\ast &\le & 2 \pi, \\[1mm]
  -1 & \le &\cos\theta^\ast &\le &1,\\[1mm]
  4 m_{\chi_1^0}^2 &\le &    s^\ast &\le &(1-x)s,\quad x = \frac{E_
\gamma}{E_{\mathrm{beam}}},\\[1mm]
   -0.99 & \le &\cos\theta & \le& 0.99.         
\end{array}     
\end{equation}

\subsection{Alternative Parametrisation in the 
Center-of-Mass System}

For the radiative production of neutralinos $ e^+e^-\to\tilde\chi_i^0
\tilde\chi_j^0\gamma$, we choose a coordinate frame in the 
center-of-mass system, such that the momentum of the photon ${\bf
p}_\gamma$ points in the $z$-direction. The scattering angle is
$\theta_\gamma \angle ({\mathbf p}_{e^-},{\mathbf p}_\gamma)$, whereas
the azimuthal angle $\phi_\gamma$ can be chosen zero. The four-momenta
are
\begin{eqnarray} 
p_{e^-}^{\mu} &=& E_b(1,\sin\theta_\gamma,0,
\cos\theta_\gamma),\qquad p_{e^+}^{\mu} =
E_b(1,-\sin\theta_\gamma,0,-\cos\theta_\gamma),\\ [3mm]
 p_\gamma^{\mu} &=&
E_\gamma(1,0,0,1), \qquad p_{\chi_i}^{\mu} = |{\bf p}_i| (E_i/|{\bf
p}_i|,\sin\theta_i\cos\phi_i,\sin\theta_i\sin\phi_i,\cos\theta_i),\\[3mm]
p_{\chi_j}^{\mu}&=&(E_j,{\bf p}_j),\qquad \;\;\; \;\;\;E_j=2E_b-E_\gamma-E_i,\qquad
{\bf p}_j=-{\bf p}_\gamma-{\bf p}_i,
\label{momentumEj}
\end{eqnarray}
with the beam energy $E_b=\sqrt{s}/2$,
and the neutralino angle $\theta_i \angle ({\mathbf p}_\gamma,{\mathbf p}_i)$  
fixed due to momentum conservation~(\ref{momentumEj})
\begin{eqnarray}
\cos\theta_i = 
\frac{m_i^2-m_j^2+E_j^2-E_i^2-E_\gamma^2}{2E_\gamma\sqrt{E_i^2-m_i^2}}.
\end{eqnarray}
The phase space integration for the cross section~(\ref{phasespace}) can then be written 
as~\cite{Fraas:1991ky,Bayer,Franke:thesis,Franke:1994ph}
\begin{eqnarray}
\sigma &=& (1-\frac{1}{2}\delta_{ij})\frac{1}{16 s (2 \pi)^4}
           \int \sin\theta_\gamma {\rm d} \theta_\gamma {\rm d} E_\gamma
           \int {\rm d} \phi_i  {\rm d} E_i \;
           |{\mathcal M}|^2.
\end{eqnarray}
For $m_i=m_j$ the integration bounds of the photon and neutralino energy are~\cite{Fraas:1991ky,Bayer}
\begin{eqnarray}
E_\gamma^{\rm max}&=&E_b\left(  1 - \frac{m_i^2}{E_b^2} \right),\\
E_i^{\rm min,max}&=& E_b-\frac{E_\gamma}{2}
      \left[ 1 \pm \sqrt{1-\frac{m_i^2}{E_b(E_b-E_\gamma)}}\;\right],
\end{eqnarray}
whereas the scattering angle $\theta_\gamma$ and the minimal photon energy $E_\gamma^{\rm min}$
have to be cut to regularise infrared and collinear divergencies,
see Eq.~(\ref{cuts}).
Similar formulas are obtained for the radiative production of
neutrinos~(\ref{productionNu}) and sneutrinos~(\ref{productionSneut}).

\end{appendix}

\end{document}